\documentclass[fleqn,usenatbib]{mnras}

\usepackage{newtxtext,newtxmath}

\usepackage[T1]{fontenc}
\usepackage{xspace,amsmath}

\DeclareRobustCommand{\VAN}[3]{#2}
\let\VANthebibliography\thebibliography
\def\thebibliography{\DeclareRobustCommand{\VAN}[3]{##3}\VANthebibliography}

\newcommand\textlcsc[1]{\textsc{\MakeLowercase{#1}}}

\newcommand{\angstrom}{\textup{\AA}\xspace}
\newcommand{\JJ}{$\rm DLA0817g1$\xspace}
\newcommand{\BB}{$\rm ^{3D}$Barolo\xspace}
\newcommand{\pyneb}{\textlcsc{PyNeb}\xspace}

\newcommand{\oiiab}{[\ion{O}{II}]$\lambda\lambda3726{,}3729$\xspace}
\newcommand{\oiiic}{[\ion{O}{III}]$\lambda4363$\xspace}
\newcommand{\hb}{$\rm H\beta$\xspace}
\newcommand{\oiiib}{[\ion{O}{III}]$\lambda5007$\xspace}
\newcommand{\niic}{[\ion{N}{II}]$\lambda5755$\xspace}
\newcommand{\hei}{HeI$\lambda 5875$\xspace}
\newcommand{\siiic}{[\ion{S}{III}]$\lambda6312$\xspace}
\newcommand{\niiab}{[\ion{N}{II}]$\lambda\lambda 6548,6584$\xspace}
\newcommand{\niia}{[\ion{N}{II}]$\lambda6548$\xspace}
\newcommand{\niib}{[\ion{N}{II}]$\lambda6584$\xspace}
\newcommand{\ha}{$\rm H\alpha$\xspace}
\newcommand{\heib}{\ion{He}{I}$\lambda 6678$\xspace}
\newcommand{\siiab}{[\ion{S}{II}]$\lambda\lambda 6716,6731$\xspace}
\newcommand{\siia}{[\ion{S}{II}]$\lambda6716$\xspace}
\newcommand{\siib}{[\ion{S}{II}]$\lambda6731$\xspace}
\newcommand{\heic}{\ion{He}{I}$\lambda 7065$\xspace}
\newcommand{\Hninezero}{\ion{H}{I}$\lambda 9014$\xspace}
\newcommand{\siiiab}{[\ion{S}{III}]$\lambda\lambda 9069,9532$\xspace}
\newcommand{\siiia}{[\ion{S}{III}]$\lambda9069$\xspace}
\newcommand{\Hninetwo}{\ion{H}{I}$\lambda 9229$\xspace}
\newcommand{\Hninefive}{\ion{H}{I}$\lambda 9546$\xspace}
\newcommand{\siiib}{[\ion{S}{III}]$\lambda9532$\xspace}

\newcommand{\cii}{[\ion{C}{II}]\xspace}
\newcommand{\oiiialma}{[\ion{O}{III}]$88\,\mu$m\xspace}

\usepackage{graphicx}	
\usepackage{amsmath}	

\usepackage{tikz,xcolor,hyperref}
\definecolor{lime}{HTML}{A6CE39}
\DeclareRobustCommand{\orcidicon}{%
    \begin{tikzpicture}
    \draw[lime, fill=lime] (0,0) 
    circle [radius=0.16] 
    node[white] {{\fontfamily{qag}\selectfont \tiny ID}};
    \draw[white, fill=white] (-0.0625,0.095) 
    circle [radius=0.007];
    \end{tikzpicture}
    \hspace{-2mm}
}

\title[GANIFS: J0817]{GA-NIFS: A post-starburst main sequence disk galaxy undergoing ordered rotation at $\mathbf{z=4.26}$}

\newcommand{\orcidGCJ}{\href{https://orcid.org/0000-0002-0267-9024}{\orcidicon}}
\newcommand{\orcidHU}{\href{https://orcid.org/0000-0003-4891-0794}{\orcidicon}}
\newcommand{\orcidRM}{\href{https://orcid.org/0000-0002-4985-3819}{\orcidicon}}
\newcommand{\orcidAJB}{\href{https://orcid.org/0000-0002-8651-9879}{\orcidicon}}
\newcommand{\orcidSCh}{\href{https://orcid.org/0000-0003-3458-2275}{\orcidicon}}
\newcommand{\orcidGC}{\href{https://orcid.org/0000-0002-5281-1417}{\orcidicon}}
\newcommand{\orcidFDE}{\href{https://orcid.org/0000-0003-2388-8172}{\orcidicon}}

\newcommand{\orcidMP}{\href{https://orcid.org/0000-0002-0362-5941}{\orcidicon}}
\newcommand{\orcidJS}{\href{https://orcid.org/0000-0001-6010-6809}{\orcidicon}}

\newcommand{\orcidSA}{\href{https://orcid.org/0000-0001-7997-1640}{\orcidicon}}

\newcommand{\orcidTB}{\href{https://orcid.org/0000-0002-5666-7782}{\orcidicon}}
\newcommand{\orcidIL}{\href{https://orcid.org/0000-0003-3336-5498}{\orcidicon}}
\newcommand{\orcidEP}{\href{https://orcid.org/0000-0002-7392-7814}{\orcidicon}}
\newcommand{\orcidRP}{\href{https://orcid.org/0000-0001-9820-5773}{\orcidicon}}
\newcommand{\orcidBRdP}{\href{https://orcid.org/0000-0001-5171-3930}{\orcidicon}}
\newcommand{\orcidSZ}{\href{https://orcid.org/0000-0003-4546-897X}{\orcidicon}}

\author[G. C. Jones, et al.]{
\hspace{-1mm}Gareth C. Jones$^{1,2}$\thanks{E-mail: gj283@cam.ac.uk}\orcidGCJ,
Roberto Maiolino$^{1,2,3}$\orcidRM,
Francesco D'Eugenio$^{1,2}$\orcidFDE,
Santiago Arribas$^{4}$\orcidSA,
\newauthor
Andrew J. Bunker$^{5}$\orcidAJB,
Stephane Charlot$^{6}$\orcidSCh,
Michele Perna$^{4}$\orcidMP,
Bruno Rodriguez del Pino$^{4}$\orcidBRdP,
\newauthor
Hannah Übler$^{7}$\orcidHU,
Torsten B\"{o}ker$^{8}$\orcidTB,
Giovanni Cresci$^{9}$\orcidGC,
Isabella Lamperti$^{10,9}$\orcidIL,
Eleonora Parlanti$^{11}$\orcidEP,
\newauthor
Robert Pascalau$^{1,2}$\orcidRP,
Jan Scholtz$^{1,2}$\orcidJS,
Sandra Zamora$^{11}$\orcidSZ
\\
$^{1}$Kavli Institute for Cosmology, University of Cambridge, Madingley Road, Cambridge CB3 0HA, UK\\
$^{2}$Cavendish Laboratory, University of Cambridge, 19 JJ Thomson Avenue, Cambridge CB3 0HE, UK\\
$^{3}$Department of Physics and Astronomy, University College London, Gower Street, London WC1E 6BT, UK\\
$^{4}$Centro de Astrobiolog\'{i}a (CAB), CSIC-INTA, Ctra. de Ajalvir km 4, Torrej\'on de Ardoz, E-28850, Madrid, Spain\\
$^{5}$Department of Physics, University of Oxford, Denys Wilkinson Building, Keble Road, Oxford OX1 3RH, UK\\
$^{6}$Sorbonne Universit\'e, CNRS, UMR 7095, Institut d'Astrophysique de Paris, 98 bis bd Arago, 75014 Paris, France\\
$^{7}$Max-Planck-Institut f\"ur extraterrestrische Physik, Gie{\ss}enbachstra{\ss}e 1, 85748 Garching, Germany\\
$^{8}$European Space Agency, c/o Space Telescope Science Institute, 3700 San Martin Drive, Baltimore MD 21218, USA\\
$^{9}$INAF - Osservatorio Astrofisco di Arcetri, largo E. Fermi 5, 50127 Firenze, Italy\\
$^{10}$Dipartimento di Fisica e Astronomia, Universit\`a di Firenze, Via G. Sansone 1, 50019, Sesto F.no (Firenze), Italy\\
$^{11}$Scuola Normale Superiore, Piazza dei Cavalieri 7, I-56126 Pisa, Italy\\
}

\date{Accepted XXX. Received YYY; in original form ZZZ}

\pubyear{\the\year{}}

\begin{document}
\label{firstpage}
\pagerange{\pageref{firstpage}--\pageref{lastpage}}
\maketitle

\begin{abstract}
Rotating galaxies with relaxed gaseous disks have been discovered across cosmic time, from the local Universe to high redshift. But few such sources have been confirmed at $z>4$, making them a precious sample to examine the conditions required for such ordered kinematics in an early, more chaotic Universe. One of the best examples of this sample is the galaxy \JJ ($z=4.2603$), which shows remarkably clear rotation in ALMA \cii data. We present JWST/NIRSpec IFU data ($R\sim2700$) of \JJ, which we combine with archival ALMA \cii observations to place constraints on its ISM conditions and morpho-kinematics. From a combination of line ratios, we find a high gas-phase metallicity ($Z_{\rm gas}\sim0.8\,Z_{\odot}$), high fraction of obscured star formation, low ionisation (compared to other high-redshift galaxies observed with JWST), and no significant evidence for AGN. Dynamical modelling with \BB reveals nearly identical rotation in \ha and \cii. Using our metallicity estimate and previous CO and \cii detections, we derive a new estimate of the molecular gas mass, relieving a previous strain in the mass budget. Altogether, we suggest that this is a post-burst, main-sequence galaxy, where high levels of past star formation resulted in significant chemical enrichment (i.e., $Z_{\rm gas}$ and $M_{\rm dust}$), but the current activity is lower (i.e., lower ionisation parameter and lack of starburst activity). These new observations have opened a window into the interplay of gas, metallicity, star formation, and kinematics in a prototypical early disk galaxy.
\end{abstract}

\begin{keywords}
galaxies: high-redshift -- galaxies: kinematics and dynamics -- galaxies: disc  -- galaxies: evolution -- galaxies: ISM
\end{keywords}



\section{Introduction}\label{intro}

The study of rotating galaxies is a historic field, with observations dating back more than a century. Following the detection of rotation in the Andromeda Galaxy (e.g., \citealt{slip14,peas18}), it was soon found that our Galaxy is also rotating (e.g., \citealt{lind27,oort27}). Similar observations of nearby galaxies showed that they feature rotation as well (e.g., \citealt{maya42,burb59,burb60,deva61,deva62,taka67,case83}). Each study resulted in a `rotation curve', or a radial profile of rotational velocity (usually taken along the kinematic major axis).

A major step forward came with the work of \citet{rubi80}, who found that the rotation curves of their observed galaxies did not rise and fall with radius, as expected from the observed baryonic matter distribution. Instead, the curves rose quickly and then were either flat or rose slowly, implying the presence of an additional, unobserved mass component (i.e., non-baryonic or dark matter). By observing the baryonic mass distribution of galaxies (i.e., gas, dust, stars), calculating the total dynamical mass from the observed kinematics, and subtracting the two, the dark matter component of a galaxy may be found (e.g., \citealt{vana85,bege91,sofu01,fors09,wuyt16,frat21}).

Large samples of galaxy rotation curves for sources beyond the local Universe have now been measured using the \ha line, including the programs
`Assessing the Mass-Abundance redshift Evolution' (AMAZE; $3.0<z<5.2$; \citealt{maio08}), 
`Lyman-break galaxies Stellar populations and Dynamics' (LSD; $2.4<z<3.4$; \citealt{mann09}), 
`Spectroscopic Imaging survey in the Near-infrared with SINFONI' (SINS; $1\lesssim z\lesssim3$; \citealt{fors09}), 
`The KMOS Redshift One Spectroscopic Survey' (KROSS; $z\sim1$; \citealt{stot16,tile16}),
and KMOS$^{\rm 3D}$ ($0.6\lesssim z\lesssim2.7$; \citealt{wisn19})
.

Historically, the kinematics of galaxies at higher redshift ($z\gtrsim3$) have been studied using far-infrared (FIR) lines (e.g., CO, [\ion{C}{II}]\,158$\mu$m; hereafter \cii) rather than rest-optical lines due to instrumental restrictions and atmospheric effects. Large surveys of \cii emission include 
`The Atacama Large Millimeter Array (ALMA) Large Program to INvestigate \cii at Early times' (ALPINE; $4<z<6$; \citealt{lefe20}) with the follow-up program `\cii Resolved ISM in Star-forming Galaxies with ALMA' (CRISTAL; \citealt{herr25}), the `Reionization Era Bright Emission Line Survey' (REBELS; $6.6<z<8.6$; \citealt{bouw22}), and `Tracing Rotation with Ionized Carbon in Early Primeval Systems' (TRICEPS; $4<z<5$; \citealt{lell26}).  

While rest-frame optical and FIR lines are both strong indicators of the underlying kinematics of a galaxy (i.e., rotation, merging, in-/outflows), 
their different emission conditions mean that they trace different phases of the interstellar medium (ISM). For example, since the ionisation potential of ionised carbon (11.3\,eV; e.g., \citealt{accu17}) is lower than that of neutral hydrogen (13.6\,eV), \cii may emanate from the cold or warm neutral medium, warm and dense molecular gas, or warm ionised medium (e.g., \citealt{wolf03,pine13}). On the other hand, doubly ionised oxygen has a higher ionisation potential (35.1\,eV; e.g., \citealt{ramo23}), meaning that \oiiib traces hot ionised gas. This can also be seen in temperature-density plots of cosmological simulations (e.g., \citealt{schi24,chou25}), where \oiiib and \oiiialma only come from warmer gas, while \cii emanates from warm and cold gas, and CO originates from colder gas.

Observations have found that ionised gas features a higher velocity dispersion than colder gas (e.g., \citealt{cort17,uble19,gira21,rizz24,fuji25}). \citet{parl24} found a higher ionised velocity dispersion in the centre of ALESS73.1, but ascribe this to feedback from an active galactic nucleus (AGN). On the other hand, some galaxies have found molecular and ionised velocity dispersions that are comparable (e.g., \citealt{genz13,uble18,moli19b,arri24}). While rotation curves of both tracers are similar, some galaxies show higher rotation velocities in the colder phase (e.g., \citealt{davi13,levy18}). Simulations have shown that molecular tracers (e.g., \cii) likely probe the gaseous disk of galaxies while ionised tracers (e.g., \ha) include contributions from extra-planar ionised gas that increase the observed dispersion (e.g., \citealt{ejde22,ejde24,koha24}). Additional observations of both the ionised and molecular gas kinematics of rotating disk galaxies are required to determine how the rotational velocity and velocity dispersion of each evolve with cosmic time - informing models of feedback from AGN and star formation, mass evolution, and galaxy formation.

Observations with the Integral Field Unit (IFU) of the JWST/NIRSpec instrument (\citealt{boke22,jako22}) have enabled detailed characterisation of galaxy morpho-kinematics and ISM conditions from the local Universe (e.g., 
\citealt{lai23,bian24,pern24,ceci25,tayl25,uliv25}) to the first $\sim500$\,Myr of cosmic time (e.g., \citealt{maio24,scho24,marc24,xu24}). The high spatial resolution of this instrument opens a clear window into the kinematics of each source. In fact, NIRSpec IFU observations of rest-optical emission lines from some galaxies that were classified as rotators based on ALMA \cii observations revealed the presence of multiple smaller galaxies (e.g., HZ10, \citealt{jone17,jone25}; COS-3018, \citealt{smit18,scho25}; HZ4, \citealt{jone21,parl25}). Other previously classified rotators were confirmed, but with a combination of strong inflows, outflows, and/or an AGN component (e.g., ALESS73.1, \citealt{lell21,parl24}; BR1202-0725-SMG, \citealt{carn13,zamo24}; GN20, \citealt{hodg12,uble24}). In this context, we will discuss JWST/NIRSpec IFU observations of the galaxy ALMA J081740.85+135138.2 ($z=4.2603$; hereafter \JJ), a well-studied \cii rotator.

\JJ was detected as the host galaxy of a high metallicity damped Lyman-alpha absorber (DLA) at an angular separation of $\sim6''$ from a background quasar (QSO J081740.52+135134.5; $z=4.398$; \citealt{rafe12}). ALMA observations at $\sim0.8''$ resolution ($\sim6.1$\,kpc) of the \cii and rest-frame FIR emission showed a strong detection with an evident velocity gradient (\citealt{jone17,neel17}). Follow-up ALMA \cii observations at higher resolution ($\sim0.19''$; $\sim1.3$\,kpc) confirmed that the galaxy kinematics are well-modelled by a rotation-dominated disk with a dynamical mass of $\log_{10}(M_{\rm dyn}/M_{\rm \odot})=10.9^{+0.2}_{-0.3}$ within the central $r\sim4$\,kpc (as derived using different dynamical models; e.g., \citealt{neel20,jone21}).

\citet{neel20} found that \JJ features a low star formation rate ($SFR$) based on rest-frame near-UV observations (HST/WFC3 F160W; $SFR_{\rm NUV}=16\pm3\,M_{\odot}\,yr^{-1}$), but a considerable \cii-based SFR ($SFR_{\rm [CII]}=420\pm260\,M_{\odot}\,yr^{-1}$) and rest-frame FIR-based SFR ($SFR_{\rm 160\mu\mathrm{m}}=118\pm14\,M_{\odot}\,yr^{-1}$). Using a detection of CO(2-1) with the
The Karl G. Jansky Very Large Array (JVLA), they estimate the molecular gas mass to be comparable to the previously determined dynamical mass: $\log_{10}(M_{\rm H_2}/M_{\odot})=10.9^{+0.1}_{-0.2}$ (assuming $r_{21}=0.81$, $\alpha_{\rm CO}=3.0\,\mathrm{M_{\odot}\,K^{-1}\,km^{-1}\,s\,pc^{-2}}$). Since this CO(2-1) detection is spatially unresolved, it represents the gas mass within a radius of $\sim7$\,kpc.

Altogether, this suggests that \JJ is a gas-rich, rotating, star-forming galaxy in the early Universe ($\mathrm{t_H\sim1.4\,Gyr}$). But there are several mysteries that still surround this source. Primarily, the dynamical and molecular gas masses are comparable to within $1\sigma$, which leaves little room for stellar mass or a dark matter component ($M_{\rm dyn} \simeq M_{\rm *}+M_{\rm gas}+M_{\rm dust}+M_{\rm DM}$\footnote{Here, $M_{\rm gas}$ encapsulates multiple components, including neutral, ionised, and molecular gas.}). This suggests that either the system contains a large amount of gas at large radii ($4\lesssim r\lesssim 7$\,kpc) or that the conversion factors used to estimate $\mathrm{M_{H_2}}$ can be incorrect, and that metallicity measurements are required to correctly calibrate the gas mass measurements (e.g., \citealt{nara12,madd20}). In addition, \citet{neel17} argue that the kinematic properties of the galaxy derived through \cii observations (i.e., $v_{\rm rot}/\sigma_{\rm v}\sim3.4$, Toomre $Q\sim1$) suggest a relaxed, rotating disk formed through filamentary accretion. But since most of the analysis of this source has been performed on rest-FIR data, the ISM properties are not well constrained (e.g., temperature, metallicity, ionization).

In this work, we present new JWST/NIRSpec IFU observations of \JJ, opening a window into the ionised gas kinematics, ISM conditions, and general nature of this galaxy. In Section \ref{datasec}, we provide details of our NIRSpec IFU observations and calibration, as well as those of archival ALMA data that we use for comparison. This is followed by a resolved and integrated spectral analysis (Section \ref{specsec}), and an application of \BB to our data (Section \ref{trmsec}). We discuss our findings in Section \ref{discsec} and conclude in Section \ref{concsec}. Throughout, we assume a concordance cosmology ($h_0=0.7$, $\Omega_{\rm m}=0.3$, $\Omega_{\rm \Lambda}=0.7$), where $\rm 1''=6.8\,kpc$ at the redshift of our source ($z=4.26$).


\section{Observations and calibration}\label{datasec}
\subsection{JWST}
The JWST/NIRSpec IFU data studied in this work originate from the `Galaxy Assembly with NIRSpec Integral Field Spectroscopy' (GA-NIFS\footnote{\url{https://ga-nifs.github.io/}}) 
Guaranteed Time Observations (GTO) program (PIs: R. Maiolino \& S. Arribas). \JJ (08h17m40.8680s $+13^{\circ}51'38.22''$) was observed as part of PID 4528 (PI: K. Isaak) for 3.53\,hours on 8 December, 2024 with the disperser/filter combination G395H/F290LP (hereafter R2700, $2.87\,{\mu \rm m}<\lambda_{\rm obs}<5.14{\mu \rm m}$, $5450\,{\angstrom}\lesssim\lambda_{\rm rest}\lesssim9770{\angstrom}$) using a 6-point medium cycling dither pattern, the NRSIRS2RAPID readout pattern \citep{raus17}, and 88 groups per integration. We note that the quasar associated with \JJ (QSO J081740.52+135134) is $\sim6''$ from our target (e.g., \citealt{neel17}) and is therefore not covered by our observations. 

The raw data from these observations were downloaded from the MAST archive\footnote{\url{https://mast.stsci.edu/portal/Mashup/Clients/Mast/Portal.html}}. Calibration was performed using a customised version of the STScI pipeline (v1.15.1; \citealt{bush24}; see \citealt{pern23} for more details of customisation) with CRDS context 1303. These customisations account for 1/f noise correction, drizzling \citep{fruc02}, the subtraction of median values from count rate maps, manually masked cosmic rays and snowballs, open MSA slits, and outlier rejection ($95\%$, \citealt{deug24}). This results in a calibrated data cube with spaxels of $0.05''$. 

Next, we performed background subtraction using the observed data cube. Using a conservative mask based on the \ha emission, we mask \JJ, and for each spectral channel of the original data cube we find the mean value of the unmasked values that lie on the same slicers (in the spatial or cross-dispersion direction) as \JJ. These are used to generate a background cube, which is subtracted from the cube. We then mask all spaxels at different slicers, as they lack a background correction.

Since we wish to compare the JWST/NIRSpec data to ALMA images, the astrometry of our data must be verified. By comparing our NIRSpec data to archival NIRCam data (aligned to the Gaia DR3 reference frame), we find a spatial offset of $\sim0.1''$, which is expected from the JWST pointing uncertainty with no target acquisition (see full details in Appendix \ref{astrosec}). We correct for this offset, resulting in a positional uncertainty of $\sim0.01''$.

It is important to account for the wavelength-dependent PSF of our data, as we will compare emission across the full wavelength range of our data. To do this, we create a PSF-matched cube by convolving each spectral channel of our data cube with a custom kernel (see Appendix \ref{psf_stuff} for details). The resulting cube has a uniform PSF that can be approximated by a Gaussian with full width at half maximum FWHM=$0.22''$.

Previous works (e.g., \citealt{uble24}) have shown that the `ERR' extension of pipeline-calibrated IFU data is useful for characterising the wavelength-dependent noise behaviour of the data, but is not necessarily representative of the absolute flux uncertainty. For each extracted spectrum (i.e., from an aperture or spaxel), the ERR extension is rescaled such that the average value matches the RMS noise level of the data.

\subsection{ALMA}\label{almasec}
The \cii emission of this source has been observed by two ALMA projects (PI: M. Neeleman) using band 7 \citep{mahi12}: 2015.1.01564.S ($\sim 0.8''$ resolution) and 2017.1.01052.S ($\sim0.1''$ resolution). The resulting data have been well studied, including detailed kinematic analyses (see \citealt{neel17,neel20,jone17,jone21,roma23}). Here, we re-reduce the data in order to enable a fair comparison to our novel JWST/NIRSpec data.

The raw data for 2015.1.01564.S and 2017.1.01052.S were downloaded from the ALMA archive\footnote{\url{https://almascience.eso.org/aq/}}, and were calibrated using the python scripts provided by ALMA staff in CASA \citep{casa22} versions 4.5.1 and 5.1.1, respectively. We explored combining data from the two projects, but since the latter features both higher spatial resolution and a longer exposure time, we proceed using only the 2017.1.01052.S data. This project includes four spectral windows (SPWs): three containing only continuum emission (each with 128\,channels of width 15.6250\,MHz) and one containing continuum and \cii emission (with 240 native channels of width 7.8125\,MHz).

An initial continuum map was created using the tclean task of CASA in multi-frequency synthesis (MFS) mode and natural weighting, excluding the SPW including \cii emission. This map was then cleaned down to $3\times$ the RMS noise level. This resulted in a synthesized beam of $0.24''\times0.16''$ (position angle PA=$130^{\circ}$)\footnote{Throughout this work, we define PA to be increasing in the counterclockwise direction from the positive declination axis.} and an RMS noise level of 24\,$\mu$Jy/beam.

Next, we subtracted the continuum emission underlying \cii using the CASA task uvcontsub. This fit a first-order polynomial model to the visibilities of the three continuum SPWs, and subtracted an interpolation of this best-fit model from the SPW containing \cii. The line-only visibilities were then imaged using tclean in `cube' mode with natural weighting, with channels of 25\,km\,s$^{-1}$. The resulting cube was cleaned down to $3\times$ the RMS noise level. This resulted in a similar synthesized beam as the continuum map and an RMS noise level per channel of 0.4\,mJy/beam.

\section{Spectral analysis}\label{specsec}

Our JWST/NIRSpec IFU observations were designed with the primary goal of detecting \ha and using this strong line to measure the morpho-kinematics of the ionised gas in \JJ. However, due to the wide spectral range of our R2700 data, we significantly detect additional emission lines. We perform a spaxel-by-spaxel fit of the \ha-\niiab complex in Section \ref{RSA}\footnote{Detailed spatial analysis of the other lines and the continuum will be performed in other works (e.g., using the JWST/NIRSpec IFU G395M/F290LP data of PID 5761, PI Neeleman), and here we are primarily interested in the morpho-kinematics of \ha.}, a full fit of the integrated spectrum in Section \ref{ISA}, and determine ISM properties in Section \ref{ISMSec}.

\subsection{Spatially resolved analysis}\label{RSA}

To begin, we extract a spectrum from each spaxel of our data cube and restrict the wavelength range to only contain emission from \niiab and \ha. Each spectrum is fit with a combined model containing continuum and line emission. The continuum is assumed to be a power law:
\begin{equation}\label{cont_eq}
F_{\lambda}=
F_{\rm cont}(\lambda_{\rm obs}/4\,\mu{\rm m})^{\rm\alpha_{opt}}
\end{equation}
We assume that each line may be described as a single Gaussian component, and force the redshift and intrinsic linewidth (in velocity space; $FWHM_{\rm V}$) of all three lines to be equal. The line spread function (LSF\footnote{We determine the LSF using the fiducial resolving power curves recorded at \url{https://jwst-docs.stsci.edu/jwst-near-infrared-spectrograph/nirspec-instrumentation/nirspec-dispersers-and-filters}.}) is accounted for by adding the intrinsic linewidth and LSF in quadrature. The ratio between \niia/\niib=0.340 is derived using \pyneb \citep{luri15}\footnote{We adopt the default atomic data library for [NII] used by \pyneb \citep{froe04}}, and we do not account for dust attenuation (see Section \ref{ebvte} for discussion of this assumption). This model is fit to the spectrum of each spaxel using \textlcsc{lmfit} \citep{newv14}. 

An initial fit is run with all parameters free, and the results are automatically inspected. Lines with insignificant emission ($F_{\rm i}<3\times\delta F_{\rm i}$, where $\delta F_{\rm i}$ is the reported flux uncertainty) are removed from the model. If the best-fit $FWHM_{\rm V}$ is very small ($<30$\,km\,s$^{-1}$), then the linewidth is set to be identical to the instrumental LSF.

We wish to compare these rest-optical morpho-kinematics to those of ALMA \cii emission. Usually, ALMA moment maps are created using a non-parametric approach (i.e., the CASA task \textlcsc{immoments}; e.g., \citealt{hari20,guru22,rowl24}). In order to ensure a fair comparison, we fit the ALMA \cii cube in a similar fashion to the NIRSpec IFU data. Specifically, we fit the spectrum of each spaxel in the continuum-subtracted \cii data cube (see Section \ref{almasec}) using a single Gaussian model. Since we do not aim to study any rest-optical to rest-FIR line ratios, we keep the ALMA data in its native units.

\begin{figure*}
\centering
\includegraphics[trim=0cm 9.5mm 0cm 0cm,clip=true,width=\textwidth]{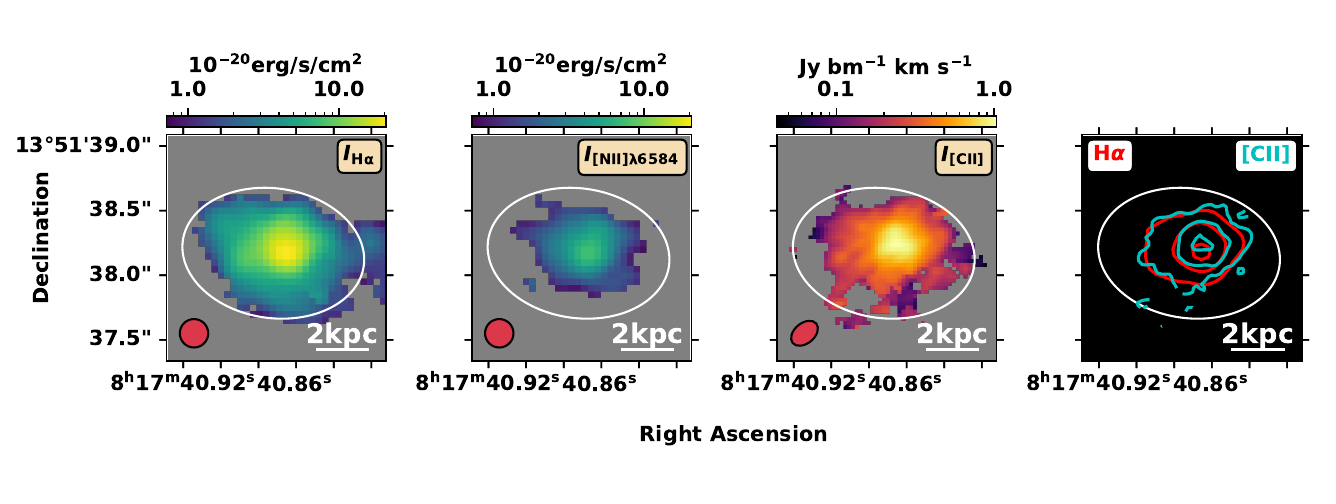}
\caption{Best fit line intensity maps of \ha, \niib, and \cii, where values are given per pixel. In the rightmost panel, we compare the \ha (red) and \cii (cyan) emission using contours of $[0.25,0.50,0.90]\times$ the peak value of each. In each panel, we show a physical scale of 2\,kpc. The PSF is represented as a red circle or ellipse to the southeast (north is up and east is to the left). Our adopted aperture is shown as a white ellipse.}
\label{linemaps}
\end{figure*}

The resulting line flux maps (Figure \ref{linemaps}) show strong emission from \JJ in \ha, \niib, and \cii. To compare the spatial distributions of each, we use \textlcsc{pysersic} \citep{pash23} to fit the \ha and \cii morphology of \JJ while accounting for the different PSF shapes. The resulting best-fit properties (listed in Table \ref{pys_results_real}) show remarkable agreement ($\lesssim2\sigma$ discrepant) between the two line tracers, with ellipticity (defined as $1-\left(b/a\right)$) of $\sim0.3$, S\'ersic index $n_{\rm s}\sim0.9$, position angle $PA\sim82^{\circ}$, and effective radius $R_{\rm e}\sim2$\,kpc. 

From the \textlcsc{pysersic} fits, we find that the spatial centroids of the \cii and \ha distributions are well aligned, with an offset of $0.07\pm0.17''$ (where the uncertainty includes both the reported uncertainty in each fit and the astrometric uncertainties derived in Appendix \ref{astrosec}).

\begin{table}
\centering
\begin{tabular}{c|ccc}
Property & \ha & \cii \\ \hline
Ellipticity	&	$0.29\pm0.02$	&	$0.26\pm0.04$	\\
$n_{\rm s}$	&	 $0.94\pm0.04$	&	$0.82\pm0.08$	\\
$PA$ [$^{\circ}$] & $82\pm2$ & $90\pm5$ &\\
$R_{\rm e}$ [$''$]	&	$0.30\pm0.01$	&	$0.28\pm0.01$	\\
$R_{\rm e}$ [kpc]	&	$2.0\pm0.1$	&	$1.9\pm0.1$	\\ \hline
\end{tabular}
\caption{Best-fit morphological properties of the \ha and \cii emission, as derived through \textlcsc{pysersic} fits.}
\label{pys_results_real}
\end{table}

\begin{figure}
\centering
\includegraphics[trim=0cm 2mm 0cm 0cm,clip=true,width=0.5\textwidth]{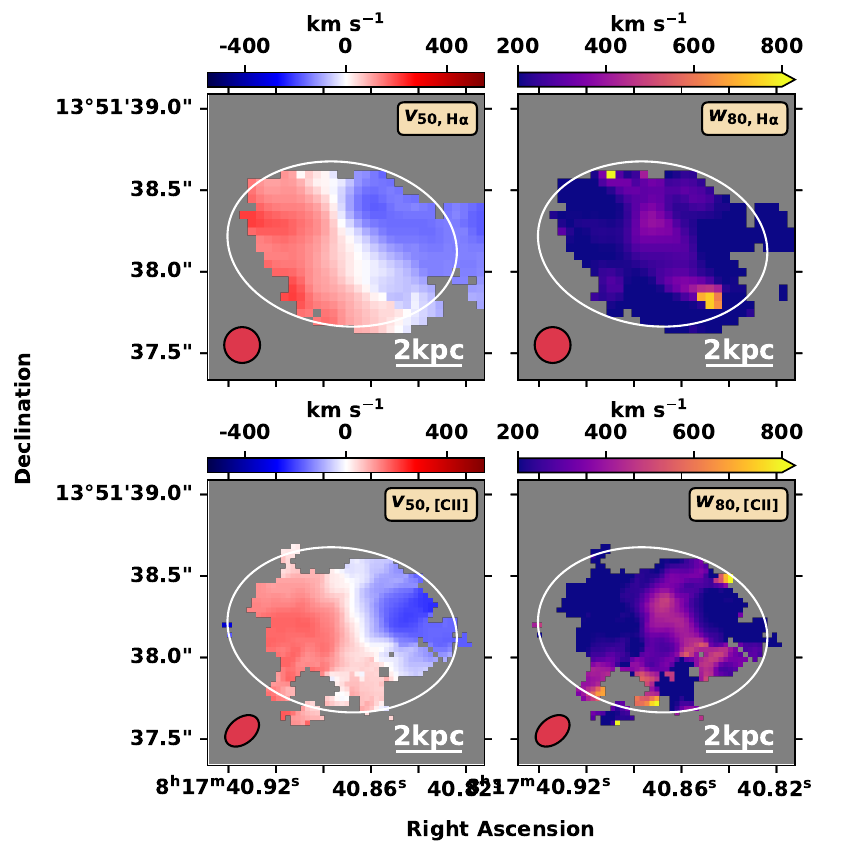}
\caption{Best fit $v_{50}$ (left column) and $w_{80}$ (right column) maps for \ha (top row) and \cii (lower row). In each panel, we show a physical scale of 2\,kpc. The PSF is represented as a red circle or ellipse to the southeast (north is up and east is to the left). Our adopted aperture is shown as a white ellipse.}
\label{velmaps}
\end{figure}

To examine the kinematics of this source, we create maps of $v_{50}$\footnote{We use the notation $v_{N}$ to represent the velocity where $N\%$ of the line flux lies at $v<v_{\rm N}$.} and $w_{80}\equiv v_{90}-v_{10}$. Since we are using Gaussian profiles, $v_{50}$ is the best-fit velocity offset of the line centroid from an assumed redshift of $z=4.2603$, while $w_{80}\simeq1.088\times$ the Gaussian FWHM. The kinematic maps (Figure \ref{velmaps}) show similar behaviour between the two tracers, which we further explore in Section \ref{trmsec}.

\subsection{Integrated spectral analysis}\label{ISA}

\subsubsection{Aperture selection}

In order to study the integrated emission of this galaxy, we must carefully choose an extraction aperture. The full emission should be captured  without missing possible weaker, diffuse emission or being biased towards bright regions.
We first generate an ellipse with the same axis ratio and position angle as the best-fit S\'ersic model of the \ha data (Section \ref{RSA}). The effective radius is then increased such that the aperture encloses $90\%$ of the best-fit S\'ersic model, following a similar approach to \citet{take26}. The resulting aperture is shown in each figure as a white ellipse. 

\subsubsection{Integrated spectral fitting}

\begin{figure*}
    \centering
    \includegraphics[trim=0cm 2mm 0cm 0cm,clip=true,width=\textwidth]{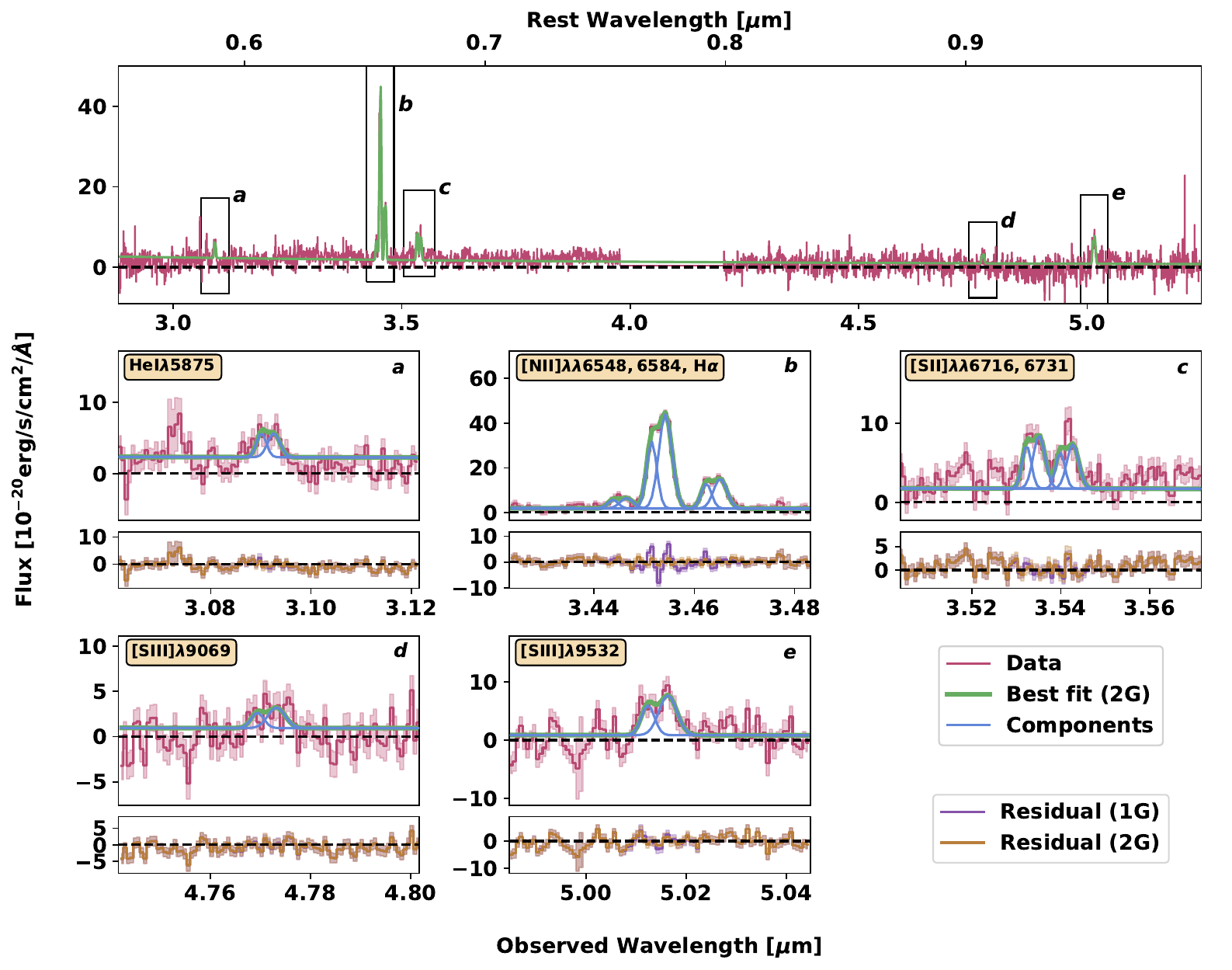}
    \caption{Results of fitting integrated spectrum of \JJ. The top panel shows a full spectrum (pink line with $1\sigma$ uncertainty shown as shaded region) along with the best-fit model (green). Panels \textit{a-e} are zoomed-in views for each line, showing the individual best-fit Gaussian components (light blue). The lower portion of each panel depicts the residual spectra for a single-Gaussian fit (purple) and out two-Gaussian fit (brown).}
    \label{intgauss}
\end{figure*}

Using this aperture, we extract an integrated spectrum from our data cube (see Figure \ref{intgauss}). Both the continuum (Equation \ref{cont_eq}) and emission lines (\hei, \niiab, \ha, \siiab, and \siiiab) are included in each fit. First, we explore whether each line is better described by a single Gaussian profile where the kinematics of each line are tied (1G) or two Gaussian profiles where each Gaussian has its own redshift and intrinsic (i.e., LSF-corrected) linewidth (2G). We find that the 2G model results in a better Bayesian inference criterion (BIC, where BIC$_{\rm 1G}$-BIC$_{\rm 2G}\simeq30$) and reduced chi-squared ($\chi_{\rm r,1G}$-$\chi_{\rm r,2G}\simeq0.04$) - both indications that the more complex 2G model is statistically a better fit. In addition, the 1G model results in significant residuals in the fit of \ha (Figure \ref{intgauss}). {Thus, we proceed with the 2G model.} We note that while this approach is a simple approximation of the morpho-kinematics of the source, it is sufficient to measure the line fluxes over the full field. A full dynamical analysis of these data are presented in Section \ref{trmsec}.

The flux of each line is set to be independent, in order to explore ISM conditions. 
The resulting fits to each spectral line are shown in Figure \ref{intgauss}, while the resulting best-fit values are listed in Table \ref{linetable}.

\begin{table}
    \centering
    \begin{tabular}{c|cc}
Property & Value & Units \\ \hline
$F_{\rm HeI\lambda5875}$&$184\pm45$&[$10^{-20}$ erg s$^{-1}$ cm$^{-2}$]\\
$F_{\rm [NII]\lambda6548}$&$248\pm32$&[$10^{-20}$ erg s$^{-1}$ cm$^{-2}$]\\
$F_{\rm H\alpha}$&$2154\pm163$&[$10^{-20}$ erg s$^{-1}$ cm$^{-2}$]\\
$F_{\rm [NII]\lambda6584}$&$720\pm32$&[$10^{-20}$ erg s$^{-1}$ cm$^{-2}$]\\
$F_{\rm [SII]\lambda6716}$&$363\pm43$&[$10^{-20}$ erg s$^{-1}$ cm$^{-2}$]\\
$F_{\rm [SII]\lambda6731}$&$304\pm43$&[$10^{-20}$ erg s$^{-1}$ cm$^{-2}$]\\
$F_{\rm [SIII]\lambda9069}$&$150\pm38$&[$10^{-20}$ erg s$^{-1}$ cm$^{-2}$]\\
$F_{\rm [SII]\lambda9532}$&$472\pm38$&[$10^{-20}$ erg s$^{-1}$ cm$^{-2}$]\\ \hline
$F_{\rm cont}$&$1.35\pm0.04$&[$10^{-20}$ erg s$^{-1}$ cm$^{-2} $ $\angstrom^{-1}$]\\
$\alpha_{\rm opt}$&$-2.04\pm0.14$&-\\ \hline
$\log_{10}(\rm S2)$&$-0.51\pm0.05$&-\\
$\log_{10}(\rm S3)$&$-0.54\pm0.05$&-\\
$\log_{10}(\rm S23)$&$-0.22\pm0.04$&-\\
$\log_{10}(\rm S32)$&$-0.03\pm0.05$&-\\
$\log_{10}(\rm N2)$&$-0.48\pm0.04$&-\\
$\log_{10}(\rm N2S2)$&$0.03\pm0.04$&-\\ \hline
$\log_{10}(n_{\rm e} {\rm [cm^{-3}]})$&$2.62\pm0.35$&-\\
12+$\log_{10}(O/H)$&$8.57\pm0.08$&-\\
$Z_{\rm gas}$&$0.75\pm0.14$&[$Z_{\odot}$]\\
$SFR_{\rm H\alpha}$&$18\pm1$&[$M_{\odot}$\,yr$^{-1}$]\\
$\log_{10}(U)$&$-3.60\pm0.06$&-\\ \hline
$\alpha_{\rm [CII]}(Z)$&$5.23\pm0.39$&[$M_{\odot}\,L_{\odot}^{-1}$]\\
$\log_{10}(M_{\rm  gas,[CII]} {\rm [M_{\odot}]})$&$10.23\pm0.04$&-\\ \hline
    \end{tabular}
    \caption{Best-fit quantities and derived properties of \JJ, as derived from fits to the integrated spectrum. No dust correction has been applied (see Section \ref{ebvte}). Definitions of line ratios are given in Table \ref{linerat}. More details on each derived property are given in Section \ref{ISMSec}.}
    \label{linetable}
\end{table}

\subsection{Resulting ISM properties}\label{ISMSec}

\subsubsection{Electron density}
We use the density-dependent ratio \siia/\siib ratio to find the electron density $n_{\rm e}$ through the \textlcsc{pyneb} task \textit{getTemDen}. Assuming $T_{\rm e}=0.8\times10^4$\,K (see Section \ref{ebvte}), we estimate $\log_{10}(n_{\rm e} {\rm [cm^{-3}]})=2.62\pm0.35$. This value is in agreement with the expected value at $z=4.2603$ from other works (e.g., \citealt{isob23,li25,li25b}). 

\subsubsection{Dust attenuation and electron temperature}\label{ebvte}
The dust attenuation of high-redshift galaxies (parametrised as $E(B-V)$ or $A_{\rm V}$) can be derived by measuring multiple strong Balmer emission lines (e.g., \ha and \hb) and comparing their intrinsic and observed flux ratios (assuming a dust attenuation curve; e.g., \citealt{calz00}). Since only one Balmer line falls within our spectral coverage, we explore two avenues to constrain the dust attenuation. First, we use our observed \niia/\niib and \siiia/\siiib ratios. The intrinsic values of these ratios depend only on the Einstein A coefficients, and are thus independent of electron temperature or density. Our \niia/\niib ratio is listed in Section \ref{RSA}, and we use the default [SIII] library of \pyneb \citep{froe06} to calculate an intrinsic ratio of \siiia/\siiib=0.405.

\begin{figure}
\centering
\includegraphics[trim=0cm 3mm 0cm 0cm,clip=true,width=0.5\textwidth]{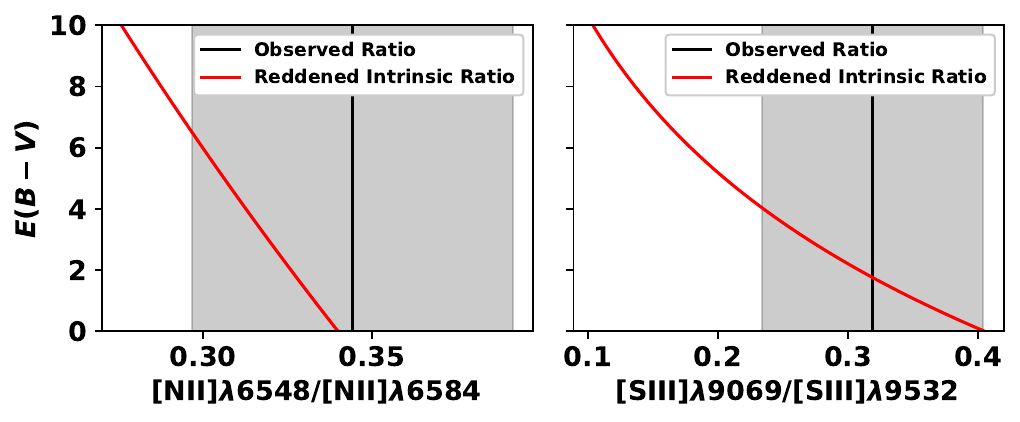}
\caption{Preliminary test to constrain $E(B-V)$ from our observed \niiab and \siiiab ratios. For each panel, we show the observed ratio and its $1\sigma$ uncertainty as a black line and shaded region, respectively. The dust-reddened intrinsic value of each ratio for a given $E(B-V)$ is shown as a red dashed line.}
\label{ebv_fig1}
\end{figure}

As shown in Figure \ref{ebv_fig1}, we explore a range of $0\leq E(B-V)\leq10$, redden each of the intrinsic line ratios (red line; as derived with \pyneb), and compare them to our observed line ratios and the associated uncertainties (black line and shaded region; see Table \ref{linetable}). Due to the large fitting uncertainties in each observed ratio, there is no definite value of $E(B-V)$ that is preferred by our data. The observed \niia/\niib$=0.34\pm0.05$ and \siiia/\siiib$=0.31\pm0.09$ argue for $E(B-V)\lesssim6.5$ and $E(B-V)\lesssim4$, respectively. These limits exclude the extreme case of hot dust-obscured galaxies, which can feature $E(B-V)>10$ (e.g., \citealt{asse16}), but encompasses the range of most JWST-derived values for $z>4$ galaxies (e.g., \citealt{fuji25,jone24,nava24,kiyo25}).

As an alternative method, we collect the fluxes of our detected lines and the upper limits of our undetected lines\footnote{Upper limits on line fluxes are calculated by creating 1000 mock spectra with the same noise level as the data, adding mock lines (adopting the kinematics of the \ha line) with different fluxes, and finding the flux limit where we recover the line $68\%$ of the time.} and use a Bayesian inference approach (\textlcsc{ultranest}; \citealt{buch21}) to constrain ISM properties through comparisons to \textlcsc{pyneb} models.

 We use \textlcsc{pyneb} to calculate the expected line ratio ${\cal R}_{\rm x}$ given a set of input parameters ${\rm x} =\{n_{\rm e},T_{\rm e},E(B-V)\}$. For each observed line ratio ($R_{\rm i}\pm\delta R_{\rm i}$), the likelihood is (e.g., \citealt{kerr24,whit25}):
\begin{equation}
\mathcal{L}_{\rm i} = \frac{1}{\sqrt{2\pi} \delta R_{\rm i}} {\rm exp} \left(\frac{-({\cal R}_{\rm x}-R_{\rm i})^2}{2\delta R_{\rm i}^2}\right)
\end{equation}
For non-detections, we calculate the likelihood given the $3\sigma$ upper limit ($<3\delta R_{\rm j}$):
\begin{equation}
\mathcal{L}_{\rm j} = 0.5\times{\rm erfc}\left(\frac{{\cal R}_{\rm x}-3\delta R_{\rm j}}{\sqrt{2}\delta R_{\rm j}}
\right)
\end{equation}
The total likelihood for $N_{\rm D}$ detected ratios (\niia/\niib, \siia/\siib, \siiia/\siiib) and $N_{\rm U}$ upper limits on ratios (including the detected lines listed in Table \ref{linetable} and upper limits on \niic, \siiic, \heib, \heic, \Hninezero, \Hninefive, and \Hninetwo) is then given as:
\begin{equation}
\mathcal{L}_{\rm total} = \prod_i^{N_{\rm D}}\mathcal{L}_{\rm i}\prod_j^{N_{\rm U}}\mathcal{L}_{\rm j}
\end{equation}

While we explore fitting $n_{\rm e}$, $T_{\rm e}$, and $E(B-V)$ simultaneously, this results in poor constraints on $n_{\rm e}$ and $T_{\rm e}$. Thus, we adopt the best-fit $n_{\rm e}$ value from our \siiab ratio (which is not strongly affected by temperature or dust attenuation), $T_{\rm e}=0.8\times 10^4$\,K (based on the metallicity-dependent relation of \citealt{mend23}, see Section \ref{metall}), and a uniform prior of $0<E(B-V)<5$.

Using \textlcsc{ultranest} with the above priors and likelihood function, we find a best-fit value of $E(B-V) = 0.90^{+0.95}_{-0.66}$ (giving the 16, 50, and 84$\%$ percentile locations). But when combined with the posterior distribution shape (which decreases from the lowest explored bin towards higher E(B-V), see Figure \ref{cornerplotsfig} in Appendix \ref{cornerplots}), this suggests a low dust attenuation with a $3\sigma$ upper limit of $E(B-V)<3.1$. Further observations (e.g., \hb, \oiiic) are needed to better constrain these properties. For the rest of this Section, we present results assuming $T_{\rm e}=0.8\times10^4$\,K and no dust correction ($E(B-V)=0$).

We note that the values of ${\cal R}_{\rm x}$ in this analysis are dependent on the intrinsic values of each line ratio as predicted by \pyneb. For each atom, we must adopt a library of Einstein A coefficients (as previously noted for \niia/\niib and \siiia/\siiib), and assume that the two emission lines originate from the same ISM conditions ($n_{\rm e}$ and $T_{\rm e}$). While this is a common assumption, recent works have found galaxies better described by two or more ISM zones (e.g., \citealt{marc24_HR,hari25,usui25}). More line detections (e.g., \oiiab, \oiiialma) are required to test if this is the case for \JJ as well.

Furthermore, ${\cal R}_{\rm x}$ is dependent on the assumed dust attenuation law (e.g., \citealt{calz00,gord03,sali18,redd20}). By design, each attenuation law is anchored near the rest-frame wavelength of \oiiib, so our choice of a \citet{calz00} law has a stronger effect on ratios containing longer-wavelength emission lines (e.g., \ha/\Hninetwo). Since the analysis of the dust attenuation law in \JJ is deferred to future work, we simply note that these estimated $T_{\rm e}$ and $E(B-V)$ values will be improved through continuum studies and further rest-UV and/or rest-optical emission line detections.


\subsubsection{Metallicities}\label{metall}
The gas-phase metallicity is derived by applying the diagnostics of \citet{sand25} to the line ratios listed in Table \ref{linerat}. Specifically, we use the chi-squared minimization approach of \citet[][see their equation 1]{curt20} to find a gas-phase metallicity of $0.75\pm0.14$ solar, assuming a solar metallicity of $12+\log(O/H)=8.69$ \citep{alle01}. 

\begin{table}
\centering
\begin{tabular}{c|c}
Name & Ratio \\ \hline
S2 & \siiab/\ha \\ 
S3 & \siiiab/\ha \\
S23 & (\siiab+\siiiab)/\ha \\
S32 & \siiiab/\siiab \\
N2 & \niib/\ha \\
N2S2 & \niib/\siiab
\end{tabular}
\caption{Definitions of line ratios used in this work (e.g., \citealt{sand25}).}
\label{linerat}
\end{table}

\subsubsection{Star formation rate}\label{sfrsec}
Next, we infer the star formation rate (SFR) using the best-fit \ha flux. Using the calibration of \citet{kenn98}:
\begin{equation}
\frac{SFR_{\rm H\alpha}}{[M_{\odot}\,yr^{-1}]} = \frac{4\pi C_{\rm SFR} F_{\rm H\alpha}}{[erg\,s^{-1}\,cm^{-2}]}\left(\frac{D_{\rm L}}{[cm]}\right)^2\frac{[erg\,s^{-1} M_{\odot}^{-1}\,yr]}{1.26\times10^{41}}
\end{equation}
where the factor of $C_{\rm SFR}$ is included to convert from a \citet{salp55} IMF to a \citet{chab03} IMF. Previous works used a range of $C_{\rm SFR}\sim0.56-0.65$ (e.g., \citealt{pant07,long09,dutt10,gonz10,beth12,driv13,mada14,capu17,figu22,hsia24}). We briefly note that other investigations of high redshift galaxies ($z>6$) with JWST have instead used conversion factors based on stellar population synthesis models (\citealt{redd18,redd22}). These models, which assume a low metallicity, result in lower factors (e.g., $C_{\rm SFR}=0.27$; \citealt{saxe23,curt24}, $C_{\rm SFR}\sim0.40$; \citealt{hsai23}). In this work, we adopt  a \citet{chab03} IMF and $C_{\rm SFR}\sim0.61$, as in another high-redshift JWST work \citep{hsia24}.

We find $SFR_{\rm H\alpha}=17\pm1\,M_{\odot}\,yr^{-1}$ for the case of no dust correction, which is $<1\sigma$ discrepant from the $SFR_{\rm NUV}=16\pm3\,M_{\odot}\,yr^{-1}$ found by \citet{neel20}. However, these values are both much smaller than the FIR-based $SFR_{\rm 160\mu\mathrm{m}}=118\pm14\,M_{\odot}\,yr^{-1}$, suggesting a considerable fraction of obscured star formation. While a high $SFR_{\rm [CII]}=420\pm260\,M_{\odot}\,yr^{-1}$ value is reported by \citet{neel20}, the low significance ($<2\sigma$) makes it difficult to compare.

From a dynamical analysis, \citet{roma24} estimated a stellar mass for \JJ of $\log_{10}(M_{\rm *}/M_{\odot})=10.6\pm0.2$. For this galaxy to lie on the star forming main sequence (SFMS; adopting the form of \citealt{spea14}), this stellar mass corresponds to a $SFR_{\rm MS}=145^{+190}_{-82}\,M_{\odot}\,yr^{-1}$. Since this agrees with the combination of the obscured and unobscured SFR, \JJ likely lies on the SFMS.

\subsubsection{Ionisation parameter}
We estimate the ionisation parameter $U$ using a relation of \citet{kewl19}, which is dependent on S32, metallicity, and ISM pressure ($P_{\rm ISM}$), parametrised as $\log_{10}(P_{\rm ISM}/k_B)$ (where $k_{\rm B}$ is the Boltzmann constant). Using our best-fit S32 and metallicity, we find that the high-pressure ($\log_{10}(P_{\rm ISM}/k_B)=7$) and low pressure ($\log_{10}(P_{\rm ISM}/k_B)=5$) diagnostics return similar $\log_{10}(U)$ values (i.e., $<1\sigma$ discrepant), and adopt the high-pressure case. The S32 value and resulting ionisation parameter ($\log_{10}(U)=-3.60\pm0.06$) lie at the weak edge of the distribution of high-redshift values (e.g., \citealt{redd23,shap25})

\subsubsection{WHAN diagram}
Due to our spectral setup, we lack the wavelength coverage necessary to study the placement of \JJ on several common line ratio diagnostic diagrams that differentiate between excitation by AGN and star formation (e.g.; BPT, \citealt{bald81}; VO-87, \citealt{veil87}; R3O1, \citealt{kewl01}). However, our high spectral resolution and strong detections of \ha and \niib enable the use of the equivalent width of \ha \textit{vs.} \niib/\ha (WHAN; \citealt{cidf10,cidf11}) diagram. 

The WHAN diagram has been applied to other $z>4$ galaxies, including the JWST/NIRSpec IFU observations of GN20 from GA-NIFS \citep{uble24}. This previous analysis showed that the majority of the spaxels are classified as strong AGN (or `Seyfert-like'), while some spaxels on the outskirts are instead classified as likely star-forming. Similarly, \citet{deug25} use this diagnostic to classify the $z=4.6348$ galaxy Ulema as an AGN. 

\begin{figure}
\centering
\includegraphics[width=0.5\textwidth]{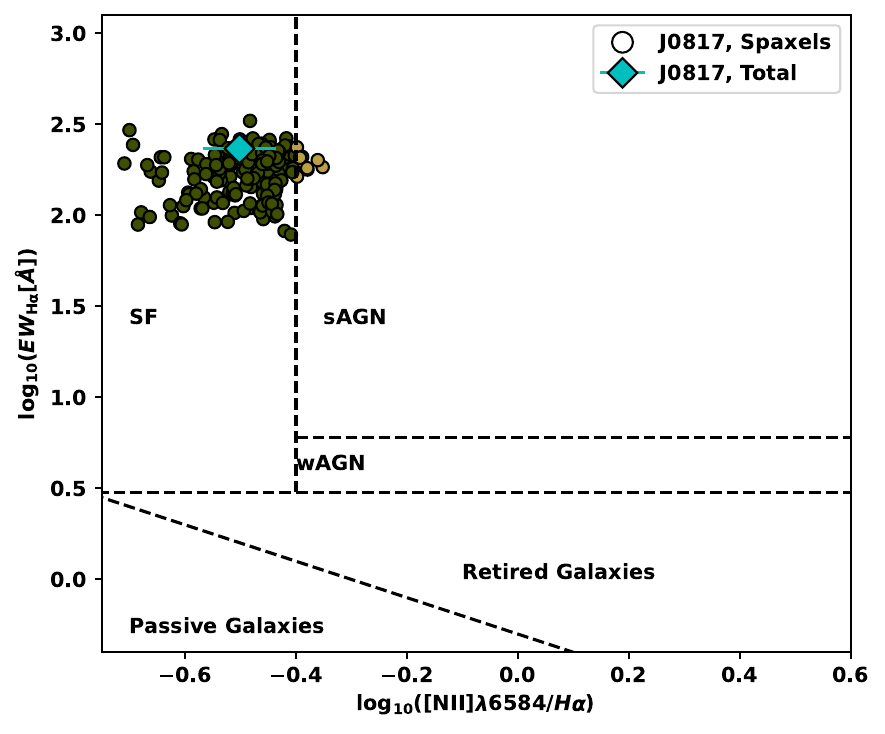}
\includegraphics[trim=0cm 2.5mm 0cm 0cm,clip=true,width=0.5\textwidth]{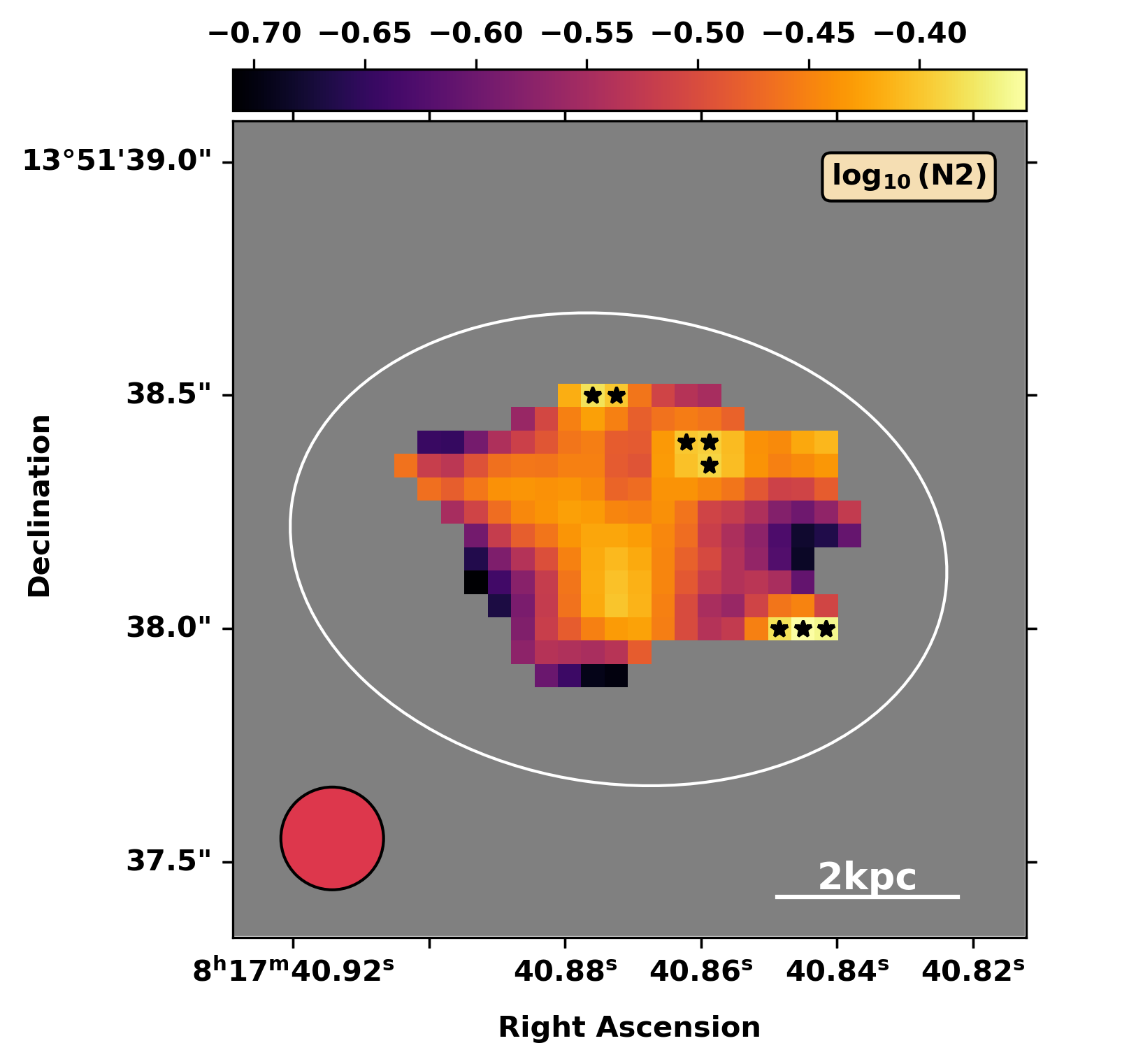}
\caption{Results of WHAN diagnostics. The top panel show the distribution of spaxel values in the $\log_{10}(EW_{\rm H\alpha}/\angstrom)$-$\log_{10}({\rm N2})$ space, with the demarcation lines and classifications of \citet{cidf10} included. The circles and blue diamond show the values found through a fit to the spaxel-by-spaxel spectra and the integrated spectrum, respectively. In the bottom panel, we show the spatial distribution of $\log_{10}({\rm N2})$ (i.e., the x-axis of the upper panel), with spaxels that would be classified as `strong AGN' marked with black stars (and all other coloured spaxels classified as `SF'). The PSF is represented as a red circle to the southeast (north is up and east is to the left), and we include a physical scale of 2\,kpc. Our adopted aperture is shown as the white ellipse.}
\label{whanfig}
\end{figure}

In the top panel of Figure \ref{whanfig}, we show the location of each spaxel-based spectrum in the WHAN diagnostic plot. All of the points for \JJ meet the $\log_{10}(EW_{\rm H\alpha}[\angstrom])>0.5$ criterion necessary to rule out classification as a passive (i.e., no significant ongoing star formation) or retired galaxy (i.e., passive galaxies that are ionised by low-mass stars). This equivalent width could be higher in the case of significant dust attenuation, and nearly all of the spaxels meet the $\log_{10}({\rm N2})<-0.4$ criterion for excitation by star formation rather than an AGN (green points). To investigate the few spaxels that lie in the `strong AGN' regime of this diagnostic (yellow points), we show the spatial distribution of spaxel-by-spaxel WHAN classification in the bottom panel of Figure \ref{whanfig}. The high-N2 points (shown as black stars) lie in the outskirts of the galaxy, and are likely affected by a lower-significance \niib detection. While this suggests that \JJ is likely excited by star formation, further observations will enable tests to search for the presence of an AGN (e.g., the additional diagnostic plots of \citealt{mazz24}, the WHaO diagnostic of \citealt{sanc25}).

\section{Tilted ring models}\label{trmsec}

\subsection{Overview}\label{trm_over}

Previous studies of the \cii emission of \JJ showed evidence for a regular rotating galaxy. Our new JWST/NIRSpec IFU data allow us to perform a similar analysis of the ionised gas, using \ha as a kinematic tracer. To do this, we use the tilted ring fitting code \BB \citep{dite15} to perform a parallel analysis of the \cii and \ha data cubes. In this way, we can compare the molecular and ionised gas morpho-kinematics in a similar fashion.

As a tilted ring fitter, \BB works by decomposing the morphokinematics of a galaxy into a set of rings with independent properties. This code takes as input the desired ring width, ring height, and the number of rings desired. The user provides a set of initial estimates for the morphological (i.e., central position, inclination, position angle, brightness) and kinematic parameters (i.e., rotational velocity, velocity dispersion) for each ring. The code then creates a three-dimensional cube with spatial axes, and populates the cube with gas particles by sampling from the input parameters from each ring. This physical model is converted to a observation-like data cube, with the same geometry as the observed data cube (i.e., right ascension, declination, and velocity). At this point, the mock data are convolved with the PSF (or synthesised beam for interferometric observations) of the observations. The model flux is normalised by comparison to the observed data cube. This allows the data and model cubes to be compared on a pixel-by-pixel basis. 

\subsection{Implementation}
Due to its fitting approach, \BB is able to fit the geometric and kinematic properties of a galaxy simultaneously. Here, we describe the assumptions that go into our modelling.

First, we assume that the galaxy may be modelled by a thin disk with a Gaussian vertical mass distribution (scale height $0.01''$) with maximum radius $r=3.5$\,kpc (e.g., \citealt{neel20}). The ring width is calculated by dividing the FWHM of the minor axis of the PSF (or synthesised beam for ALMA) by 1.5, a conservative middle ground between the values from high-redshift ($\gtrsim2$; e.g., \citealt{shao17,tali18,fan19,jone20}) and low-redshift studies ($\lesssim1$; e.g., \citealt{shel20,manc20,sala20}). This results in five $0.10''$-wide rings for the \cii data and four $0.13''$-wide rings for the \ha data. The kinematic centroid of each model is fixed to the best-fit S\'ersic model centroid of the \ha data (see Section \ref{RSA}).

Because \BB compares the data and model directly, it is vital to first mask the input data cube, to avoid fitting a model to noise. We do this by using the \BB task \textlcsc{SEARCH}, which searches each data cube for emission above some high threshold (which we set to $5\times$ the RMS noise level of the cube) and creates a mask by expanding these regions in both the spatial and spectral directions until it encounters emission lower than a given threshold (which we set to $2\times$ the RMS noise level). We apply this mask to our data.

While \BB is able to fit for the inclination of the  galaxy, there are known degeneracies between the best-fit kinematics and inclination (e.g., \citealt{kamp15}). To avoid this, we fix the inclination of our models to the average value from our S\'ersic fits to the \ha and \cii data ($44\pm2^{\circ}$; see Section \ref{RSA}). This value is in agreement with the result of the morphological fitting code \textlcsc{cannubi} \citealt{roma23} for the \cii data cube ($46\pm4^{\circ}$).

We explore fixing the position angle of our models to the average of the best-fit position angles from the S\'ersic fits ($83\pm3^{\circ}$)\footnote{This agrees with the \textlcsc{cannubi} estimate for the \ha cube ($85\pm6^{\circ}$)}, but this results in a poor result (i.e., a misaligned kinematic major axis). To better capture the kinematic behaviour of \JJ, we use the the best-fit position angle of the S\'ersic models as an initial estimate and allow the value to vary by $\pm25^{\circ}$.


The initial estimate of the systemic redshift is $z=4.2603$. We provide a rotational velocity and velocity dispersion estimate for all rings of $10^2$\,km\,s$^{-1}$, and constrain each value to be $<10^3$\,km\,s$^{-1}$. Our model does not include any radial motion.

We also consider the effects of both spectral resolution and binning. The two data cubes feature different spectral channel widths (25\,km\,s$^{-1}$ for the \cii data cube and 58\,km\,s$^{-1}$ for the \ha cube), which is accounted for by \BB during the fitting process. More subtly, the two datasets have different LSFs. ALMA features a spectral resolution comparable to twice the native channel size\footnote{See Section 6.5 of ALMA Technical Handbook: \url{https://almascience.eso.org/proposing/technical-handbook}} ($\sim12$\,km\,s$^{-1}$), while the \ha cube has a broader resolution of $\sim128$\,km\,s$^{-1}$ (based on a fiducial LSF value of $R=2340$ at $\lambda_{\rm obs}=3.45\,\mu$m, see Section \ref{RSA}). These are accounted for in \BB through the instrumental LSF parameter \textit{LINEAR}.

We first perform an initial tilted ring model fit by allowing all parameters for each ring (i.e., rotational velocity, velocity dispersion, position angle, systemic velocity) to be free. The resulting free parameters are then smoothed with a Bezier function, and a second fit is performed where the position angle and systemic velocity of the galaxy are fixed. 

Throughout the fitting procedure, we fit both sides of the disk simultaneously and apply uniform weighting to all azimuthal angles. The model is normalised on a pixel-by-pixel basis such that the integrated intensity matches that of the masked cube (i.e., `LOCAL' normalisation). We also present the results for `AZIM' normalisation (which enforces azimuthal symmetry in the model flux) in Appendix \ref{azapp}.

\subsection{Results}

\begin{figure*}
\centering
\includegraphics[width=0.49\textwidth]{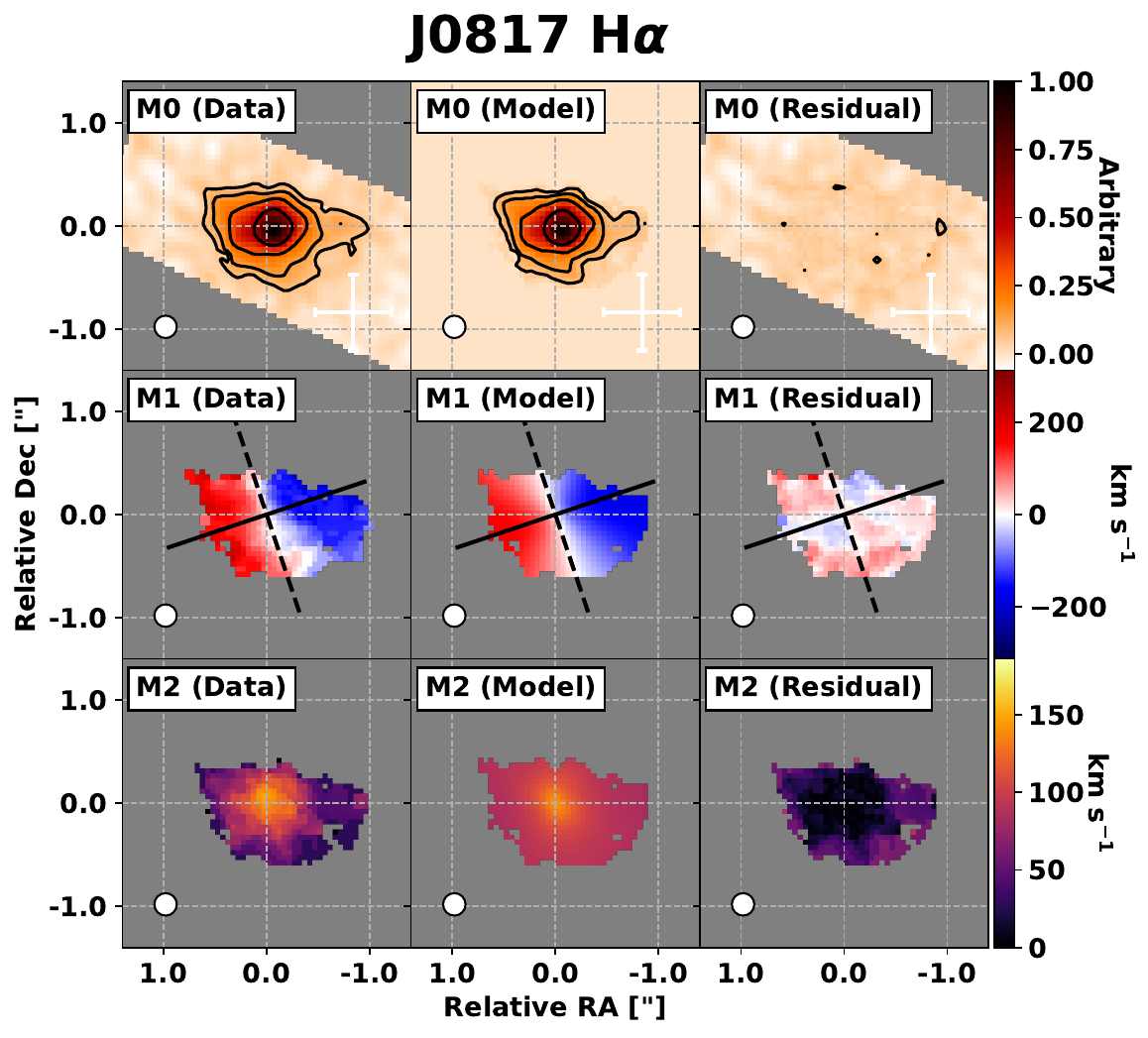}
\includegraphics[width=0.49\textwidth]{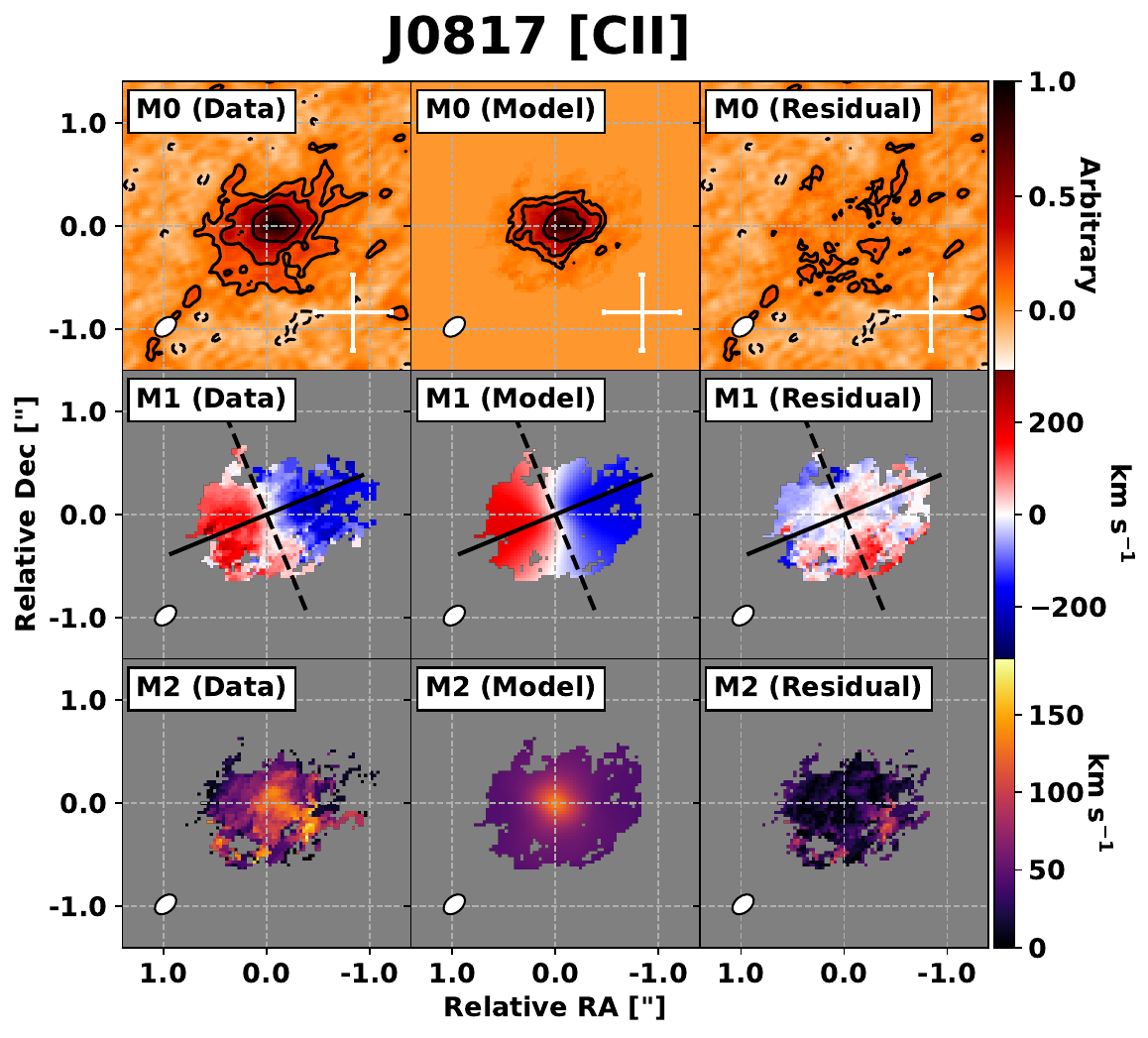}
\\ \hspace{-11.5mm}\includegraphics[width=0.33\textwidth]{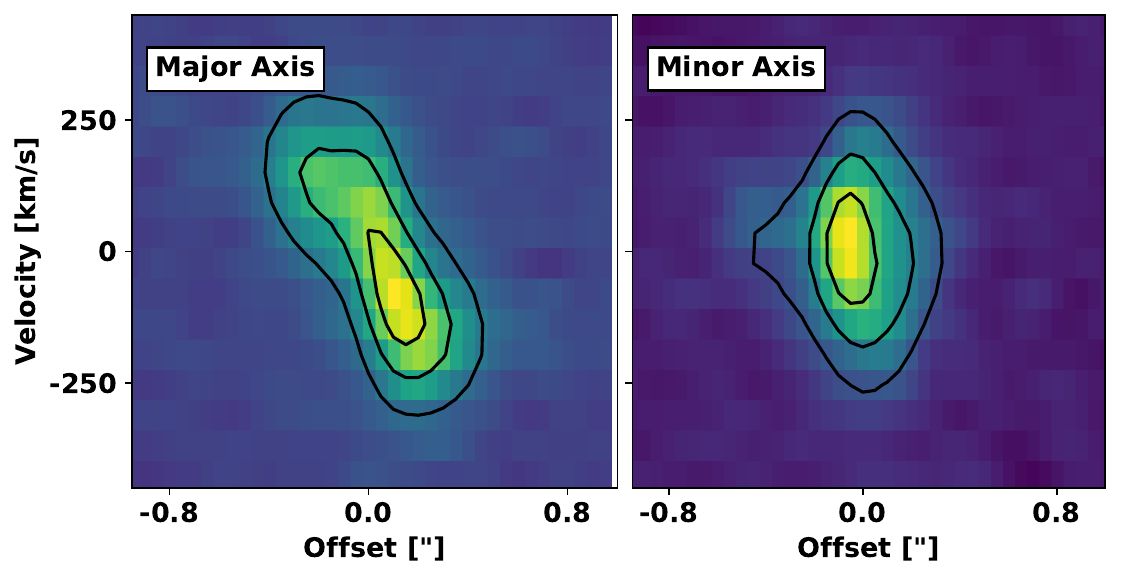}
\hspace{28.5mm}\includegraphics[width=0.33\textwidth]{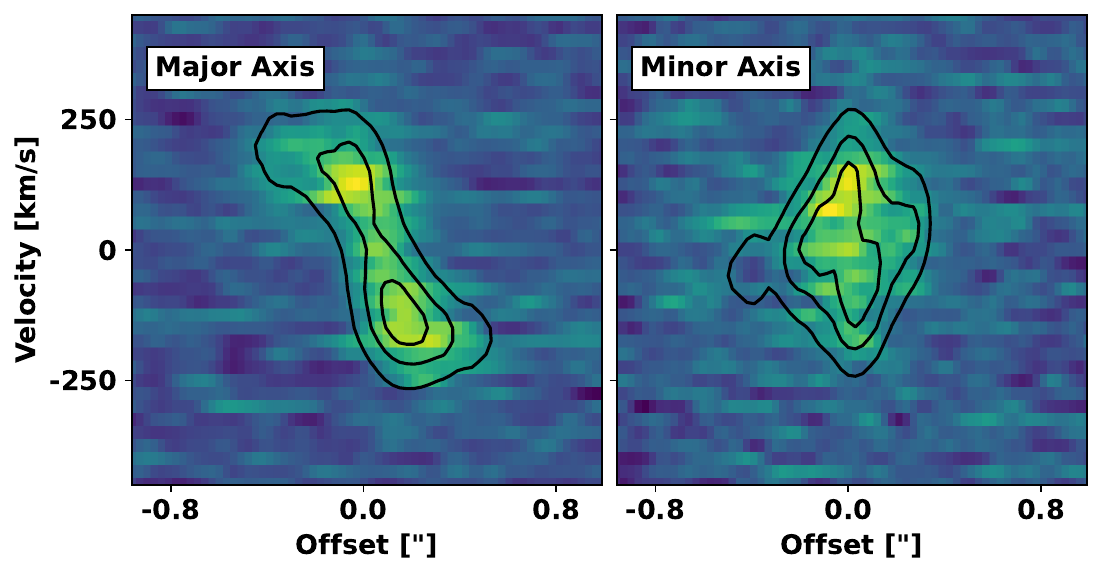}
\caption{Comparison of \BB fits to the \ha (left) and \cii (right) data cubes assuming `LOCAL' normalisation. For each, we present the moment 0 (first row), moment 1 (second row), and moment 2 (third row) for the observed data (left column), model (centre column), and the difference between the two (right column). The white crosses in the top row show a scale of $\rm5\,kpc\times5\,kpc$. The bottom row includes PVDs extracted from the observed data cube (colours) and model (contours) along the major (lower left) and minor (lower right) axes. }
\label{mommaps}
\end{figure*}

The results of this procedure are inspected by comparing the data and model cubes in multiple ways, which we display in Figure \ref{mommaps}. First, we consider the integrated intensity (moment 0; M0), line-of-sight velocity (moment 1; M1), and velocity dispersion (moment 2; M2) maps, which we create using the \textlcsc{moment} task of spectral-cube \citep{robi16}. Next, we use the best-fit position angle to define the kinematic major and minor axes, and extract position-velocity diagrams (PVDs) along both, averaging over three spaxels in the cross-slit direction. The best-fit morphological and kinematic properties are listed in Table \ref{pys_results}.

\begin{figure*}
\centering
\includegraphics[width=\textwidth]{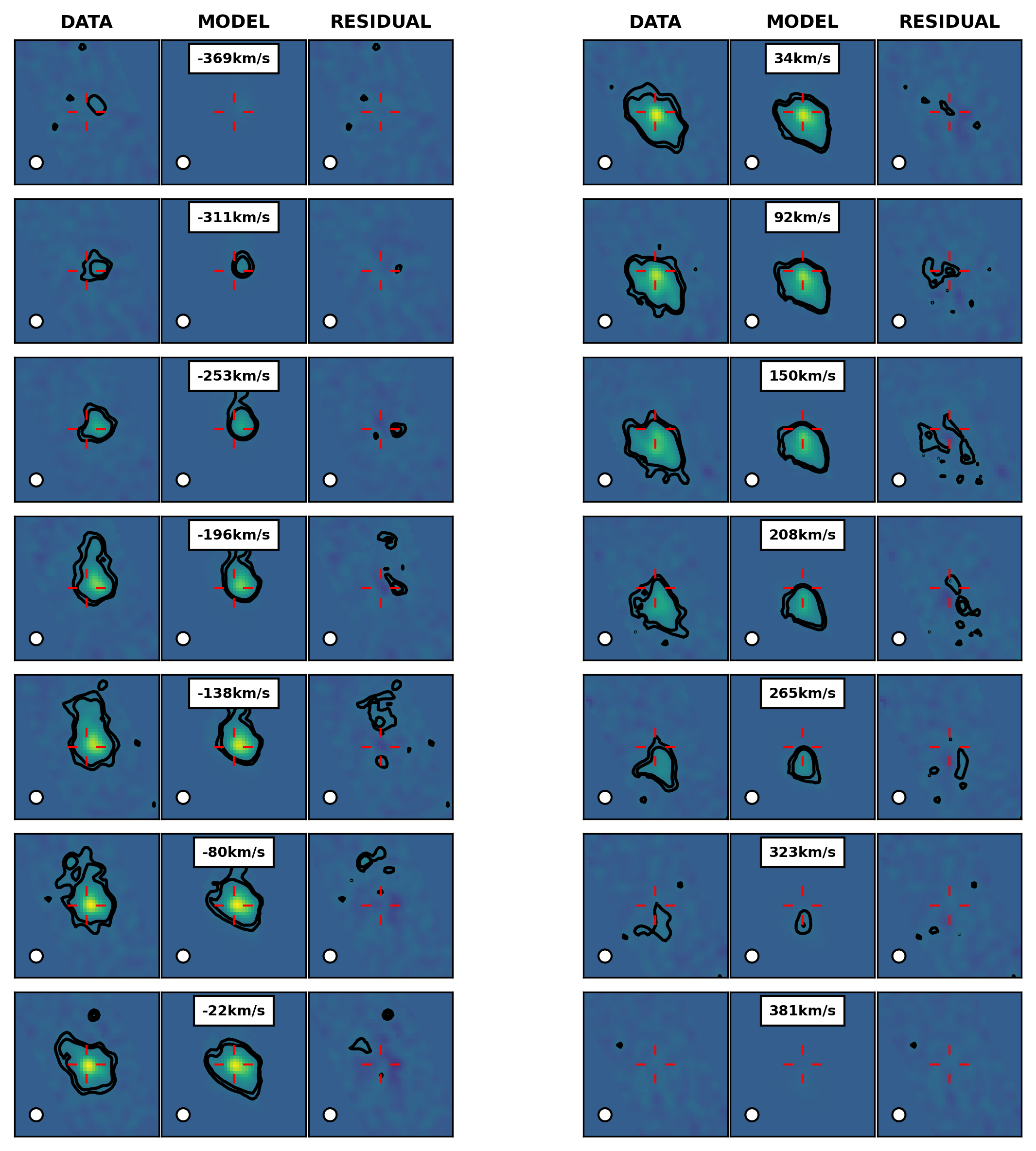}
\caption{Channel maps of the observed \ha data (left columns), the best-fit model (central columns) and the residual (right columns) for all channels used in the \BB fit assuming `LOCAL' normalisation. Each panel features the same colourmap limits and contours (3 and $5\times$ the mean channel RMS noise level). The PSF is included as a white ellipse to the lower left, and the velocity of the channel is included in the `MODEL' panel.}
\label{cm_ha}
\end{figure*}
\begin{figure*}
\centering
\includegraphics[width=\textwidth]{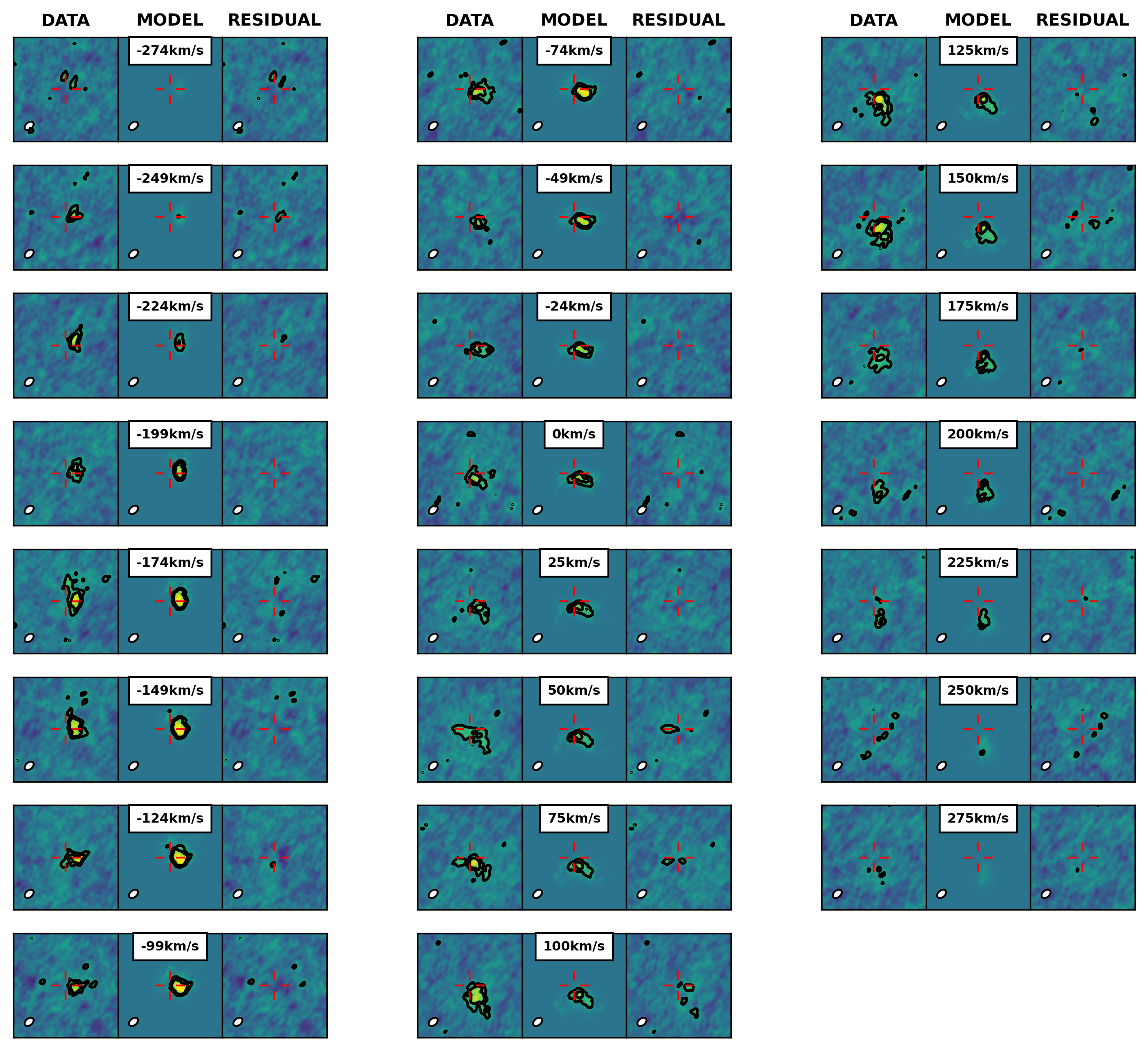}
\caption{Channel maps of the observed \cii data (left columns), the best-fit model (central columns) and the residual (right columns) for all channels used in the \BB fit assuming `LOCAL' normalisation. Each panel features the same colourmap limits and contours (3 and $5\times$ the mean channel RMS noise level). The PSF is included as a white ellipse to the lower left, and the velocity of the channel is included in the `MODEL' panel.}
\label{cm_cii}
\end{figure*}

From Figure \ref{mommaps}, it is clear that the morphology and kinematics are similar between the \cii and \ha data (as already seen in Section \ref{RSA}). The kinematic position angles are in agreement ($<1\sigma$ discrepant) between \ha and \cii. Since the datasets were fit separately, this agreement provides strong evidence for co-spatial emission. Both tracers are well fit by \BB, with a strong east-west velocity gradient (M1) and coincident peaks of intensity and velocity dispersion (M0 and M2, respectively). 
We note that both tracers show red velocity residuals to the southwest (see M1 residual panels of Figure \ref{mommaps}), which are also visible in the minor axis PVDs as positive-velocity clumps not covered by the \BB model contours. Since these are not associated with a higher velocity dispersion, they likely indicate a redshifted gas cloud falling into the galaxy rather than an outflow. The nature of these features will be investigated using JWST/NIRCam imaging in future work.

\begin{table}
\centering
\begin{tabular}{c|ccc}
Property & \ha & \cii \\ \hline
$PA_{\rm kin}$ [$^{\circ}$]	                        &	$109\pm8$	&	$112\pm26$	\\ \hline
$v_{\rm rot}(R_{\rm e})$ [km\,s$^{-1}$]	            &	$230\pm15$	&	$269\pm25$	\\
$\sigma_{\rm o}(R_{\rm e})$ [km\,s$^{-1}$]	        &	$87\pm11$	&	$55\pm10$	\\
$v_{\rm rot}(R_{\rm e})/\sigma_{\rm o}(R_{\rm e})$  &	$2.6\pm0.4$	&	$4.9\pm1.1$	\\ \hline
$\log_{10}(M_{\rm dyn}/M_{\odot})$	                &	$10.8\pm0.1$	&	$10.9\pm0.1$	\\ \hline
\end{tabular}
\caption{Best-fit morpho-kinematic properties of the \ha and \cii emission, as derived through our \BB fits using `LOCAL' normalisation. Details of the dynamical mass estimation are given in Section \ref{dynmasssec}.}
\label{pys_results}
\end{table}

Since \BB fits the data on a pixel-by-pixel basis, we may also compare the data and model for each spectral channel. Figures \ref{cm_ha} and \ref{cm_cii} show the channel maps (data, model, and residual) for the \ha and \cii data cubes, respectively. While some minor residuals are present, there is no large-scale disagreement.

\begin{figure*}
\centering
\includegraphics[width=0.49\textwidth]{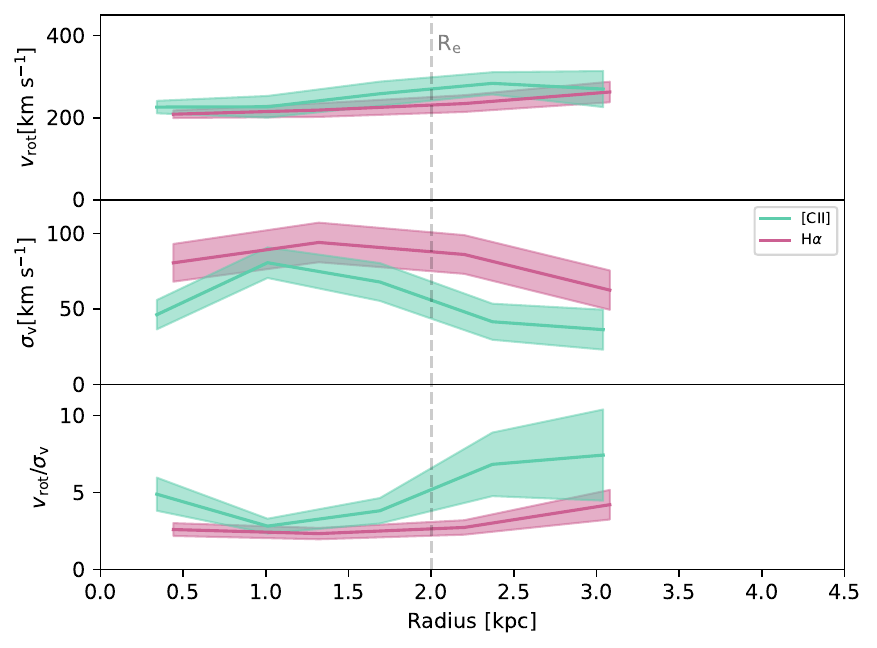}
\includegraphics[width=0.49\textwidth]{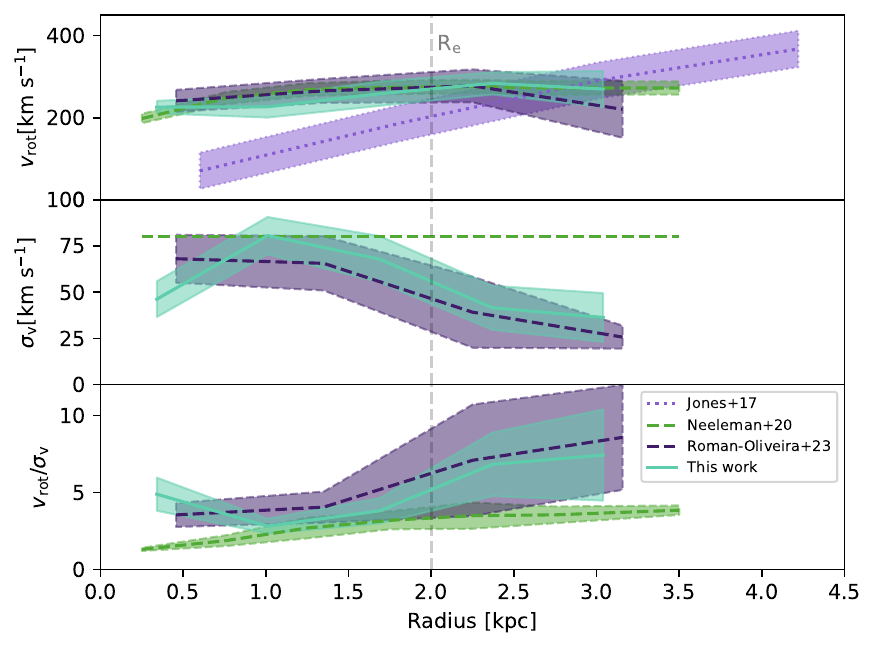}
\caption{Comparison of best-fit rotation curves (top panels), velocity dispersion profiles (centre panels), and the ratio of the two (lower panels) for the \cii and \ha data studied in this work (left column). In the right column, we compare our \cii values for \JJ to those of previous works that studies the same galaxy (\citealt{jone17,neel20,roma23}). For each panel, we mark the effective radius of the \ha emission (as derived through a S\'ersic fit; Section \ref{RSA}) with a dashed vertical line.}
\label{rotcurves}
\end{figure*}

While all of the panels in Figure \ref{mommaps} include observational effects (e.g., synthesised beam or PSF smearing, spectral and spatial binning), we may also compare the best-fit parameters directly. One of the primary ways that we may compare the \BB fits is to plot the best-fit curves of rotation velocity, velocity dispersion, and the ratio of the two. The top left panel of Figure \ref{rotcurves} shows a remarkable agreement between the best-fit rotation curves of the two datasets, despite the fact that the modelling was done separately (i.e., without linked priors). On the other hand, the velocity dispersion of \ha is higher than that of \cii (central left panel), resulting in a higher degree of rotational support based on the \cii data (lower left panel). 

This apparent discrepancy originates from the different phases probed by the two emission lines: molecular gas for \cii and ionised gas for \ha. Observations and simulations have demonstrated that while ionised and molecular gas kinematics trace the same rotation profile, the velocity dispersion of ionised gas can reach $\sim2-3$ that of molecular gas (e.g., \citealt{fuji25,koha24,parl24,rizz24}). However, we find that the two dispersions differ by less than a factor of 2. Since they feature similar morphologies (based on S\'ersic model fits; Section \ref{RSA}), kinematic position angles, rotational velocity profiles, and inferred dynamical masses, they likely originate from the same disk (i.e., \ha is not primarily from large-scale extraplanar gas).

As noted in Section \ref{almasec}, the \cii data for this galaxy have already been analysed by multiple works. This opens a useful avenue to verify our approach through comparisons to previous analyses of the same galaxy that used a variety of dynamical fitting codes, as shown in the right column of Figure \ref{rotcurves}. The first (\citealt{jone17}; purple dotted line) used the \textit{rotcur} task from the `The Groningen Image Processing SYstem' package (\textlcsc{GIPSY}; \citealt{vand92}) to fit a tilted ring model to the velocity field of the low-resolution \cii data (2015.1.01564.S; $\sim0.8''$ resolution; see Section \ref{almasec}). The beam size was not included in the modelling, resulting in significant beam smearing, as seen in a steeper profile than later analyses. 

On the other hand, \citet{neel20} and \citet{roma23} used the codes \textlcsc{qubefit} and \BB, respectively. While both perform a fit to the full data cube (i.e., not only moment maps) and account for the PSF, \BB uses a model composed of several rings (see Section \ref{trm_over}) while \textlcsc{qubefit} is a parametric code and fits for the best parameters of a brightness profile (e.g., exponential disk) and rotation curve (e.g., linear or arctangent, with constant velocity dispersion). The different approaches yield similar rotation curves, but different velocity dispersion profiles. Since we use \BB as well, our results are in agreement with those of \citet{roma24}.


\section{Discussion}\label{discsec}

\subsection{Mass of \JJ}

\subsubsection{Dynamical mass}\label{dynmasssec}

Our \BB analysis allows us to estimate the total dynamical mass of the system. While some works estimate the dynamical mass of a rotating disk galaxy using only the observed rotational velocity ($M_{\rm dyn}(r<R_{\rm e})=R_{\rm e}v_{\rm rot}(R_{\rm e})^2/G$; e.g., \citealt{debr14,jone17}), this approach is only valid for sources with extremely high rotational support (i.e., $v_{\rm rot}/\sigma>10$; \citealt{phil25}). More generally, the contribution of support from random motion must be taken into account - an effect generally labelled as the asymmetric drift correction. 

This correction is included by replacing the rotational velocity with a `circular velocity' of $v_{\rm circ}=\sqrt{v_{\rm rot}^2+{v_{\rm A}^2}}$ (e.g., \citealt{genz17,fors18,pric20,danh25}). We adopt the form of $v_{\rm A}$ of \citet{roma24}:
\begin{equation}
v_{\rm A}(R_{\rm e})^2 = 
\sigma_{\rm v}(R_{\rm e})^2 \times \left[1.678 -2R_{\rm e} \frac{\partial ln(\sigma_{\rm v})}{\partial R}\bigg|_{R_{\rm e}} \right]
\end{equation}
which assumes that the gas in the disk (as traced by \cii) is in an exponential disk ($n_{\rm s}=1$, similar to the best-fit values of this work; Table \ref{pys_results_real}). This circular velocity may be used to estimate the total dynamical mass:
\begin{equation}\label{mdyneq}
M_{\rm dyn} = k_{\rm tot}\frac{R_{\rm e}v_{\rm circ}(R_{\rm e})^2}{G}
\end{equation}
where the virial coefficient $k_{\rm tot}$ is dependent on the geometry of the galaxy (i.e., $n_{\rm s}$ and intrinsic spheroid axis ratio $q_{\rm 0}\equiv c/a$; e.g., \citealt{pric22}).


To apply equation \ref{mdyneq} to our data, we use the \textlcsc{pysersic} fits of the \ha and \cii data performed in Section \ref{RSA}. The results of \citet{pric22} may then be used to calculate $k_{\rm tot}=2.1\pm0.3$ (assuming a nearly flat disk of $q_{\rm 0}=0.4\pm0.2$). By interpolating the best-fit rotation curves and velocity dispersion curves (Figure \ref{rotcurves}), we find a total dynamical mass of $\log_{10}(M_{\rm dyn}/M_{\odot})\sim10.9$ for both tracers (Table \ref{pys_results}), in agreement with previous results (e.g., \citealt{neel20})\footnote{We do not discuss the \cii-based dynamical mass estimates of \citet{jone17} and \citet{jone20} due to a lack of PSF treatment and asymmetric drift correction, respectively.}.

\subsubsection{Gas mass}\label{gasmass}
Based on CO(2-1) observations, \citet{neel20} estimate the molecular gas mass of \JJ to be $\log_{10}(M_{\rm H_2}/M_{\odot})=10.9^{+0.1}_{-0.2}$ (assuming $r_{21}=0.81$, $\alpha_{\rm CO}=3.0\,\mathrm{M_{\odot}\,K^{-1}\,km^{-1}\,s\,pc^{-2}}$). However, the necessary factors to convert from $L_{\rm CO(2-1)}$ to $M_{\rm H_2}$ (i.e., $r_{21}$, $\alpha_{\rm CO}$) have been found to vary between galaxies (e.g., \citealt{hu22,teng22,anir25}), with $\alpha_{\rm CO}$ featuring a strong dependence on metallicity (e.g., \citealt{nara12,madd20}). The fact that this mass is comparable to the dynamical mass introduces some doubt as to the applicability of the conversion factors. 

While the factor $r_{\rm 21}$ has been observed to feature a significant variation within a galaxy (e.g., \citealt{yaji21,lero22}), the spatially integrated factors usually only vary by a factor of $\lesssim2$ (e.g., \citealt{cari13,maed22}). Increasing $r_{21}$ from $0.81$ (as assumed by \citealt{neel20}) to an extreme value of unity (appropriate for quasar host galaxies; \citealt{cari13}) results in a $<1\sigma$ decrease in $M_{\rm mol}$, so the assumed $r_{\rm 21}$ value is likely not the source of this high gas mass. 

A number of metallicity-dependent formulations of $\alpha_{\rm CO}$ have been measured (e.g., \citealt{glov11,schr12,bola13,genz15,amor16,accu17,tacc18}, as collected by \citealt{madd20}), which yield a large range of $\alpha_{\rm CO}\sim3-20\,\mathrm{M_{\odot}\,K^{-1}\,km^{-1}\,s\,pc^{-2}}$ for $Z/Z_{\odot}\sim0.8$. Since a higher $\alpha_{\rm CO}$ for a given $L'_{\rm CO}$ will result in a higher estimated gas mass, these values would only result in an estimated gas mass greater than the dynamical mass. On the other hand, the \citet{nara12} calibration (which includes an additional dependence on the integrated CO luminosity) yields $\alpha_{\rm CO}\sim0.5\,\mathrm{M_{\odot}\,K^{-1}\,km^{-1}\,s\,pc^{-2}}$ and $M_{\rm H_2}\sim10^{10.0}\,M_{\odot}$. So while a high $\alpha_{\rm CO}$ may be the culprit behind the large gas mass estimate, further analysis of this conversion factor at high redshift is needed.

While \cii line has historically been used a tracer of SFR (e.g., \citealt{delo14,scha20}) and $M_{\rm HI}$ (e.g., \citealt{hein21}), it has seen great recent success as a tracer of $M_{\rm H_2}$ (e.g., \citealt{zane18,dess20,vizg22}). Here, we adopt the metallicity-dependent $L_{\rm [CII]}$-$M_{\rm gas}$ conversion of \citet{vall25}, where $M_{\rm gas}$ includes both atomic and molecular cold gas. Using the gas-phase metallicity derived from our data (Table \ref{linetable}) results in $\alpha_{\rm [CII]}\equiv M_{\rm gas}/L_{\rm [CII]}=5.2\,M_{\odot}/L_{\rm \odot}$, and $\log_{10}(M_{\rm gas,[CII]}/M_{\odot})=10.23\pm0.04$ (see Table \ref{linetable}; assuming the $L_{\rm [CII]}$ value of \citealt{neel20} which was extracted using a circular aperture of radius $1.5''\simeq10.5$\,kpc). This is $\sim0.6$\,dex lower than the previously determined $M_{\rm H_2,CO}$ and $M_{\rm dyn}$, relieving a discrepancy in the mass budget.

We note that the previous work of \citet{zane18} found a much higher $\alpha_{\rm [CII]}\sim30\,M_{\odot}/L_{\rm \odot}$, which would result in a $\sim0.8$\,dex higher gas mass estimate. However, \citet{rizz21} found that for a sample of $4\lesssim z\lesssim5$ dusty, rotating SFGs (i.e., similar galaxies to \JJ), a lower conversion factor of $\alpha_{\rm [CII]}\sim5-20\,M_{\odot}/L_{\rm \odot}$ was preferred. Simulations have also found a lower value, in agreement with our estimate (e.g., \citealt{guru24,casa25}).

\subsubsection{Mass budget}
As noted in Section \ref{intro}, the dynamical mass is the sum of both the baryonic (stars, gas, and dust) and the non-baryonic (dark matter) components of a galaxy. If we subtract the gas mass (Section \ref{gasmass}), expected stellar mass (\citealt{roma24}), and dust mass (Appendix \ref{FIRSED}) from the dynamical mass (Section \ref{dynmasssec}), we may constrain the dark matter content of this source.

To examine the dark matter component of \JJ, we consider a dark matter halo whose mass density follows an NFW profile \citep{nava97}. The mass within a given radius takes the form (e.g., \citealt{debl08}):
\begin{equation}
M_{\rm NFW}(<R) = 4\pi \rho_{\rm crit}\delta_{\rm c}\int_{0}^{R}\frac{r^2}{(1+r/R_{\rm S})^2(r/R_{\rm S})}\delta r
\end{equation}
where $\rho_{\rm crit}=3H^2/8\pi G$ is the critical density of the Universe at a given redshift, $R_{\rm S}$ is the characteristic radius of the halo, and $\delta_{\rm c}$ is a function of the concentration $c_{200}\equiv R_{200}/R_{\rm S}$ \citep{ludl13}. Following other works (e.g., \citealt{mits20,frat21,roma24,neel25}), we adopt $c_{200}=3.4$ (valid for halos of mass $\sim10^{10.0-12.5}\,M_{\odot}$ at $z\sim4$; \citealt{dutt14}) and $R_{200}=100$\,kpc (valid for a halo of mass $\sim10^{12}\,M_{\odot}$). This results in an enclosed mass of $M_{\rm halo}(r<{\rm 3.5kpc})\sim10^{10.7}\,M_{\odot}$. We note that this value is highly dependent on the assumed scale radius, with $R_{200}=70$\,kpc yielding an enclosed mass that is lower by $0.2$\,dex. 

Combining the estimated stellar and dark matter masses results in a value of $\sim10^{11.0}\,M_{\odot}$, which is $\sim1\sigma$ larger than the dynamical mass of \JJ. But if we lower the estimated $M_*$ by $1\sigma$ and assume $R_{200}=70$\,kpc, we find an estimate of $\sim10^{10.75}\,M_{\odot}$. This is comparable to the difference between the dynamical mass and sum of the \cii-based gas mass and dust mass, implying the lack of a discrepancy in this budget. Precise estimates of $M_*$ and $M_{\rm dust}$ will be determined through spectral energy distribution (SED) fitting in a future work, but we note that the dynamical mass estimate ($\sim10^{10.9}\,M_{\odot}$) of this work supports a decomposition into contributions from dark matter ($\sim10^{10.4}\,M_{\odot}$), stars ($\sim10^{10.4}\,M_{\odot}$), gas ($\sim10^{10.2}\,M_{\odot}$), and dust ($\sim10^{8.0}\,M_{\odot}$).

\subsection{Nature of \JJ}

By combining the results of this work with those of previous analyses, we may consider a more complete picture of \JJ. The remarkable agreement of the \cii and \ha morpho-kinematics confirms that this is a relaxed rotating disk, with no significant evidence for ongoing major mergers, outflows, or AGN (from the WHAN diagram). It is \cii-luminous, with a FIR-based SFR that far exceeds the NUV- or \ha-based values - suggesting a high fraction of obscured star formation. This is further supported by a preliminary FIR SED analysis, which shows evidence for a significant dust reservoir. We also find a relatively high gas-phase metallicity ($0.8\pm0.2$\,solar), implying a history of chemical enrichment.

It is clear that this source is not undergoing a major starburst episode, as evidenced by the low ionization parameter and likely placement on the SFMS. Its electron density is similar to `normal' (i.e., representative of the most numerous population of observed sources) galaxies at $z\sim4$. Our estimates of the molecular gas mass show that this is not a gas-dominated source ($f_{\rm gas}\sim1$), as previously suggested, but instead contains a large gas reservoir, with room in the mass budget for a significant stellar population.

From this, we label \JJ as a post-burst, main-sequence disk galaxy - one that has a history of significant star formation (resulting in a high metallicity and a dust reservoir that obscures ongoing star formation), but is not currently highly excited (based on the low $U$ and $SFR$). This maturity is noteworthy, as its high redshift means that it only had $\lesssim1.4$\,Gyr to evolve. Our dynamical analysis implies that the material in this galaxy has relaxed into a disc, with any signatures of major disturbances already faded. Further properties (e.g., new constraints on stellar mass, star formation history, possible stellar rotation) are deferred to future analyses

\section{Conclusions}\label{concsec}

In this work, we present new high spectral resolution ($R\sim2700$) JWST/NIRSpec IFU observations of the rotating disk galaxy \JJ. Combining our data with archival ALMA observations allows us to explore the multi-phase morphology, kinematics, and ISM conditions of this $z\sim4$ galaxy.

By combining the fluxes of our detected emission lines with the upper limits of undetected lines and the code \pyneb, we find evidence for an electron density that is in agreement with the redshift-dependent correlation of previous works ($\log_{10}(n_{\rm e} {\rm [cm^{-3}]})=2.5\pm0.4$) and a conservative estimate on $E(B-V)<3.1$. A strong-line metallicity calibration yields a high metallicity for this redshift ($0.8\pm0.2$\,solar), while the S32 ratio is used to predict a relatively low ionization parameter ($\log_{10}(U)\sim-3.5$). Despite only having one observed Balmer transition, we use the WHAN diagnostic to show a lack of evidence for the presence of an AGN (but note that this is not the same as evidence for the lack of an AGN).

The \cii and \ha emission show remarkably similar morpho-kinematics. S\'ersic profile fits to the integrated maps of each result in best-fit parameters that are in agreements, while both lines show similar line of sight velocity (moment 1) and velocity dispersion (moment 2) maps. Through \BB fits to each data cube, we demonstrate that while the two tracers exhibit the same rotational velocity, \ha features a velocity dispersion that is higher than that of \cii.


We combine our new estimate of the gas-phase metallicity with metallicity-dependent conversion factors to estimate the molecular gas mass of \JJ using the CO(2-1) and \cii luminosities. With these factors, we no longer find the previously reported extreme $M_{\rm H_2}$ (which was comparable to the dynamical mass), but instead a gas mass that is $\sim20\%$ of the dynamical mass. This leaves space in the mass budget for a sizeable stellar mass (as suggested by previous mass decomposition analyses) and a dark matter halo.

Finally, we synthesise our findings to explore the nature of \JJ, finding that it is a representative galaxy (in terms of electron density and likely star formation) undergoing ordered rotation (as traced by the molecular and ionised gas), with signs of past star formation (i.e., high $Z$, evidence for warm dust emission, low $U$) and potential for more SF (sizeable molecular gas reservoir). Future works will explore additional properties (e.g., the stellar mass, the presence of AGN or a Ly$\alpha$ halo).

\section*{Acknowledgements}

We thank the anonymous referee for constructive feedback that strengthened this work.
GCJ, FDE, and JS acknowledge support by the Science and Technology Facilities Council (STFC), by the ERC through Advanced Grant 695671 ``QUENCH'', and by the UKRI Frontier Research grant RISEandFALL.
SA, MP, and BRP acknowledge support from the research projects PID2021-127718NB-I00, PID2024-159902NA-I00, PID2024-158856NA-I00, and RYC2023-044853-I of the Spanish Ministry of Science and Innovation/State Agency of Research (MCIN/AEI/10.13039/501100011033), FSE+, and by “ERDF A way of making Europe”.
AJB acknowledges funding from the ``FirstGalaxies'' Advanced Grant from the European Research Council (ERC) under the European Union’s Horizon 2020 research and innovation programme (Grant agreement No. 789056).
GC acknowledges the support of the INAF Large Grant 2022 "The metal circle: a new sharp view of the baryon cycle up to Cosmic Dawn with the latest generation IFU facilities" and the INAF GO grant ``A JWST/MIRI MIRACLE: Mid-IR Activity of Circumnuclear Line Emission''.
IL acknowledges support from PRIN-MUR project “PROMETEUS”  financed by the European Union -  Next Generation EU, Mission 4 Component 1 CUP B53D23004750006.
H\"U acknowledges funding by the European Union (ERC APEX, 101164796). Views and opinions expressed are however those of the authors only and do not necessarily reflect those of the European Union or the European Research Council Executive Agency. Neither the European Union nor the granting authority can be held responsible for them.
SZ acknowledges support from the European Union (ERC, WINGS, 101040227).
This paper makes use of the following ALMA data: ADS/JAO.ALMA\#2017.1.01052. ALMA is a partnership of ESO (representing its member states), NSF (USA) and NINS (Japan), together with NRC (Canada), MOST and ASIAA (Taiwan), and KASI (Republic of Korea), in cooperation with the Republic of Chile. The Joint ALMA Observatory is operated by ESO, AUI/NRAO and NAOJ.
GCJ thanks Sandro Tacchella for useful discussion.

\section*{Data Availability}

The JWST/NIRSpec IFU data studied here (Project 4528, observation 7) is available from the MAST archive. The ALMA data studied in this work is available from the ALMA archive (\url{https://almascience.eso.org/aq/}) under project code 2017.1.01052.



\bibliographystyle{mnras}
\bibliography{example} 

@ARTICLE{rafe12,
       author = {{Rafelski}, Marc and {Wolfe}, Arthur M. and {Prochaska}, J. Xavier and {Neeleman}, Marcel and {Mendez}, Alexander J.},
        title = "{Metallicity Evolution of Damped Ly{\ensuremath{\alpha}} Systems Out to z \raisebox{-0.5ex}\textasciitilde 5}",
      journal = {\apj},
         year = 2012,
        month = aug,
       volume = {755},
       number = {2},
          eid = {89},
        pages = {89},
          doi = {10.1088/0004-637X/755/2/89},
archivePrefix = {arXiv},
       eprint = {1205.5047},
 primaryClass = {astro-ph.CO},
       adsurl = {https://ui.adsabs.harvard.edu/abs/2012ApJ...755...89R}
}

@ARTICLE{neel17,
       author = {{Neeleman}, Marcel and {Kanekar}, Nissim and {Prochaska}, J. Xavier and {Rafelski}, Marc and {Carilli}, Chris L. and {Wolfe}, Arthur M.},
        title = "{[C II] 158-{\ensuremath{\mu}}m emission from the host galaxies of damped Lyman-alpha systems}",
      journal = {Science},
         year = 2017,
        month = mar,
       volume = {355},
       number = {6331},
        pages = {1285-1288},
          doi = {10.1126/science.aal1737},
archivePrefix = {arXiv},
       eprint = {1703.07797},
 primaryClass = {astro-ph.GA},
       adsurl = {https://ui.adsabs.harvard.edu/abs/2017Sci...355.1285N}
}

@ARTICLE{jone17,
       author = {{Jones}, G.~C. and {Carilli}, C.~L. and {Shao}, Y. and {Wang}, R. and {Capak}, P.~L. and {Pavesi}, R. and {Riechers}, D.~A. and {Karim}, A. and {Neeleman}, M. and {Walter}, F.},
        title = "{Dynamical Characterization of Galaxies at z {\ensuremath{\sim}} 4-6 via Tilted Ring Fitting to ALMA [C II] Observations}",
      journal = {\apj},
         year = 2017,
        month = dec,
       volume = {850},
       number = {2},
          eid = {180},
        pages = {180},
          doi = {10.3847/1538-4357/aa8df2},
archivePrefix = {arXiv},
       eprint = {1709.04954},
 primaryClass = {astro-ph.GA},
       adsurl = {https://ui.adsabs.harvard.edu/abs/2017ApJ...850..180J}
}

@ARTICLE{jone21,
       author = {{Jones}, G.~C. and {Vergani}, D. and {Romano}, M. and {Ginolfi}, M. and {Fudamoto}, Y. and {B{\'e}thermin}, M. and {Fujimoto}, S. and {Lemaux}, B.~C. and {Morselli}, L. and {Capak}, P. and {Cassata}, P. and {Faisst}, A. and {Le F{\`e}vre}, O. and {Schaerer}, D. and {Silverman}, J.~D. and {Yan}, Lin and {Boquien}, M. and {Cimatti}, A. and {Dessauges-Zavadsky}, M. and {Ibar}, E. and {Maiolino}, R. and {Rizzo}, F. and {Talia}, M. and {Zamorani}, G.},
        title = "{The ALPINE-ALMA [C II] Survey: kinematic diversity and rotation in massive star-forming galaxies at z 4.4-5.9}",
      journal = {\mnras},
         year = 2021,
        month = nov,
       volume = {507},
       number = {3},
        pages = {3540-3563},
          doi = {10.1093/mnras/stab2226},
archivePrefix = {arXiv},
       eprint = {2104.03099},
 primaryClass = {astro-ph.GA},
       adsurl = {https://ui.adsabs.harvard.edu/abs/2021MNRAS.507.3540J}
}

@ARTICLE{neel20,
       author = {{Neeleman}, Marcel and {Prochaska}, J. Xavier and {Kanekar}, Nissim and {Rafelski}, Marc},
        title = "{A cold, massive, rotating disk galaxy 1.5 billion years after the Big Bang}",
      journal = {\nat},
         year = 2020,
        month = may,
       volume = {581},
       number = {7808},
        pages = {269-272},
          doi = {10.1038/s41586-020-2276-y},
archivePrefix = {arXiv},
       eprint = {2005.09661},
 primaryClass = {astro-ph.GA},
       adsurl = {https://ui.adsabs.harvard.edu/abs/2020Natur.581..269N}
}

@ARTICLE{spea14,
       author = {{Speagle}, J.~S. and {Steinhardt}, C.~L. and {Capak}, P.~L. and {Silverman}, J.~D.},
        title = "{A Highly Consistent Framework for the Evolution of the Star-Forming ``Main Sequence'' from z \raisebox{-0.5ex}\textasciitilde 0-6}",
      journal = {\apjs},
         year = 2014,
        month = oct,
       volume = {214},
       number = {2},
          eid = {15},
        pages = {15},
          doi = {10.1088/0067-0049/214/2/15},
archivePrefix = {arXiv},
       eprint = {1405.2041},
 primaryClass = {astro-ph.GA},
       adsurl = {https://ui.adsabs.harvard.edu/abs/2014ApJS..214...15S}
}

@ARTICLE{nara12,
       author = {{Narayanan}, Desika and {Krumholz}, Mark R. and {Ostriker}, Eve C. and {Hernquist}, Lars},
        title = "{A general model for the CO-H$_{2}$ conversion factor in galaxies with applications to the star formation law}",
      journal = {\mnras},
         year = 2012,
        month = apr,
       volume = {421},
       number = {4},
        pages = {3127-3146},
          doi = {10.1111/j.1365-2966.2012.20536.x},
archivePrefix = {arXiv},
       eprint = {1110.3791},
 primaryClass = {astro-ph.GA},
       adsurl = {https://ui.adsabs.harvard.edu/abs/2012MNRAS.421.3127N}
}

@ARTICLE{dite15,
       author = {{Di Teodoro}, E.~M. and {Fraternali}, F.},
        title = "{$^{3D}$ BAROLO: a new 3D algorithm to derive rotation curves of galaxies}",
      journal = {\mnras},
         year = 2015,
        month = aug,
       volume = {451},
       number = {3},
        pages = {3021-3033},
          doi = {10.1093/mnras/stv1213},
archivePrefix = {arXiv},
       eprint = {1505.07834},
 primaryClass = {astro-ph.GA},
       adsurl = {https://ui.adsabs.harvard.edu/abs/2015MNRAS.451.3021D}
}

@ARTICLE{luri15,
       author = {{Luridiana}, V. and {Morisset}, C. and {Shaw}, R.~A.},
        title = "{PyNeb: a new tool for analyzing emission lines. I. Code description and validation of results}",
      journal = {\aap},
         year = 2015,
        month = jan,
       volume = {573},
          eid = {A42},
        pages = {A42},
          doi = {10.1051/0004-6361/201323152},
archivePrefix = {arXiv},
       eprint = {1410.6662},
 primaryClass = {astro-ph.IM},
       adsurl = {https://ui.adsabs.harvard.edu/abs/2015A&A...573A..42L}
}

@software{newv14,
       author = {{Newville}, Matthew and {Stensitzki}, Till and {Allen}, Daniel B. and {Ingargiola}, Antonino},
        title = "{LMFIT: Non-Linear Least-Square Minimization and Curve-Fitting for Python}",
         year = 2014,
        month = sep,
          eid = {10.5281/zenodo.11813},
          doi = {10.5281/zenodo.11813},
      version = {0.8.0},
    publisher = {Zenodo},
       adsurl = {https://ui.adsabs.harvard.edu/abs/2014zndo.....11813N}
}

@ARTICLE{casa22,
       author = {{CASA Team} and {Bean}, Ben and {Bhatnagar}, Sanjay and {Castro}, Sandra and {Donovan Meyer}, Jennifer and {Emonts}, Bjorn and {Garcia}, Enrique and {Garwood}, Robert and {Golap}, Kumar and {Gonzalez Villalba}, Justo and {Harris}, Pamela and {Hayashi}, Yohei and {Hoskins}, Josh and {Hsieh}, Mingyu and {Jagannathan}, Preshanth and {Kawasaki}, Wataru and {Keimpema}, Aard and {Kettenis}, Mark and {Lopez}, Jorge and {Marvil}, Joshua and {Masters}, Joseph and {McNichols}, Andrew and {Mehringer}, David and {Miel}, Renaud and {Moellenbrock}, George and {Montesino}, Federico and {Nakazato}, Takeshi and {Ott}, Juergen and {Petry}, Dirk and {Pokorny}, Martin and {Raba}, Ryan and {Rau}, Urvashi and {Schiebel}, Darrell and {Schweighart}, Neal and {Sekhar}, Srikrishna and {Shimada}, Kazuhiko and {Small}, Des and {Steeb}, Jan-Willem and {Sugimoto}, Kanako and {Suoranta}, Ville and {Tsutsumi}, Takahiro and {van Bemmel}, Ilse M. and {Verkouter}, Marjolein and {Wells}, Akeem and {Xiong}, Wei and {Szomoru}, Arpad and {Griffith}, Morgan and {Glendenning}, Brian and {Kern}, Jeff},
        title = "{CASA, the Common Astronomy Software Applications for Radio Astronomy}",
      journal = {\pasp},
         year = 2022,
        month = nov,
       volume = {134},
       number = {1041},
          eid = {114501},
        pages = {114501},
          doi = {10.1088/1538-3873/ac9642},
archivePrefix = {arXiv},
       eprint = {2210.02276},
 primaryClass = {astro-ph.IM},
       adsurl = {https://ui.adsabs.harvard.edu/abs/2022PASP..134k4501C}
}

@ARTICLE{curt20,
       author = {{Curti}, Mirko and {Mannucci}, Filippo and {Cresci}, Giovanni and {Maiolino}, Roberto},
        title = "{The mass-metallicity and the fundamental metallicity relation revisited on a fully T$_{e}$-based abundance scale for galaxies}",
      journal = {\mnras},
         year = 2020,
        month = jan,
       volume = {491},
       number = {1},
        pages = {944-964},
          doi = {10.1093/mnras/stz2910},
archivePrefix = {arXiv},
       eprint = {1910.00597},
 primaryClass = {astro-ph.GA},
       adsurl = {https://ui.adsabs.harvard.edu/abs/2020MNRAS.491..944C}
}

@ARTICLE{alle01,
       author = {{Allende Prieto}, Carlos and {Lambert}, David L. and {Asplund}, Martin},
        title = "{The Forbidden Abundance of Oxygen in the Sun}",
      journal = {\apjl},
         year = 2001,
        month = jul,
       volume = {556},
       number = {1},
        pages = {L63-L66},
          doi = {10.1086/322874},
archivePrefix = {arXiv},
       eprint = {astro-ph/0106360},
 primaryClass = {astro-ph},
       adsurl = {https://ui.adsabs.harvard.edu/abs/2001ApJ...556L..63A}
}

@ARTICLE{veil87,
       author = {{Veilleux}, Sylvain and {Osterbrock}, Donald E.},
        title = "{Spectral Classification of Emission-Line Galaxies}",
      journal = {\apjs},
         year = 1987,
        month = feb,
       volume = {63},
        pages = {295},
          doi = {10.1086/191166},
       adsurl = {https://ui.adsabs.harvard.edu/abs/1987ApJS...63..295V}
}

@ARTICLE{kewl01,
       author = {{Kewley}, L.~J. and {Heisler}, C.~A. and {Dopita}, M.~A. and {Lumsden}, S.},
        title = "{Optical Classification of Southern Warm Infrared Galaxies}",
      journal = {\apjs},
         year = 2001,
        month = jan,
       volume = {132},
       number = {1},
        pages = {37-71},
          doi = {10.1086/318944},
       adsurl = {https://ui.adsabs.harvard.edu/abs/2001ApJS..132...37K}
}

@ARTICLE{cidf10,
       author = {{Cid Fernandes}, R. and {Stasi{\'n}ska}, G. and {Schlickmann}, M.~S. and {Mateus}, A. and {Vale Asari}, N. and {Schoenell}, W. and {Sodr{\'e}}, L.},
        title = "{Alternative diagnostic diagrams and the `forgotten' population of weak line galaxies in the SDSS}",
      journal = {\mnras},
         year = 2010,
        month = apr,
       volume = {403},
       number = {2},
        pages = {1036-1053},
          doi = {10.1111/j.1365-2966.2009.16185.x},
archivePrefix = {arXiv},
       eprint = {0912.1643},
 primaryClass = {astro-ph.CO},
       adsurl = {https://ui.adsabs.harvard.edu/abs/2010MNRAS.403.1036C}
}

@ARTICLE{cidf11,
       author = {{Cid Fernandes}, R. and {Stasi{\'n}ska}, G. and {Mateus}, A. and {Vale Asari}, N.},
        title = "{A comprehensive classification of galaxies in the Sloan Digital Sky Survey: how to tell true from fake AGN?}",
      journal = {\mnras},
         year = 2011,
        month = may,
       volume = {413},
       number = {3},
        pages = {1687-1699},
          doi = {10.1111/j.1365-2966.2011.18244.x},
archivePrefix = {arXiv},
       eprint = {1012.4426},
 primaryClass = {astro-ph.CO},
       adsurl = {https://ui.adsabs.harvard.edu/abs/2011MNRAS.413.1687C}
}

@ARTICLE{bald81,
       author = {{Baldwin}, J.~A. and {Phillips}, M.~M. and {Terlevich}, R.},
        title = "{Classification parameters for the emission-line spectra of extragalactic objects.}",
      journal = {\pasp},
         year = 1981,
        month = feb,
       volume = {93},
        pages = {5-19},
          doi = {10.1086/130766},
       adsurl = {https://ui.adsabs.harvard.edu/abs/1981PASP...93....5B}
}

@ARTICLE{li25,
       author = {{Li}, Sijia and {Wang}, Xin and {Chen}, Yuguang and {Jones}, Tucker and {Treu}, Tommaso and {Glazebrook}, Karl and {He}, Xianlong and {Henry}, Alaina and {Meng}, Xiao-Lei and {Morishita}, Takahiro and {Roberts-Borsani}, Guido and {Yang}, Lilan and {Yu}, Hao-Ran and {Calabr{\`o}}, Antonello and {Castellano}, Marco and {Leethochawalit}, Nicha and {Metha}, Benjamin and {Nanayakkara}, Themiya and {Roy}, Namrata and {Vulcani}, Benedetta},
        title = "{Early Results from GLASS-JWST. XXV. Electron Density in the Interstellar Medium at 0.7 {\ensuremath{\lesssim}} z {\ensuremath{\lesssim}} 9.3 with NIRSpec High-resolution Spectroscopy}",
      journal = {\apjl},
         year = 2025,
        month = jan,
       volume = {979},
       number = {1},
          eid = {L13},
        pages = {L13},
          doi = {10.3847/2041-8213/ad9eac},
archivePrefix = {arXiv},
       eprint = {2412.08382},
 primaryClass = {astro-ph.GA},
       adsurl = {https://ui.adsabs.harvard.edu/abs/2025ApJ...979L..13L}
}

@ARTICLE{li25b,
       author = {{Li}, Sijia and {Yu}, Si-Yue and {Ho}, Luis C. and {Silverman}, John D. and {Wang}, Jing and {Saintonge}, Am{\'e}lie and {Yu}, Niankun and {Fei}, Qinyue and {Kashino}, Daichi and {Yu}, Hao-ran},
        title = "{Linking Electron Density with Elevated Star Formation Activity from z = 0 to z = 10}",
      journal = {\apjl},
         year = 2025,
        month = nov,
       volume = {993},
       number = {2},
          eid = {L51},
        pages = {L51},
          doi = {10.3847/2041-8213/ae1695},
archivePrefix = {arXiv},
       eprint = {2510.18764},
 primaryClass = {astro-ph.GA},
       adsurl = {https://ui.adsabs.harvard.edu/abs/2025ApJ...993L..51L}
}

@ARTICLE{tayl25,
       author = {{Taylor}, Matthew A. and {Tahmasebzadeh}, Behzad and {Thompson}, Solveig and {Vasiliev}, Eugene and {Valluri}, Monica and {Drinkwater}, Michael J. and {C{\^o}t{\'e}}, Patrick and {Ferrarese}, Laura and {Roediger}, Joel and {Baumgardt}, Holger and {Bentz}, Misty C. and {Dage}, Kristen and {Peng}, Eric W. and {Lapeer}, Drew and {Liu}, Chengze and {Sumners}, Zach and {Wang}, Kaixiang and {Baldassare}, Vivienne and {Blakeslee}, John P. and {Ko}, Youkyung and {Woods}, Tyrone E.},
        title = "{A Supermassive Black Hole in a Diminutive Ultracompact Dwarf Galaxy Discovered with JWST/NIRSpec+IFU}",
      journal = {\apjl},
         year = 2025,
        month = sep,
       volume = {991},
       number = {1},
          eid = {L24},
        pages = {L24},
          doi = {10.3847/2041-8213/ae028e},
archivePrefix = {arXiv},
       eprint = {2503.00113},
 primaryClass = {astro-ph.GA},
       adsurl = {https://ui.adsabs.harvard.edu/abs/2025ApJ...991L..24T}
}

@ARTICLE{lai23,
       author = {{Lai}, Thomas S. -Y. and {Armus}, Lee and {Bianchin}, Marina and {D{\'\i}az-Santos}, Tanio and {Linden}, Sean T. and {Privon}, George C. and {Inami}, Hanae and {U}, Vivian and {Bohn}, Thomas and {Evans}, Aaron S. and {Larson}, Kirsten L. and {Hensley}, Brandon S. and {Smith}, J. -D.~T. and {Malkan}, Matthew A. and {Song}, Yiqing and {Stierwalt}, Sabrina and {van der Werf}, Paul P. and {McKinney}, Jed and {Aalto}, Susanne and {Buiten}, Victorine A. and {Rich}, Jeff and {Charmandaris}, Vassilis and {Appleton}, Philip and {Barcos-Mu{\~n}oz}, Loreto and {B{\"o}ker}, Torsten and {Finnerty}, Luke and {Kader}, Justin A. and {Law}, David R. and {Medling}, Anne M. and {Brown}, Michael J.~I. and {Hayward}, Christopher C. and {Howell}, Justin and {Iwasawa}, Kazushi and {Kemper}, Francisca and {Marshall}, Jason and {Mazzarella}, Joseph M. and {M{\"u}ller-S{\'a}nchez}, Francisco and {Murphy}, Eric J. and {Sanders}, David and {Surace}, Jason},
        title = "{GOALS-JWST: Small Neutral Grains and Enhanced 3.3 {\ensuremath{\mu}}m PAH Emission in the Seyfert Galaxy NGC 7469}",
      journal = {\apjl},
         year = 2023,
        month = nov,
       volume = {957},
       number = {2},
          eid = {L26},
        pages = {L26},
          doi = {10.3847/2041-8213/ad0387},
archivePrefix = {arXiv},
       eprint = {2307.15169},
 primaryClass = {astro-ph.GA},
       adsurl = {https://ui.adsabs.harvard.edu/abs/2023ApJ...957L..26L}
}

@ARTICLE{bian24,
       author = {{Bianchin}, Marina and {U}, Vivian and {Song}, Yiqing and {Lai}, Thomas S. -Y. and {Remigio}, Raymond P. and {Barcos-Mu{\~n}oz}, Loreto and {D{\'\i}az-Santos}, Tanio and {Armus}, Lee and {Inami}, Hanae and {Larson}, Kirsten L. and {Evans}, Aaron S. and {B{\"o}ker}, Torsten and {Kader}, Justin A. and {Linden}, Sean T. and {Charmandaris}, Vassilis and {Malkan}, Matthew A. and {Rich}, Jeff and {Bohn}, Thomas and {Medling}, Anne M. and {Stierwalt}, Sabrina and {Mazzarella}, Joseph M. and {Law}, David R. and {Privon}, George C. and {Aalto}, Susanne and {Appleton}, Philip and {Brown}, Michael J.~I. and {Buiten}, Victorine A. and {Finnerty}, Luke and {Hayward}, Christopher C. and {Howell}, Justin and {Iwasawa}, Kazushi and {Kemper}, Francisca and {Marshall}, Jason and {McKinney}, Jed and {M{\"u}ller-S{\'a}nchez}, Francisco and {Murphy}, Eric J. and {van der Werf}, Paul P. and {Sanders}, David B. and {Surace}, Jason},
        title = "{GOALS-JWST: Gas Dynamics and Excitation in NGC 7469 Revealed by NIRSpec}",
      journal = {\apj},
         year = 2024,
        month = apr,
       volume = {965},
       number = {2},
          eid = {103},
        pages = {103},
          doi = {10.3847/1538-4357/ad2a50},
archivePrefix = {arXiv},
       eprint = {2308.00209},
 primaryClass = {astro-ph.GA},
       adsurl = {https://ui.adsabs.harvard.edu/abs/2024ApJ...965..103B}
}

@ARTICLE{ceci25,
       author = {{Ceci}, M. and {Cresci}, G. and {Arribas}, S. and {B{\"o}ker}, T. and {Bunker}, A.~J. and {Charlot}, S. and {Fahrion}, K. and {Lamperti}, I. and {Marconi}, A. and {Perna}, M. and {Tozzi}, G. and {Ulivi}, L.},
        title = "{The JWST/NIRSpec view of the nuclear region in the prototypical merging galaxy NGC 6240}",
      journal = {\aap},
         year = 2025,
        month = mar,
       volume = {695},
          eid = {A116},
        pages = {A116},
          doi = {10.1051/0004-6361/202452207},
archivePrefix = {arXiv},
       eprint = {2412.14685},
 primaryClass = {astro-ph.GA},
       adsurl = {https://ui.adsabs.harvard.edu/abs/2025A&A...695A.116C}
}

@ARTICLE{pern24,
       author = {{Perna}, Michele and {Arribas}, Santiago and {Lamperti}, Isabella and {Pereira-Santaella}, Miguel and {Ulivi}, Lorenzo and {B{\"o}ker}, Torsten and {Maiolino}, Roberto and {Bunker}, Andrew J. and {Charlot}, St{\'e}phane and {Cresci}, Giovanni and {Rodr{\'\i}guez Del Pino}, Bruno and {D'Eugenio}, Francesco and {{\"U}bler}, Hannah and {Fahrion}, Katja and {Ceci}, Matteo},
        title = "{No evidence of active galactic nucleus features in the nuclei of Arp 220 from JWST/NIRSpec IFS}",
      journal = {\aap},
         year = 2024,
        month = oct,
       volume = {690},
          eid = {A171},
        pages = {A171},
          doi = {10.1051/0004-6361/202450094},
archivePrefix = {arXiv},
       eprint = {2403.13948},
 primaryClass = {astro-ph.GA},
       adsurl = {https://ui.adsabs.harvard.edu/abs/2024A&A...690A.171P}
}

@ARTICLE{uliv25,
       author = {{Ulivi}, Lorenzo and {Perna}, Michele and {Lamperti}, Isabella and {Arribas}, Santiago and {Cresci}, Giovanni and {Marconcini}, Cosimo and {Rodr{\'\i}guez Del Pino}, Bruno and {B{\"o}ker}, Torsten and {Bunker}, Andrew J. and {Ceci}, Matteo and {Charlot}, St{\'e}phane and {D'Eugenio}, Francesco and {Fahrion}, Katja and {Maiolino}, Roberto and {Marconi}, Alessandro and {Pereira-Santaella}, Miguel},
        title = "{JWST/NIRSpec insights into the circumnuclear region of Arp 220: A detailed kinematic study}",
      journal = {\aap},
         year = 2025,
        month = jan,
       volume = {693},
          eid = {A36},
        pages = {A36},
          doi = {10.1051/0004-6361/202451442},
archivePrefix = {arXiv},
       eprint = {2407.08505},
 primaryClass = {astro-ph.GA},
       adsurl = {https://ui.adsabs.harvard.edu/abs/2025A&A...693A..36U}
}

@ARTICLE{xu24,
       author = {{Xu}, Yi and {Ouchi}, Masami and {Yajima}, Hidenobu and {Fukushima}, Hajime and {Harikane}, Yuichi and {Isobe}, Yuki and {Nakajima}, Kimihiko and {Nakane}, Minami and {Ono}, Yoshiaki and {Umeda}, Hiroya and {Yanagisawa}, Hiroto and {Zhang}, Yechi},
        title = "{Dynamics of a Galaxy at z > 10 Explored by JWST Integral Field Spectroscopy: Hints of Rotating Disk Suggesting Weak Feedback}",
      journal = {\apj},
         year = 2024,
        month = nov,
       volume = {976},
       number = {1},
          eid = {142},
        pages = {142},
          doi = {10.3847/1538-4357/ad82dd},
archivePrefix = {arXiv},
       eprint = {2404.16963},
 primaryClass = {astro-ph.GA},
       adsurl = {https://ui.adsabs.harvard.edu/abs/2024ApJ...976..142X}
}

@ARTICLE{scho24,
       author = {{Scholtz}, Jan and {Witten}, Callum and {Laporte}, Nicolas and {{\"U}bler}, Hannah and {Perna}, Michele and {Maiolino}, Roberto and {Arribas}, Santiago and {Baker}, William M. and {Bennett}, Jake S. and {D'Eugenio}, Francesco and {Simmonds}, Charlotte and {Tacchella}, Sandro and {Witstok}, Joris and {Bunker}, Andrew J. and {Carniani}, Stefano and {Charlot}, St{\'e}phane and {Cresci}, Giovanni and {Curtis-Lake}, Emma and {Eisenstein}, Daniel J. and {Kumari}, Nimisha and {Robertson}, Brant and {Rodr{\'\i}guez Del Pino}, Bruno and {Smit}, Renske and {Venturi}, Giacomo and {Williams}, Christina C. and {Willmer}, Christopher N.~A.},
        title = "{GN-z11: The environment of an active galactic nucleus at z = 10.603. New insights into the most distant Ly{\ensuremath{\alpha}} detection}",
      journal = {\aap},
         year = 2024,
        month = jul,
       volume = {687},
          eid = {A283},
        pages = {A283},
          doi = {10.1051/0004-6361/202347187},
archivePrefix = {arXiv},
       eprint = {2306.09142},
 primaryClass = {astro-ph.GA},
       adsurl = {https://ui.adsabs.harvard.edu/abs/2024A&A...687A.283S}
}

@ARTICLE{maio24,
       author = {{Maiolino}, Roberto and {{\"U}bler}, Hannah and {Perna}, Michele and {Scholtz}, Jan and {D'Eugenio}, Francesco and {Witten}, Callum and {Laporte}, Nicolas and {Witstok}, Joris and {Carniani}, Stefano and {Tacchella}, Sandro and {Baker}, William M. and {Arribas}, Santiago and {Nakajima}, Kimihiko and {Eisenstein}, Daniel J. and {Bunker}, Andrew J. and {Charlot}, St{\'e}phane and {Cresci}, Giovanni and {Curti}, Mirko and {Curtis-Lake}, Emma and {de Graaff}, Anna and {Egami}, Eiichi and {Ji}, Zhiyuan and {Johnson}, Benjamin D. and {Kumari}, Nimisha and {Looser}, Tobias J. and {Maseda}, Michael and {Nelson}, Erica and {Robertson}, Brant and {Rodr{\'\i}guez Del Pino}, Bruno and {Sandles}, Lester and {Simmonds}, Charlotte and {Smit}, Renske and {Sun}, Fengwu and {Venturi}, Giacomo and {Williams}, Christina C. and {Willmer}, Christopher N.~A.},
        title = "{JADES. Possible Population III signatures at z = 10.6 in the halo of GN-z11}",
      journal = {\aap},
         year = 2024,
        month = jul,
       volume = {687},
          eid = {A67},
        pages = {A67},
          doi = {10.1051/0004-6361/202347087},
archivePrefix = {arXiv},
       eprint = {2306.00953},
 primaryClass = {astro-ph.GA},
       adsurl = {https://ui.adsabs.harvard.edu/abs/2024A&A...687A..67M}
}

@ARTICLE{marc24,
       author = {{Marconcini}, C. and {D'Eugenio}, F. and {Maiolino}, R. and {Arribas}, S. and {Bunker}, A. and {Carniani}, S. and {Charlot}, S. and {Perna}, M. and {Rodr{\'\i}guez Del Pino}, B. and {{\"U}bler}, H. and {Willott}, C.~J. and {B{\"o}ker}, T. and {Cresci}, G. and {Curti}, M. and {Jones}, G.~C. and {Lamperti}, I. and {Parlanti}, E. and {Venturi}, G.},
        title = "{GA-NIFS: the interplay between merger, star formation, and chemical enrichment in MACS1149-JD1 at z = 9.11 with JWST/NIRSpec}",
      journal = {\mnras},
         year = 2024,
        month = sep,
       volume = {533},
       number = {2},
        pages = {2488-2501},
          doi = {10.1093/mnras/stae1971},
archivePrefix = {arXiv},
       eprint = {2407.08616},
 primaryClass = {astro-ph.GA},
       adsurl = {https://ui.adsabs.harvard.edu/abs/2024MNRAS.533.2488M}
}

@ARTICLE{jone25,
       author = {{Jones}, Gareth C. and {Bunker}, Andrew J. and {Telikova}, Kseniia and {Arribas}, Santiago and {Carniani}, Stefano and {Charlot}, Stephane and {D'Eugenio}, Francesco and {Maiolino}, Roberto and {Perna}, Michele and {Rodr{\'\i}guez Del Pino}, Bruno and {{\"U}bler}, Hannah and {Willott}, Chris and {Aravena}, Manuel and {B{\"o}ker}, Torsten and {Cresci}, Giovanni and {Curti}, Mirko and {Gonz{\'a}lez-L{\'o}pez}, Jorge and {Herrera-Camus}, Rodrigo and {Lamperti}, Isabella and {Parlanti}, Eleonora and {P{\'e}rez-Gonz{\'a}lez}, Pablo G. and {Villanueva}, Vicente},
        title = "{GA-NIFS: witnessing the complex assembly of a star-forming system at z = 5.7}",
      journal = {\mnras},
         year = 2025,
        month = jul,
       volume = {540},
       number = {4},
        pages = {3311-3329},
          doi = {10.1093/mnras/staf899},
archivePrefix = {arXiv},
       eprint = {2405.12955},
 primaryClass = {astro-ph.GA},
       adsurl = {https://ui.adsabs.harvard.edu/abs/2025MNRAS.540.3311J}
}

@ARTICLE{smit18,
       author = {{Smit}, Renske and {Bouwens}, Rychard J. and {Carniani}, Stefano and {Oesch}, Pascal A. and {Labb{\'e}}, Ivo and {Illingworth}, Garth D. and {van der Werf}, Paul and {Bradley}, Larry D. and {Gonzalez}, Valentino and {Hodge}, Jacqueline A. and {Holwerda}, Benne W. and {Maiolino}, Roberto and {Zheng}, Wei},
        title = "{Rotation in [C II]-emitting gas in two galaxies at a redshift of 6.8}",
      journal = {\nat},
         year = 2018,
        month = jan,
       volume = {553},
       number = {7687},
        pages = {178-181},
          doi = {10.1038/nature24631},
archivePrefix = {arXiv},
       eprint = {1706.04614},
 primaryClass = {astro-ph.GA},
       adsurl = {https://ui.adsabs.harvard.edu/abs/2018Natur.553..178S}
}

@ARTICLE{scho25,
       author = {{Scholtz}, J. and {Curti}, M. and {D'Eugenio}, F. and {{\"U}bler}, H. and {Maiolino}, R. and {Marconcini}, C. and {Smit}, R. and {Perna}, M. and {Witstok}, J. and {Arribas}, S. and {B{\"o}ker}, T. and {Bunker}, A.~J. and {Carniani}, S. and {Charlot}, S. and {Cresci}, G. and {Lamperti}, I. and {Parlanti}, E. and {P{\'e}rez-Gonz{\'a}lez}, P.~G. and {Rodr{\'\i}guez Del Pino}, B. and {Venturi}, G.},
        title = "{GA-NIFS: ISM properties and metal enrichment in a merger-driven starburst during the epoch of reionization probed with JWST and ALMA}",
      journal = {\mnras},
         year = 2025,
        month = may,
       volume = {539},
       number = {3},
        pages = {2463-2484},
          doi = {10.1093/mnras/staf518},
archivePrefix = {arXiv},
       eprint = {2411.07695},
 primaryClass = {astro-ph.GA},
       adsurl = {https://ui.adsabs.harvard.edu/abs/2025MNRAS.539.2463S}
}

@ARTICLE{parl25,
       author = {{Parlanti}, Eleonora and {Carniani}, Stefano and {Venturi}, Giacomo and {Herrera-Camus}, Rodrigo and {Arribas}, Santiago and {Bunker}, Andrew J. and {Charlot}, St{\'e}phane and {D'Eugenio}, Francesco and {Maiolino}, Roberto and {Perna}, Michele and {{\"U}bler}, Hannah and {B{\"o}ker}, Torsten and {Cresci}, Giovanni and {Curti}, Mirko and {Jones}, Gareth C. and {Lamperti}, Isabella and {P{\'e}rez-Gonz{\'a}lez}, Pablo G. and {Del Pino}, Bruno Rodr{\'\i}guez and {Zamora}, Sandra},
        title = "{GA-NIFS: Multiphase analysis of a star-forming galaxy at z {\ensuremath{\sim}} 5.5}",
      journal = {\aap},
         year = 2025,
        month = mar,
       volume = {695},
          eid = {A6},
        pages = {A6},
          doi = {10.1051/0004-6361/202451692},
archivePrefix = {arXiv},
       eprint = {2407.19008},
 primaryClass = {astro-ph.GA},
       adsurl = {https://ui.adsabs.harvard.edu/abs/2025A&A...695A...6P}
}

@ARTICLE{parl24,
       author = {{Parlanti}, Eleonora and {Carniani}, Stefano and {{\"U}bler}, Hannah and {Venturi}, Giacomo and {Circosta}, Chiara and {D'Eugenio}, Francesco and {Arribas}, Santiago and {Bunker}, Andrew J. and {Charlot}, St{\'e}phane and {L{\"u}tzgendorf}, Nora and {Maiolino}, Roberto and {Perna}, Michele and {Rodr{\'\i}guez Del Pino}, Bruno and {Willott}, Chris J. and {B{\"o}ker}, Torsten and {Cameron}, Alex J. and {Chevallard}, Jacopo and {Cresci}, Giovanni and {Jones}, Gareth C. and {Kumari}, Nimisha and {Lamperti}, Isabella and {Scholtz}, Jan},
        title = "{GA-NIFS: Early-stage feedback in a heavily obscured active galactic nucleus at z = 4.76}",
      journal = {\aap},
         year = 2024,
        month = apr,
       volume = {684},
          eid = {A24},
        pages = {A24},
          doi = {10.1051/0004-6361/202347914},
archivePrefix = {arXiv},
       eprint = {2309.05713},
 primaryClass = {astro-ph.GA},
       adsurl = {https://ui.adsabs.harvard.edu/abs/2024A&A...684A..24P}
}

@ARTICLE{lell21,
       author = {{Lelli}, Federico and {Di Teodoro}, Enrico M. and {Fraternali}, Filippo and {Man}, Allison W.~S. and {Zhang}, Zhi-Yu and {De Breuck}, Carlos and {Davis}, Timothy A. and {Maiolino}, Roberto},
        title = "{A massive stellar bulge in a regularly rotating galaxy 1.2 billion years after the Big Bang}",
      journal = {Science},
         year = 2021,
        month = feb,
       volume = {371},
       number = {6530},
        pages = {713-716},
          doi = {10.1126/science.abc1893},
archivePrefix = {arXiv},
       eprint = {2102.05957},
 primaryClass = {astro-ph.GA},
       adsurl = {https://ui.adsabs.harvard.edu/abs/2021Sci...371..713L}
}

@ARTICLE{carn13,
       author = {{Carniani}, S. and {Marconi}, A. and {Biggs}, A. and {Cresci}, G. and {Cupani}, G. and {D'Odorico}, V. and {Humphreys}, E. and {Maiolino}, R. and {Mannucci}, F. and {Molaro}, P. and {Nagao}, T. and {Testi}, L. and {Zwaan}, M.~A.},
        title = "{Strongly star-forming rotating disks in a complex merging system at z = 4.7 as revealed by ALMA}",
      journal = {\aap},
         year = 2013,
        month = nov,
       volume = {559},
          eid = {A29},
        pages = {A29},
          doi = {10.1051/0004-6361/201322320},
archivePrefix = {arXiv},
       eprint = {1308.5113},
 primaryClass = {astro-ph.CO},
       adsurl = {https://ui.adsabs.harvard.edu/abs/2013A&A...559A..29C}
}

@ARTICLE{zamo24,
       author = {{Zamora}, Sandra and {Venturi}, Giacomo and {Carniani}, Stefano and {Bertola}, Elena and {Parlanti}, Eleonora and {Perna}, Michele and {Arribas}, Santiago and {B{\"o}ker}, Torsten and {Bunker}, Andrew J. and {Charlot}, St{\'e}phane and {D'Eugenio}, Francesco and {Maiolino}, Roberto and {Del Pino}, Bruno Rodr{\'\i}guez and {{\"U}bler}, Hannah and {Cresci}, Giovanni and {Jones}, Gareth C. and {Lamperti}, Isabella},
        title = "{GA-NIFS: The highly overdense system BR1202-0725 at z {\ensuremath{\sim}} 4.7: A double active galactic nucleus with fast outflows plus eight companion galaxies}",
      journal = {\aap},
         year = 2025,
        month = oct,
       volume = {702},
          eid = {A102},
        pages = {A102},
          doi = {10.1051/0004-6361/202453236},
archivePrefix = {arXiv},
       eprint = {2412.02751},
 primaryClass = {astro-ph.GA},
       adsurl = {https://ui.adsabs.harvard.edu/abs/2025A&A...702A.102Z}
}

@ARTICLE{uble24,
       author = {{{\"U}bler}, Hannah and {D'Eugenio}, Francesco and {Perna}, Michele and {Arribas}, Santiago and {Jones}, Gareth C. and {Bunker}, Andrew J. and {Carniani}, Stefano and {Charlot}, St{\'e}phane and {Maiolino}, Roberto and {Rodr{\'\i}guez del Pino}, Bruno and {Willott}, Chris J. and {B{\"o}ker}, Torsten and {Cresci}, Giovanni and {Kumari}, Nimisha and {Lamperti}, Isabella and {Parlanti}, Eleonora and {Scholtz}, Jan and {Venturi}, Giacomo},
        title = "{GA-NIFS: NIRSpec reveals evidence for non-circular motions and AGN feedback in GN20}",
      journal = {\mnras},
         year = 2024,
        month = oct,
       volume = {533},
       number = {4},
        pages = {4287-4299},
          doi = {10.1093/mnras/stae1993},
archivePrefix = {arXiv},
       eprint = {2403.03192},
 primaryClass = {astro-ph.GA},
       adsurl = {https://ui.adsabs.harvard.edu/abs/2024MNRAS.533.4287U}
}

@ARTICLE{hodg12,
       author = {{Hodge}, J.~A. and {Carilli}, C.~L. and {Walter}, F. and {de Blok}, W.~J.~G. and {Riechers}, D. and {Daddi}, E. and {Lentati}, L.},
        title = "{Evidence for a Clumpy, Rotating Gas Disk in a Submillimeter Galaxy at z = 4}",
      journal = {\apj},
         year = 2012,
        month = nov,
       volume = {760},
       number = {1},
          eid = {11},
        pages = {11},
          doi = {10.1088/0004-637X/760/1/11},
archivePrefix = {arXiv},
       eprint = {1209.2418},
 primaryClass = {astro-ph.CO},
       adsurl = {https://ui.adsabs.harvard.edu/abs/2012ApJ...760...11H}
}

@ARTICLE{rizz24,
       author = {{Rizzo}, F. and {Bacchini}, C. and {Kohandel}, M. and {Di Mascolo}, L. and {Fraternali}, F. and {Roman-Oliveira}, F. and {Zanella}, A. and {Popping}, G. and {Valentino}, F. and {Magdis}, G. and {Whitaker}, K.},
        title = "{The ALMA-ALPAKA survey: II. Evolution of turbulence in galaxy disks across cosmic time: Difference between cold and warm gas}",
      journal = {\aap},
         year = 2024,
        month = sep,
       volume = {689},
          eid = {A273},
        pages = {A273},
          doi = {10.1051/0004-6361/202450455},
archivePrefix = {arXiv},
       eprint = {2407.06261},
 primaryClass = {astro-ph.GA},
       adsurl = {https://ui.adsabs.harvard.edu/abs/2024A&A...689A.273R}
}

@ARTICLE{koha24,
       author = {{Kohandel}, M. and {Pallottini}, A. and {Ferrara}, A. and {Zanella}, A. and {Rizzo}, F. and {Carniani}, S.},
        title = "{Dynamically cold disks in the early Universe: Myth or reality?}",
      journal = {\aap},
         year = 2024,
        month = may,
       volume = {685},
          eid = {A72},
        pages = {A72},
          doi = {10.1051/0004-6361/202348209},
archivePrefix = {arXiv},
       eprint = {2311.05832},
 primaryClass = {astro-ph.GA},
       adsurl = {https://ui.adsabs.harvard.edu/abs/2024A&A...685A..72K}
}

@ARTICLE{peas18,
       author = {{Pease}, F.~G.},
        title = "{The Rotation and Radial Velocity of the Central Part of the Andromeda Nebula}",
      journal = {Proceedings of the National Academy of Science},
         year = 1918,
        month = jan,
       volume = {4},
       number = {1},
        pages = {21-24},
          doi = {10.1073/pnas.4.1.21},
       adsurl = {https://ui.adsabs.harvard.edu/abs/1918PNAS....4...21P}
}

@ARTICLE{oort27,
       author = {{Oort}, J.~H.},
        title = "{Observational evidence confirming Lindblad's hypothesis of a rotation of the galactic system}",
      journal = {\bain},
         year = 1927,
        month = apr,
       volume = {3},
        pages = {275},
       adsurl = {https://ui.adsabs.harvard.edu/abs/1927BAN.....3..275O}
}

@ARTICLE{lind27,
       author = {{Lindblad}, B.},
        title = "{On the state of motion in the galactic system}",
      journal = {\mnras},
         year = 1927,
        month = may,
       volume = {87},
        pages = {553-564},
          doi = {10.1093/mnras/87.7.553},
       adsurl = {https://ui.adsabs.harvard.edu/abs/1927MNRAS..87..553L}
}

@ARTICLE{slip14,
       author = {{Slipher}, V.~M.},
        title = "{The Radial Velocity of the Andromeda Nebula}",
      journal = {Popular Astronomy},
         year = 1914,
        month = jan,
       volume = {22},
        pages = {19-21},
       adsurl = {https://ui.adsabs.harvard.edu/abs/1914PA.....22...19S}
}

@ARTICLE{rubi80,
       author = {{Rubin}, V.~C. and {Ford}, Jr., W.~K. and {Thonnard}, N.},
        title = "{Rotational properties of 21 SC galaxies with a large range of luminosities and radii, from NGC 4605 (R=4kpc) to UGC 2885 (R=122kpc).}",
      journal = {\apj},
         year = 1980,
        month = jun,
       volume = {238},
        pages = {471-487},
          doi = {10.1086/158003},
       adsurl = {https://ui.adsabs.harvard.edu/abs/1980ApJ...238..471R}
}

@ARTICLE{bege91,
       author = {{Begeman}, K.~G. and {Broeils}, A.~H. and {Sanders}, R.~H.},
        title = "{Extended rotation curves of spiral galaxies : dark haloes and modified dynamics.}",
      journal = {\mnras},
         year = 1991,
        month = apr,
       volume = {249},
        pages = {523},
          doi = {10.1093/mnras/249.3.523},
       adsurl = {https://ui.adsabs.harvard.edu/abs/1991MNRAS.249..523B}
}

@ARTICLE{frat21,
       author = {{Fraternali}, F. and {Karim}, A. and {Magnelli}, B. and {G{\'o}mez-Guijarro}, C. and {Jim{\'e}nez-Andrade}, E.~F. and {Posses}, A.~C.},
        title = "{Fast rotating and low-turbulence discs at z ≃ 4.5: Dynamical evidence of their evolution into local early-type galaxies}",
      journal = {\aap},
         year = 2021,
        month = mar,
       volume = {647},
          eid = {A194},
        pages = {A194},
          doi = {10.1051/0004-6361/202039807},
archivePrefix = {arXiv},
       eprint = {2011.05340},
 primaryClass = {astro-ph.GA},
       adsurl = {https://ui.adsabs.harvard.edu/abs/2021A&A...647A.194F}
}

@ARTICLE{case83,
       author = {{Casertano}, S.},
        title = "{Rotation curve of the edge-on spiral galaxy NGC 5907 : disc and halo masses.}",
      journal = {\mnras},
         year = 1983,
        month = may,
       volume = {203},
        pages = {735-747},
          doi = {10.1093/mnras/203.3.735},
       adsurl = {https://ui.adsabs.harvard.edu/abs/1983MNRAS.203..735C}
}

@ARTICLE{maya42,
       author = {{Mayall}, N.~U. and {Aller}, L.~H.},
        title = "{The Rotation of the Spiral Nebula Messier 33.}",
      journal = {\apj},
         year = 1942,
        month = jan,
       volume = {95},
        pages = {5},
          doi = {10.1086/144369},
       adsurl = {https://ui.adsabs.harvard.edu/abs/1942ApJ....95....5M}
}

@ARTICLE{burb59,
       author = {{Burbidge}, E. Margaret and {Burbidge}, G.~R.},
        title = "{Rotation and Internal Motions in NGC 5128.}",
      journal = {\apj},
         year = 1959,
        month = mar,
       volume = {129},
        pages = {271},
          doi = {10.1086/146617},
       adsurl = {https://ui.adsabs.harvard.edu/abs/1959ApJ...129..271B}
}

@ARTICLE{burb60,
       author = {{Burbidge}, E. Margaret and {Burbidge}, G.~R. and {Prendergast}, K.~H.},
        title = "{The Rotation Mass Distribution, and Mass of NGC 2903.}",
      journal = {\apj},
         year = 1960,
        month = nov,
       volume = {132},
        pages = {640},
          doi = {10.1086/146967},
       adsurl = {https://ui.adsabs.harvard.edu/abs/1960ApJ...132..640B}
}

@ARTICLE{deva61,
       author = {{de Vaucouleurs}, Gerard},
        title = "{Southern Galaxies. I. Luminosity, Rotation, and Mass of the Magellanic System NGC 55.}",
      journal = {\apj},
         year = 1961,
        month = mar,
       volume = {133},
        pages = {405},
          doi = {10.1086/147043},
       adsurl = {https://ui.adsabs.harvard.edu/abs/1961ApJ...133..405D}
}

@ARTICLE{deva62,
       author = {{de Vaucouleurs}, Gerard and {de Vaucouleurs}, Antoinette},
        title = "{Rotation and Mass of the Magellanic-Type Galaxy NGC 4631.}",
      journal = {\aj},
         year = 1962,
        month = mar,
       volume = {67},
        pages = {113},
          doi = {10.1086/108614},
       adsurl = {https://ui.adsabs.harvard.edu/abs/1962AJ.....67..113D}
}

@ARTICLE{taka67,
       author = {{Takase}, B.},
        title = "{Distribution of Mass, Angular Momentum, and Rotational Energy in the Galaxy and NGC 224}",
      journal = {\pasj},
         year = 1967,
        month = jan,
       volume = {19},
        pages = {427},
       adsurl = {https://ui.adsabs.harvard.edu/abs/1967PASJ...19..427T}
}

@ARTICLE{fors09,
       author = {{F{\"o}rster Schreiber}, N.~M. and {Genzel}, R. and {Bouch{\'e}}, N. and {Cresci}, G. and {Davies}, R. and {Buschkamp}, P. and {Shapiro}, K. and {Tacconi}, L.~J. and {Hicks}, E.~K.~S. and {Genel}, S. and {Shapley}, A.~E. and {Erb}, D.~K. and {Steidel}, C.~C. and {Lutz}, D. and {Eisenhauer}, F. and {Gillessen}, S. and {Sternberg}, A. and {Renzini}, A. and {Cimatti}, A. and {Daddi}, E. and {Kurk}, J. and {Lilly}, S. and {Kong}, X. and {Lehnert}, M.~D. and {Nesvadba}, N. and {Verma}, A. and {McCracken}, H. and {Arimoto}, N. and {Mignoli}, M. and {Onodera}, M.},
        title = "{The SINS Survey: SINFONI Integral Field Spectroscopy of z \raisebox{-0.5ex}\textasciitilde 2 Star-forming Galaxies}",
      journal = {\apj},
         year = 2009,
        month = dec,
       volume = {706},
       number = {2},
        pages = {1364-1428},
          doi = {10.1088/0004-637X/706/2/1364},
archivePrefix = {arXiv},
       eprint = {0903.1872},
 primaryClass = {astro-ph.CO},
       adsurl = {https://ui.adsabs.harvard.edu/abs/2009ApJ...706.1364F}
}

@ARTICLE{wisn19,
       author = {{Wisnioski}, E. and {F{\"o}rster Schreiber}, N.~M. and {Fossati}, M. and {Mendel}, J.~T. and {Wilman}, D. and {Genzel}, R. and {Bender}, R. and {Wuyts}, S. and {Davies}, R.~L. and {{\"U}bler}, H. and {Bandara}, K. and {Beifiori}, A. and {Belli}, S. and {Brammer}, G. and {Chan}, J. and {Davies}, R.~I. and {Fabricius}, M. and {Galametz}, A. and {Lang}, P. and {Lutz}, D. and {Nelson}, E.~J. and {Momcheva}, I. and {Price}, S. and {Rosario}, D. and {Saglia}, R. and {Seitz}, S. and {Shimizu}, T. and {Tacconi}, L.~J. and {Tadaki}, K. and {van Dokkum}, P.~G. and {Wuyts}, E.},
        title = "{The KMOS$^{3D}$ Survey: Data Release and Final Survey Paper}",
      journal = {\apj},
         year = 2019,
        month = dec,
       volume = {886},
       number = {2},
          eid = {124},
        pages = {124},
          doi = {10.3847/1538-4357/ab4db8},
archivePrefix = {arXiv},
       eprint = {1909.11096},
 primaryClass = {astro-ph.GA},
       adsurl = {https://ui.adsabs.harvard.edu/abs/2019ApJ...886..124W}
}

@ARTICLE{lefe20,
       author = {{Le F{\`e}vre}, O. and {B{\'e}thermin}, M. and {Faisst}, A. and {Jones}, G.~C. and {Capak}, P. and {Cassata}, P. and {Silverman}, J.~D. and {Schaerer}, D. and {Yan}, L. and {Amorin}, R. and {Bardelli}, S. and {Boquien}, M. and {Cimatti}, A. and {Dessauges-Zavadsky}, M. and {Giavalisco}, M. and {Hathi}, N.~P. and {Fudamoto}, Y. and {Fujimoto}, S. and {Ginolfi}, M. and {Gruppioni}, C. and {Hemmati}, S. and {Ibar}, E. and {Koekemoer}, A. and {Khusanova}, Y. and {Lagache}, G. and {Lemaux}, B.~C. and {Loiacono}, F. and {Maiolino}, R. and {Mancini}, C. and {Narayanan}, D. and {Morselli}, L. and {M{\'e}ndez-Hern{\`a}ndez}, Hugo and {Oesch}, P.~A. and {Pozzi}, F. and {Romano}, M. and {Riechers}, D. and {Scoville}, N. and {Talia}, M. and {Tasca}, L.~A.~M. and {Thomas}, R. and {Toft}, S. and {Vallini}, L. and {Vergani}, D. and {Walter}, F. and {Zamorani}, G. and {Zucca}, E.},
        title = "{The ALPINE-ALMA [CII] survey. Survey strategy, observations, and sample properties of 118 star-forming galaxies at 4 < z < 6}",
      journal = {\aap},
         year = 2020,
        month = nov,
       volume = {643},
          eid = {A1},
        pages = {A1},
          doi = {10.1051/0004-6361/201936965},
archivePrefix = {arXiv},
       eprint = {1910.09517},
 primaryClass = {astro-ph.CO},
       adsurl = {https://ui.adsabs.harvard.edu/abs/2020A&A...643A...1L}
}

@ARTICLE{bouw22,
       author = {{Bouwens}, R.~J. and {Smit}, R. and {Schouws}, S. and {Stefanon}, M. and {Bowler}, R. and {Endsley}, R. and {Gonzalez}, V. and {Inami}, H. and {Stark}, D. and {Oesch}, P. and {Hodge}, J. and {Aravena}, M. and {da Cunha}, E. and {Dayal}, P. and {de Looze}, I. and {Ferrara}, A. and {Fudamoto}, Y. and {Graziani}, L. and {Li}, C. and {Nanayakkara}, T. and {Pallottini}, A. and {Schneider}, R. and {Sommovigo}, L. and {Topping}, M. and {van der Werf}, P. and {Algera}, H. and {Barrufet}, L. and {Hygate}, A. and {Labb{\'e}}, I. and {Riechers}, D. and {Witstok}, J.},
        title = "{Reionization Era Bright Emission Line Survey: Selection and Characterization of Luminous Interstellar Medium Reservoirs in the z > 6.5 Universe}",
      journal = {\apj},
         year = 2022,
        month = jun,
       volume = {931},
       number = {2},
          eid = {160},
        pages = {160},
          doi = {10.3847/1538-4357/ac5a4a},
archivePrefix = {arXiv},
       eprint = {2106.13719},
 primaryClass = {astro-ph.GA},
       adsurl = {https://ui.adsabs.harvard.edu/abs/2022ApJ...931..160B}
}

@ARTICLE{maio08,
       author = {{Maiolino}, R. and {Nagao}, T. and {Grazian}, A. and {Cocchia}, F. and {Marconi}, A. and {Mannucci}, F. and {Cimatti}, A. and {Pipino}, A. and {Ballero}, S. and {Calura}, F. and {Chiappini}, C. and {Fontana}, A. and {Granato}, G.~L. and {Matteucci}, F. and {Pastorini}, G. and {Pentericci}, L. and {Risaliti}, G. and {Salvati}, M. and {Silva}, L.},
        title = "{AMAZE. I. The evolution of the mass-metallicity relation at z > 3}",
      journal = {\aap},
         year = 2008,
        month = sep,
       volume = {488},
       number = {2},
        pages = {463-479},
          doi = {10.1051/0004-6361:200809678},
archivePrefix = {arXiv},
       eprint = {0806.2410},
 primaryClass = {astro-ph},
       adsurl = {https://ui.adsabs.harvard.edu/abs/2008A&A...488..463M}
}

@ARTICLE{mann09,
       author = {{Mannucci}, F. and {Cresci}, G. and {Maiolino}, R. and {Marconi}, A. and {Pastorini}, G. and {Pozzetti}, L. and {Gnerucci}, A. and {Risaliti}, G. and {Schneider}, R. and {Lehnert}, M. and {Salvati}, M.},
        title = "{LSD: Lyman-break galaxies Stellar populations and Dynamics - I. Mass, metallicity and gas at z \raisebox{-0.5ex}\textasciitilde 3.1}",
      journal = {\mnras},
         year = 2009,
        month = oct,
       volume = {398},
       number = {4},
        pages = {1915-1931},
          doi = {10.1111/j.1365-2966.2009.15185.x},
archivePrefix = {arXiv},
       eprint = {0902.2398},
 primaryClass = {astro-ph.CO},
       adsurl = {https://ui.adsabs.harvard.edu/abs/2009MNRAS.398.1915M}
}

@ARTICLE{fruc02,
       author = {{Fruchter}, A.~S. and {Hook}, R.~N.},
        title = "{Drizzle: A Method for the Linear Reconstruction of Undersampled Images}",
      journal = {\pasp},
         year = 2002,
        month = feb,
       volume = {114},
       number = {792},
        pages = {144-152},
          doi = {10.1086/338393},
archivePrefix = {arXiv},
       eprint = {astro-ph/9808087},
 primaryClass = {astro-ph},
       adsurl = {https://ui.adsabs.harvard.edu/abs/2002PASP..114..144F}
}

@ARTICLE{sofu01,
       author = {{Sofue}, Yoshiaki and {Rubin}, Vera},
        title = "{Rotation Curves of Spiral Galaxies}",
      journal = {\araa},
         year = 2001,
        month = jan,
       volume = {39},
        pages = {137-174},
          doi = {10.1146/annurev.astro.39.1.137},
archivePrefix = {arXiv},
       eprint = {astro-ph/0010594},
 primaryClass = {astro-ph},
       adsurl = {https://ui.adsabs.harvard.edu/abs/2001ARA&A..39..137S}
}

@ARTICLE{vana85,
       author = {{van Albada}, T.~S. and {Bahcall}, J.~N. and {Begeman}, K. and {Sancisi}, R.},
        title = "{Distribution of dark matter in the spiral galaxy NGC 3198.}",
      journal = {\apj},
         year = 1985,
        month = aug,
       volume = {295},
        pages = {305-313},
          doi = {10.1086/163375},
       adsurl = {https://ui.adsabs.harvard.edu/abs/1985ApJ...295..305V}
}

@ARTICLE{wuyt16,
       author = {{Wuyts}, Stijn and {F{\"o}rster Schreiber}, Natascha M. and {Wisnioski}, Emily and {Genzel}, Reinhard and {Burkert}, Andreas and {Bandara}, Kaushala and {Beifiori}, Alessandra and {Belli}, Sirio and {Bender}, Ralf and {Brammer}, Gabriel B. and {Chan}, Jeffrey and {Davies}, Ric and {Fossati}, Matteo and {Galametz}, Audrey and {Kulkarni}, Sandesh K. and {Lang}, Philipp and {Lutz}, Dieter and {Mendel}, J. Trevor and {Momcheva}, Ivelina G. and {Naab}, Thorsten and {Nelson}, Erica J. and {Saglia}, Roberto P. and {Seitz}, Stella and {Tacconi}, Linda J. and {Tadaki}, Ken-ichi and {{\"U}bler}, Hannah and {van Dokkum}, Pieter G. and {Wilman}, David J. and {Wuyts}, Eva},
        title = "{KMOS3D: Dynamical Constraints on the Mass Budget in Early Star-forming Disks}",
      journal = {\apj},
         year = 2016,
        month = nov,
       volume = {831},
       number = {2},
          eid = {149},
        pages = {149},
          doi = {10.3847/0004-637X/831/2/149},
archivePrefix = {arXiv},
       eprint = {1603.03432},
 primaryClass = {astro-ph.GA},
       adsurl = {https://ui.adsabs.harvard.edu/abs/2016ApJ...831..149W}
}

@ARTICLE{herr25,
       author = {{Herrera-Camus}, R. and {Gonz{\'a}lez-L{\'o}pez}, J. and {F{\"o}rster Schreiber}, N. and {Aravena}, M. and {de Looze}, I. and {Spilker}, J. and {Tadaki}, K. and {Barcos-Mu{\~n}oz}, L. and {Assef}, R.~J. and {Birkin}, J.~E. and {Bolatto}, A.~D. and {Bouwens}, R. and {Bovino}, S. and {Bowler}, R.~A.~A. and {Calistro Rivera}, G. and {da Cunha}, E. and {Davies}, R.~I. and {Davies}, R.~L. and {D{\'\i}az-Santos}, T. and {Ferrara}, A. and {Fisher}, D. and {Genzel}, R. and {Hodge}, J. and {Ikeda}, R. and {Killi}, M. and {Lee}, L. and {Li}, Y. and {Li}, J. and {Liu}, D. and {Lutz}, D. and {Mitsuhashi}, I. and {Narayanan}, D. and {Naab}, T. and {Palla}, M. and {Price}, S.~H. and {Posses}, A. and {Rela{\~n}o}, M. and {Smit}, R. and {Solimano}, M. and {Sternberg}, A. and {Tacconi}, L. and {Telikova}, K. and {{\"U}bler}, H. and {van der Giessen}, S.~A. and {Veilleux}, S. and {Villanueva}, V. and {Baeza-Garay}, M.},
        title = "{The ALMA-CRISTAL survey: Gas, dust, and stars in star-forming galaxies when the Universe was {\ensuremath{\sim}}1 Gyr old: I. Survey overview and case studies}",
      journal = {\aap},
         year = 2025,
        month = jul,
       volume = {699},
          eid = {A80},
        pages = {A80},
          doi = {10.1051/0004-6361/202553896},
archivePrefix = {arXiv},
       eprint = {2505.06340},
 primaryClass = {astro-ph.GA},
       adsurl = {https://ui.adsabs.harvard.edu/abs/2025A&A...699A..80H}
}

@ARTICLE{guru22,
       author = {{Gururajan}, G. and {B{\'e}thermin}, M. and {Theul{\'e}}, P. and {Spilker}, J.~S. and {Aravena}, M. and {Archipley}, M.~A. and {Chapman}, S.~C. and {De Breuck}, C. and {Gonzalez}, A. and {Hayward}, C.~C. and {Hezaveh}, Y. and {Hill}, R. and {Jarugula}, S. and {Litke}, K.~C. and {Malkan}, M. and {Marrone}, D.~P. and {Narayanan}, D. and {Phadke}, K.~A. and {Reuter}, C. and {Vieira}, J.~D. and {Vizgan}, D. and {Wei{\ss}}, A.},
        title = "{High resolution spectral imaging of CO(7-6), [CI](2-1), and continuum of three high-z lensed dusty star-forming galaxies using ALMA}",
      journal = {\aap},
         year = 2022,
        month = jul,
       volume = {663},
          eid = {A22},
        pages = {A22},
          doi = {10.1051/0004-6361/202142172},
archivePrefix = {arXiv},
       eprint = {2109.03450},
 primaryClass = {astro-ph.GA},
       adsurl = {https://ui.adsabs.harvard.edu/abs/2022A&A...663A..22G}
}

@ARTICLE{rowl24,
       author = {{Rowland}, Lucie E. and {Hodge}, Jacqueline and {Bouwens}, Rychard and {Mancera Pi{\~n}a}, Pavel E. and {Hygate}, Alexander and {Algera}, Hiddo and {Aravena}, Manuel and {Bowler}, Rebecca and {da Cunha}, Elisabete and {Dayal}, Pratika and {Ferrara}, Andrea and {Herard-Demanche}, Thomas and {Inami}, Hanae and {van Leeuwen}, Ivana and {de Looze}, Ilse and {Oesch}, Pascal and {Pallottini}, Andrea and {Phillips}, Si{\^a}n and {Rybak}, Matus and {Schouws}, Sander and {Smit}, Renske and {Sommovigo}, Laura and {Stefanon}, Mauro and {van der Werf}, Paul},
        title = "{REBELS-25: discovery of a dynamically cold disc galaxy at z = 7.31}",
      journal = {\mnras},
         year = 2024,
        month = dec,
       volume = {535},
       number = {3},
        pages = {2068-2091},
          doi = {10.1093/mnras/stae2217},
archivePrefix = {arXiv},
       eprint = {2405.06025},
 primaryClass = {astro-ph.GA},
       adsurl = {https://ui.adsabs.harvard.edu/abs/2024MNRAS.535.2068R}
}

@ARTICLE{hari20,
       author = {{Harikane}, Yuichi and {Ouchi}, Masami and {Inoue}, Akio K. and {Matsuoka}, Yoshiki and {Tamura}, Yoichi and {Bakx}, Tom and {Fujimoto}, Seiji and {Moriwaki}, Kana and {Ono}, Yoshiaki and {Nagao}, Tohru and {Tadaki}, Ken-ichi and {Kojima}, Takashi and {Shibuya}, Takatoshi and {Egami}, Eiichi and {Ferrara}, Andrea and {Gallerani}, Simona and {Hashimoto}, Takuya and {Kohno}, Kotaro and {Matsuda}, Yuichi and {Matsuo}, Hiroshi and {Pallottini}, Andrea and {Sugahara}, Yuma and {Vallini}, Livia},
        title = "{Large Population of ALMA Galaxies at z > 6 with Very High [O III] 88 {\ensuremath{\mu}}m to [C II] 158 {\ensuremath{\mu}}m Flux Ratios: Evidence of Extremely High Ionization Parameter or PDR Deficit?}",
      journal = {\apj},
         year = 2020,
        month = jun,
       volume = {896},
       number = {2},
          eid = {93},
        pages = {93},
          doi = {10.3847/1538-4357/ab94bd},
archivePrefix = {arXiv},
       eprint = {1910.10927},
 primaryClass = {astro-ph.GA},
       adsurl = {https://ui.adsabs.harvard.edu/abs/2020ApJ...896...93H}
}

@ARTICLE{pern23,
       author = {{Perna}, M. and {Arribas}, S. and {Marshall}, M. and {D'Eugenio}, F. and {{\"U}bler}, H. and {Bunker}, A. and {Charlot}, S. and {Carniani}, S. and {Jakobsen}, P. and {Maiolino}, R. and {Rodr{\'\i}guez Del Pino}, B. and {Willott}, C.~J. and {B{\"o}ker}, T. and {Circosta}, C. and {Cresci}, G. and {Curti}, M. and {Husemann}, B. and {Kumari}, N. and {Lamperti}, I. and {P{\'e}rez-Gonz{\'a}lez}, P.~G. and {Scholtz}, J.},
        title = "{GA-NIFS: The ultra-dense, interacting environment of a dual AGN at z {\ensuremath{\sim}} 3.3 revealed by JWST/NIRSpec IFS}",
      journal = {\aap},
         year = 2023,
        month = nov,
       volume = {679},
          eid = {A89},
        pages = {A89},
          doi = {10.1051/0004-6361/202346649},
archivePrefix = {arXiv},
       eprint = {2304.06756},
 primaryClass = {astro-ph.GA},
       adsurl = {https://ui.adsabs.harvard.edu/abs/2023A&A...679A..89P}
}

@ARTICLE{deug24,
       author = {{D'Eugenio}, Francesco and {P{\'e}rez-Gonz{\'a}lez}, Pablo G. and {Maiolino}, Roberto and {Scholtz}, Jan and {Perna}, Michele and {Circosta}, Chiara and {{\"U}bler}, Hannah and {Arribas}, Santiago and {B{\"o}ker}, Torsten and {Bunker}, Andrew J. and {Carniani}, Stefano and {Charlot}, Stephane and {Chevallard}, Jacopo and {Cresci}, Giovanni and {Curtis-Lake}, Emma and {Jones}, Gareth C. and {Kumari}, Nimisha and {Lamperti}, Isabella and {Looser}, Tobias J. and {Parlanti}, Eleonora and {Rix}, Hans-Walter and {Robertson}, Brant and {Rodr{\'\i}guez Del Pino}, Bruno and {Tacchella}, Sandro and {Venturi}, Giacomo and {Willott}, Chris J.},
        title = "{A fast-rotator post-starburst galaxy quenched by supermassive black-hole feedback at z = 3}",
      journal = {Nature Astronomy},
         year = 2024,
        month = nov,
       volume = {8},
        pages = {1443-1456},
          doi = {10.1038/s41550-024-02345-1},
archivePrefix = {arXiv},
       eprint = {2308.06317},
 primaryClass = {astro-ph.GA},
       adsurl = {https://ui.adsabs.harvard.edu/abs/2024NatAs...8.1443D}
}

@ARTICLE{asse16,
       author = {{Assef}, R.~J. and {Walton}, D.~J. and {Brightman}, M. and {Stern}, D. and {Alexander}, D. and {Bauer}, F. and {Blain}, A.~W. and {Diaz-Santos}, T. and {Eisenhardt}, P.~R.~M. and {Finkelstein}, S.~L. and {Hickox}, R.~C. and {Tsai}, C. -W. and {Wu}, J.~W.},
        title = "{Hot Dust Obscured Galaxies with Excess Blue Light: Dual AGN or Single AGN Under Extreme Conditions?}",
      journal = {\apj},
         year = 2016,
        month = mar,
       volume = {819},
       number = {2},
          eid = {111},
        pages = {111},
          doi = {10.3847/0004-637X/819/2/111},
archivePrefix = {arXiv},
       eprint = {1511.05155},
 primaryClass = {astro-ph.GA},
       adsurl = {https://ui.adsabs.harvard.edu/abs/2016ApJ...819..111A}
}

@ARTICLE{nava24,
       author = {{Navarro-Carrera}, Rafael and {Caputi}, Karina I. and {Iani}, Edoardo and {Rinaldi}, Pierluigi and {Kokorev}, Vasily and {Kerutt}, Josephine},
        title = "{The Interstellar Medium Conditions of a Strong Ly{\ensuremath{\alpha}} Emitter at z = 8.279 from JWST: A Robust Lyman Continuum Leaker Candidate at the Epoch of Reionization}",
      journal = {\apj},
         year = 2025,
        month = nov,
       volume = {993},
       number = {2},
          eid = {194},
        pages = {194},
          doi = {10.3847/1538-4357/adca35},
archivePrefix = {arXiv},
       eprint = {2407.14201},
 primaryClass = {astro-ph.GA},
       adsurl = {https://ui.adsabs.harvard.edu/abs/2025ApJ...993..194N}
}

@ARTICLE{fuji25,
       author = {{Fujimoto}, S. and {Ouchi}, M. and {Kohno}, K. and {Valentino}, F. and {Gim{\'e}nez-Arteaga}, C. and {Brammer}, G.~B. and {Furtak}, L.~J. and {Kohandel}, M. and {Oguri}, M. and {Pallottini}, A. and {Richard}, J. and {Zitrin}, A. and {Bauer}, F.~E. and {Boylan-Kolchin}, M. and {Dessauges-Zavadsky}, M. and {Egami}, E. and {Finkelstein}, S.~L. and {Ma}, Z. and {Smail}, I. and {Watson}, D. and {Hutchison}, T.~A. and {Rigby}, J.~R. and {Welch}, B.~D. and {Ao}, Y. and {Bradley}, L.~D. and {Caminha}, G.~B. and {Caputi}, K.~I. and {Espada}, D. and {Endsley}, R. and {Fudamoto}, Y. and {Gonz{\'a}lez-L{\'o}pez}, J. and {Hatsukade}, B. and {Koekemoer}, A.~M. and {Kokorev}, V. and {Laporte}, N. and {Lee}, M. and {Magdis}, G.~E. and {Ono}, Y. and {Rizzo}, F. and {Shibuya}, T. and {Shimasaku}, K. and {Sun}, F. and {Toft}, S. and {Umehata}, H. and {Wang}, T. and {Yajima}, H.},
        title = "{Primordial rotating disk composed of at least 15 dense star-forming clumps at cosmic dawn}",
      journal = {Nature Astronomy},
         year = 2025,
        month = oct,
       volume = {9},
        pages = {1553-1567},
          doi = {10.1038/s41550-025-02592-w},
archivePrefix = {arXiv},
       eprint = {2402.18543},
 primaryClass = {astro-ph.GA},
       adsurl = {https://ui.adsabs.harvard.edu/abs/2025NatAs...9.1553F}
}

@ARTICLE{kiyo25,
       author = {{Kiyota}, Tomokazu and {Ouchi}, Masami and {Xu}, Yi and {Nakazato}, Yurina and {Soga}, Kenta and {Yajima}, Hidenobu and {Fujimoto}, Seiji and {Harikane}, Yuichi and {Nakajima}, Kimihiko and {Ono}, Yoshiaki and {Sun}, Dongsheng and {Kusakabe}, Haruka and {Ceverino}, Daniel and {Hatsukade}, Bunyo and {Iono}, Daisuke and {Kohno}, Kotaro and {Nakanishi}, Kouichiro},
        title = "{Comprehensive JWST+ALMA Study on the Extended Ly{\ensuremath{\alpha}} Emitters, Himiko, and CR7 at z {\ensuremath{\sim}} 7: Blue Major Merger Systems in Stark Contrast to Submillimeter Galaxies}",
      journal = {\apj},
         year = 2025,
        month = dec,
       volume = {995},
       number = {2},
          eid = {150},
        pages = {150},
          doi = {10.3847/1538-4357/ae1cc3},
archivePrefix = {arXiv},
       eprint = {2504.03156},
 primaryClass = {astro-ph.GA},
       adsurl = {https://ui.adsabs.harvard.edu/abs/2025ApJ...995..150K}
}

@ARTICLE{jone24,
       author = {{Jones}, Gareth C. and {Bowler}, Rebecca A.~A. and {Bunker}, Andrew J. and {Curti}, Mirko and {Arribas}, Santiago and {Carniani}, Stefano and {Charlot}, Stephane and {Perna}, Michele and {Rodr{\'\i}guez Del Pino}, Bruno and {{\"U}bler}, Hannah and {Willott}, Chris J. and {Chevallard}, Jacopo and {Cresci}, Giovanni and {Parlanti}, Eleonora and {Scholtz}, Jan and {Venturi}, Giacomo},
        title = "{GA-NIFS: interstellar medium properties and tidal interactions in the evolved massive merging system B14-65666 at z = 7.152}",
      journal = {\mnras},
         year = 2026,
        month = apr,
       volume = {547},
       number = {2},
          eid = {stag336},
        pages = {stag336},
          doi = {10.1093/mnras/stag336},
archivePrefix = {arXiv},
       eprint = {2412.15027},
 primaryClass = {astro-ph.GA},
       adsurl = {https://ui.adsabs.harvard.edu/abs/2026MNRAS.547ag336J}
}

@ARTICLE{gaia16,
       author = {{Gaia Collaboration} and {Prusti}, T. and {de Bruijne}, J.~H.~J. and {Brown}, A.~G.~A. and {Vallenari}, A. and {Babusiaux}, C. and {Bailer-Jones}, C.~A.~L. and {Bastian}, U. and {Biermann}, M. and {Evans}, D.~W. and {Eyer}, L. and {Jansen}, F. and {Jordi}, C. and {Klioner}, S.~A. and {Lammers}, U. and {Lindegren}, L. and {Luri}, X. and {Mignard}, F. and {Milligan}, D.~J. and {Panem}, C. and {Poinsignon}, V. and {Pourbaix}, D. and {Randich}, S. and {Sarri}, G. and {Sartoretti}, P. and {Siddiqui}, H.~I. and {Soubiran}, C. and {Valette}, V. and {van Leeuwen}, F. and {Walton}, N.~A. and {Aerts}, C. and {Arenou}, F. and {Cropper}, M. and {Drimmel}, R. and {H{\o}g}, E. and {Katz}, D. and {Lattanzi}, M.~G. and {O'Mullane}, W. and {Grebel}, E.~K. and {Holland}, A.~D. and {Huc}, C. and {Passot}, X. and {Bramante}, L. and {Cacciari}, C. and {Casta{\~n}eda}, J. and {Chaoul}, L. and {Cheek}, N. and {De Angeli}, F. and {Fabricius}, C. and {Guerra}, R. and {Hern{\'a}ndez}, J. and {Jean-Antoine-Piccolo}, A. and {Masana}, E. and {Messineo}, R. and {Mowlavi}, N. and {Nienartowicz}, K. and {Ord{\'o}{\~n}ez-Blanco}, D. and {Panuzzo}, P. and {Portell}, J. and {Richards}, P.~J. and {Riello}, M. and {Seabroke}, G.~M. and {Tanga}, P. and {Th{\'e}venin}, F. and {Torra}, J. and {Els}, S.~G. and {Gracia-Abril}, G. and {Comoretto}, G. and {Garcia-Reinaldos}, M. and {Lock}, T. and {Mercier}, E. and {Altmann}, M. and {Andrae}, R. and {Astraatmadja}, T.~L. and {Bellas-Velidis}, I. and {Benson}, K. and {Berthier}, J. and {Blomme}, R. and {Busso}, G. and {Carry}, B. and {Cellino}, A. and {Clementini}, G. and {Cowell}, S. and {Creevey}, O. and {Cuypers}, J. and {Davidson}, M. and {De Ridder}, J. and {de Torres}, A. and {Delchambre}, L. and {Dell'Oro}, A. and {Ducourant}, C. and {Fr{\'e}mat}, Y. and {Garc{\'\i}a-Torres}, M. and {Gosset}, E. and {Halbwachs}, J. -L. and {Hambly}, N.~C. and {Harrison}, D.~L. and {Hauser}, M. and {Hestroffer}, D. and {Hodgkin}, S.~T. and {Huckle}, H.~E. and {Hutton}, A. and {Jasniewicz}, G. and {Jordan}, S. and {Kontizas}, M. and {Korn}, A.~J. and {Lanzafame}, A.~C. and {Manteiga}, M. and {Moitinho}, A. and {Muinonen}, K. and {Osinde}, J. and {Pancino}, E. and {Pauwels}, T. and {Petit}, J. -M. and {Recio-Blanco}, A. and {Robin}, A.~C. and {Sarro}, L.~M. and {Siopis}, C. and {Smith}, M. and {Smith}, K.~W. and {Sozzetti}, A. and {Thuillot}, W. and {van Reeven}, W. and {Viala}, Y. and {Abbas}, U. and {Abreu Aramburu}, A. and {Accart}, S. and {Aguado}, J.~J. and {Allan}, P.~M. and {Allasia}, W. and {Altavilla}, G. and {{\'A}lvarez}, M.~A. and {Alves}, J. and {Anderson}, R.~I. and {Andrei}, A.~H. and {Anglada Varela}, E. and {Antiche}, E. and {Antoja}, T. and {Ant{\'o}n}, S. and {Arcay}, B. and {Atzei}, A. and {Ayache}, L. and {Bach}, N. and {Baker}, S.~G. and {Balaguer-N{\'u}{\~n}ez}, L. and {Barache}, C. and {Barata}, C. and {Barbier}, A. and {Barblan}, F. and {Baroni}, M. and {Barrado y Navascu{\'e}s}, D. and {Barros}, M. and {Barstow}, M.~A. and {Becciani}, U. and {Bellazzini}, M. and {Bellei}, G. and {Bello Garc{\'\i}a}, A. and {Belokurov}, V. and {Bendjoya}, P. and {Berihuete}, A. and {Bianchi}, L. and {Bienaym{\'e}}, O. and {Billebaud}, F. and {Blagorodnova}, N. and {Blanco-Cuaresma}, S. and {Boch}, T. and {Bombrun}, A. and {Borrachero}, R. and {Bouquillon}, S. and {Bourda}, G. and {Bouy}, H. and {Bragaglia}, A. and {Breddels}, M.~A. and {Brouillet}, N. and {Br{\"u}semeister}, T. and {Bucciarelli}, B. and {Budnik}, F. and {Burgess}, P. and {Burgon}, R. and {Burlacu}, A. and {Busonero}, D. and {Buzzi}, R. and {Caffau}, E. and {Cambras}, J. and {Campbell}, H. and {Cancelliere}, R. and {Cantat-Gaudin}, T. and {Carlucci}, T. and {Carrasco}, J.~M. and {Castellani}, M. and {Charlot}, P. and {Charnas}, J. and {Charvet}, P. and {Chassat}, F. and {Chiavassa}, A. and {Clotet}, M. and {Cocozza}, G. and {Collins}, R.~S. and {Collins}, P. and {Costigan}, G. and {Crifo}, F. and {Cross}, N.~J.~G. and {Crosta}, M. and {Crowley}, C. and {Dafonte}, C. and {Damerdji}, Y. and {Dapergolas}, A. and {David}, P. and {David}, M. and {De Cat}, P. and {de Felice}, F. and {de Laverny}, P. and {De Luise}, F. and {De March}, R. and {de Martino}, D. and {de Souza}, R. and {Debosscher}, J. and {del Pozo}, E. and {Delbo}, M. and {Delgado}, A. and {Delgado}, H.~E. and {di Marco}, F. and {Di Matteo}, P. and {Diakite}, S. and {Distefano}, E. and {Dolding}, C. and {Dos Anjos}, S. and {Drazinos}, P. and {Dur{\'a}n}, J. and {Dzigan}, Y. and {Ecale}, E. and {Edvardsson}, B. and {Enke}, H. and {Erdmann}, M. and {Escolar}, D. and {Espina}, M. and {Evans}, N.~W. and {Eynard Bontemps}, G. and {Fabre}, C. and {Fabrizio}, M. and {Faigler}, S. and {Falc{\~a}o}, A.~J. and {Farr{\`a}s Casas}, M. and {Faye}, F. and {Federici}, L. and {Fedorets}, G. and {Fern{\'a}ndez-Hern{\'a}ndez}, J. and {Fernique}, P. and {Fienga}, A. and {Figueras}, F. and {Filippi}, F. and {Findeisen}, K. and {Fonti}, A. and {Fouesneau}, M. and {Fraile}, E. and {Fraser}, M. and {Fuchs}, J. and {Furnell}, R. and {Gai}, M. and {Galleti}, S. and {Galluccio}, L. and {Garabato}, D. and {Garc{\'\i}a-Sedano}, F. and {Gar{\'e}}, P. and {Garofalo}, A. and {Garralda}, N. and {Gavras}, P. and {Gerssen}, J. and {Geyer}, R. and {Gilmore}, G. and {Girona}, S. and {Giuffrida}, G. and {Gomes}, M. and {Gonz{\'a}lez-Marcos}, A. and {Gonz{\'a}lez-N{\'u}{\~n}ez}, J. and {Gonz{\'a}lez-Vidal}, J.~J. and {Granvik}, M. and {Guerrier}, A. and {Guillout}, P. and {Guiraud}, J. and {G{\'u}rpide}, A. and {Guti{\'e}rrez-S{\'a}nchez}, R. and {Guy}, L.~P. and {Haigron}, R. and {Hatzidimitriou}, D. and {Haywood}, M. and {Heiter}, U. and {Helmi}, A. and {Hobbs}, D. and {Hofmann}, W. and {Holl}, B. and {Holland}, G. and {Hunt}, J.~A.~S. and {Hypki}, A. and {Icardi}, V. and {Irwin}, M. and {Jevardat de Fombelle}, G. and {Jofr{\'e}}, P. and {Jonker}, P.~G. and {Jorissen}, A. and {Julbe}, F. and {Karampelas}, A. and {Kochoska}, A. and {Kohley}, R. and {Kolenberg}, K. and {Kontizas}, E. and {Koposov}, S.~E. and {Kordopatis}, G. and {Koubsky}, P. and {Kowalczyk}, A. and {Krone-Martins}, A. and {Kudryashova}, M. and {Kull}, I. and {Bachchan}, R.~K. and {Lacoste-Seris}, F. and {Lanza}, A.~F. and {Lavigne}, J. -B. and {Le Poncin-Lafitte}, C. and {Lebreton}, Y. and {Lebzelter}, T. and {Leccia}, S. and {Leclerc}, N. and {Lecoeur-Taibi}, I. and {Lemaitre}, V. and {Lenhardt}, H. and {Leroux}, F. and {Liao}, S. and {Licata}, E. and {Lindstr{\o}m}, H.~E.~P. and {Lister}, T.~A. and {Livanou}, E. and {Lobel}, A. and {L{\"o}ffler}, W. and {L{\'o}pez}, M. and {Lopez-Lozano}, A. and {Lorenz}, D. and {Loureiro}, T. and {MacDonald}, I. and {Magalh{\~a}es Fernandes}, T. and {Managau}, S. and {Mann}, R.~G. and {Mantelet}, G. and {Marchal}, O. and {Marchant}, J.~M. and {Marconi}, M. and {Marie}, J. and {Marinoni}, S. and {Marrese}, P.~M. and {Marschalk{\'o}}, G. and {Marshall}, D.~J. and {Mart{\'\i}n-Fleitas}, J.~M. and {Martino}, M. and {Mary}, N. and {Matijevi{\v{c}}}, G. and {Mazeh}, T. and {McMillan}, P.~J. and {Messina}, S. and {Mestre}, A. and {Michalik}, D. and {Millar}, N.~R. and {Miranda}, B.~M.~H. and {Molina}, D. and {Molinaro}, R. and {Molinaro}, M. and {Moln{\'a}r}, L. and {Moniez}, M. and {Montegriffo}, P. and {Monteiro}, D. and {Mor}, R. and {Mora}, A. and {Morbidelli}, R. and {Morel}, T. and {Morgenthaler}, S. and {Morley}, T. and {Morris}, D. and {Mulone}, A.~F. and {Muraveva}, T. and {Musella}, I. and {Narbonne}, J. and {Nelemans}, G. and {Nicastro}, L. and {Noval}, L. and {Ord{\'e}novic}, C. and {Ordieres-Mer{\'e}}, J. and {Osborne}, P. and {Pagani}, C. and {Pagano}, I. and {Pailler}, F. and {Palacin}, H. and {Palaversa}, L. and {Parsons}, P. and {Paulsen}, T. and {Pecoraro}, M. and {Pedrosa}, R. and {Pentik{\"a}inen}, H. and {Pereira}, J. and {Pichon}, B. and {Piersimoni}, A.~M. and {Pineau}, F. -X. and {Plachy}, E. and {Plum}, G. and {Poujoulet}, E. and {Pr{\v{s}}a}, A. and {Pulone}, L. and {Ragaini}, S. and {Rago}, S. and {Rambaux}, N. and {Ramos-Lerate}, M. and {Ranalli}, P. and {Rauw}, G. and {Read}, A. and {Regibo}, S. and {Renk}, F. and {Reyl{\'e}}, C. and {Ribeiro}, R.~A. and {Rimoldini}, L. and {Ripepi}, V. and {Riva}, A. and {Rixon}, G. and {Roelens}, M. and {Romero-G{\'o}mez}, M. and {Rowell}, N. and {Royer}, F. and {Rudolph}, A. and {Ruiz-Dern}, L. and {Sadowski}, G. and {Sagrist{\`a} Sell{\'e}s}, T. and {Sahlmann}, J. and {Salgado}, J. and {Salguero}, E. and {Sarasso}, M. and {Savietto}, H. and {Schnorhk}, A. and {Schultheis}, M. and {Sciacca}, E. and {Segol}, M. and {Segovia}, J.~C. and {Segransan}, D. and {Serpell}, E. and {Shih}, I. -C. and {Smareglia}, R. and {Smart}, R.~L. and {Smith}, C. and {Solano}, E. and {Solitro}, F. and {Sordo}, R. and {Soria Nieto}, S. and {Souchay}, J. and {Spagna}, A. and {Spoto}, F. and {Stampa}, U. and {Steele}, I.~A. and {Steidelm{\"u}ller}, H. and {Stephenson}, C.~A. and {Stoev}, H. and {Suess}, F.~F. and {S{\"u}veges}, M. and {Surdej}, J. and {Szabados}, L. and {Szegedi-Elek}, E. and {Tapiador}, D. and {Taris}, F. and {Tauran}, G. and {Taylor}, M.~B. and {Teixeira}, R. and {Terrett}, D. and {Tingley}, B. and {Trager}, S.~C. and {Turon}, C. and {Ulla}, A. and {Utrilla}, E. and {Valentini}, G. and {van Elteren}, A. and {Van Hemelryck}, E. and {van Leeuwen}, M. and {Varadi}, M. and {Vecchiato}, A. and {Veljanoski}, J. and {Via}, T. and {Vicente}, D. and {Vogt}, S. and {Voss}, H. and {Votruba}, V. and {Voutsinas}, S. and {Walmsley}, G. and {Weiler}, M. and {Weingrill}, K. and {Werner}, D. and {Wevers}, T. and {Whitehead}, G. and {Wyrzykowski}, {\L}. and {Yoldas}, A. and {{\v{Z}}erjal}, M. and {Zucker}, S. and {Zurbach}, C. and {Zwitter}, T. and {Alecu}, A. and {Allen}, M. and {Allende Prieto}, C. and {Amorim}, A. and {Anglada-Escud{\'e}}, G. and {Arsenijevic}, V. and {Azaz}, S. and {Balm}, P. and {Beck}, M. and {Bernstein}, H. -H. and {Bigot}, L. and {Bijaoui}, A. and {Blasco}, C. and {Bonfigli}, M. and {Bono}, G. and {Boudreault}, S. and {Bressan}, A. and {Brown}, S. and {Brunet}, P. -M. and {Bunclark}, P. and {Buonanno}, R. and {Butkevich}, A.~G. and {Carret}, C. and {Carrion}, C. and {Chemin}, L. and {Ch{\'e}reau}, F. and {Corcione}, L. and {Darmigny}, E. and {de Boer}, K.~S. and {de Teodoro}, P. and {de Zeeuw}, P.~T. and {Delle Luche}, C. and {Domingues}, C.~D. and {Dubath}, P. and {Fodor}, F. and {Fr{\'e}zouls}, B. and {Fries}, A. and {Fustes}, D. and {Fyfe}, D. and {Gallardo}, E. and {Gallegos}, J. and {Gardiol}, D. and {Gebran}, M. and {Gomboc}, A. and {G{\'o}mez}, A. and {Grux}, E. and {Gueguen}, A. and {Heyrovsky}, A. and {Hoar}, J. and {Iannicola}, G. and {Isasi Parache}, Y. and {Janotto}, A. -M. and {Joliet}, E. and {Jonckheere}, A. and {Keil}, R. and {Kim}, D. -W. and {Klagyivik}, P. and {Klar}, J. and {Knude}, J. and {Kochukhov}, O. and {Kolka}, I. and {Kos}, J. and {Kutka}, A. and {Lainey}, V. and {LeBouquin}, D. and {Liu}, C. and {Loreggia}, D. and {Makarov}, V.~V. and {Marseille}, M.~G. and {Martayan}, C. and {Martinez-Rubi}, O. and {Massart}, B. and {Meynadier}, F. and {Mignot}, S. and {Munari}, U. and {Nguyen}, A. -T. and {Nordlander}, T. and {Ocvirk}, P. and {O'Flaherty}, K.~S. and {Olias Sanz}, A. and {Ortiz}, P. and {Osorio}, J. and {Oszkiewicz}, D. and {Ouzounis}, A. and {Palmer}, M. and {Park}, P. and {Pasquato}, E. and {Peltzer}, C. and {Peralta}, J. and {P{\'e}turaud}, F. and {Pieniluoma}, T. and {Pigozzi}, E. and {Poels}, J. and {Prat}, G. and {Prod'homme}, T. and {Raison}, F. and {Rebordao}, J.~M. and {Risquez}, D. and {Rocca-Volmerange}, B. and {Rosen}, S. and {Ruiz-Fuertes}, M.~I. and {Russo}, F. and {Sembay}, S. and {Serraller Vizcaino}, I. and {Short}, A. and {Siebert}, A. and {Silva}, H. and {Sinachopoulos}, D. and {Slezak}, E. and {Soffel}, M. and {Sosnowska}, D. and {Strai{\v{z}}ys}, V. and {ter Linden}, M. and {Terrell}, D. and {Theil}, S. and {Tiede}, C. and {Troisi}, L. and {Tsalmantza}, P. and {Tur}, D. and {Vaccari}, M. and {Vachier}, F. and {Valles}, P. and {Van Hamme}, W. and {Veltz}, L. and {Virtanen}, J. and {Wallut}, J. -M. and {Wichmann}, R. and {Wilkinson}, M.~I. and {Ziaeepour}, H. and {Zschocke}, S.},
        title = "{The Gaia mission}",
      journal = {\aap},
         year = 2016,
        month = nov,
       volume = {595},
          eid = {A1},
        pages = {A1},
          doi = {10.1051/0004-6361/201629272},
archivePrefix = {arXiv},
       eprint = {1609.04153},
 primaryClass = {astro-ph.IM},
       adsurl = {https://ui.adsabs.harvard.edu/abs/2016A&A...595A...1G}
}

@ARTICLE{gaia21,
       author = {{Gaia Collaboration} and {Brown}, A.~G.~A. and {Vallenari}, A. and {Prusti}, T. and {de Bruijne}, J.~H.~J. and {Babusiaux}, C. and {Biermann}, M. and {Creevey}, O.~L. and {Evans}, D.~W. and {Eyer}, L. and {Hutton}, A. and {Jansen}, F. and {Jordi}, C. and {Klioner}, S.~A. and {Lammers}, U. and {Lindegren}, L. and {Luri}, X. and {Mignard}, F. and {Panem}, C. and {Pourbaix}, D. and {Randich}, S. and {Sartoretti}, P. and {Soubiran}, C. and {Walton}, N.~A. and {Arenou}, F. and {Bailer-Jones}, C.~A.~L. and {Bastian}, U. and {Cropper}, M. and {Drimmel}, R. and {Katz}, D. and {Lattanzi}, M.~G. and {van Leeuwen}, F. and {Bakker}, J. and {Cacciari}, C. and {Casta{\~n}eda}, J. and {De Angeli}, F. and {Ducourant}, C. and {Fabricius}, C. and {Fouesneau}, M. and {Fr{\'e}mat}, Y. and {Guerra}, R. and {Guerrier}, A. and {Guiraud}, J. and {Jean-Antoine Piccolo}, A. and {Masana}, E. and {Messineo}, R. and {Mowlavi}, N. and {Nicolas}, C. and {Nienartowicz}, K. and {Pailler}, F. and {Panuzzo}, P. and {Riclet}, F. and {Roux}, W. and {Seabroke}, G.~M. and {Sordo}, R. and {Tanga}, P. and {Th{\'e}venin}, F. and {Gracia-Abril}, G. and {Portell}, J. and {Teyssier}, D. and {Altmann}, M. and {Andrae}, R. and {Bellas-Velidis}, I. and {Benson}, K. and {Berthier}, J. and {Blomme}, R. and {Brugaletta}, E. and {Burgess}, P.~W. and {Busso}, G. and {Carry}, B. and {Cellino}, A. and {Cheek}, N. and {Clementini}, G. and {Damerdji}, Y. and {Davidson}, M. and {Delchambre}, L. and {Dell'Oro}, A. and {Fern{\'a}ndez-Hern{\'a}ndez}, J. and {Galluccio}, L. and {Garc{\'\i}a-Lario}, P. and {Garcia-Reinaldos}, M. and {Gonz{\'a}lez-N{\'u}{\~n}ez}, J. and {Gosset}, E. and {Haigron}, R. and {Halbwachs}, J. -L. and {Hambly}, N.~C. and {Harrison}, D.~L. and {Hatzidimitriou}, D. and {Heiter}, U. and {Hern{\'a}ndez}, J. and {Hestroffer}, D. and {Hodgkin}, S.~T. and {Holl}, B. and {Jan{\ss}en}, K. and {Jevardat de Fombelle}, G. and {Jordan}, S. and {Krone-Martins}, A. and {Lanzafame}, A.~C. and {L{\"o}ffler}, W. and {Lorca}, A. and {Manteiga}, M. and {Marchal}, O. and {Marrese}, P.~M. and {Moitinho}, A. and {Mora}, A. and {Muinonen}, K. and {Osborne}, P. and {Pancino}, E. and {Pauwels}, T. and {Petit}, J. -M. and {Recio-Blanco}, A. and {Richards}, P.~J. and {Riello}, M. and {Rimoldini}, L. and {Robin}, A.~C. and {Roegiers}, T. and {Rybizki}, J. and {Sarro}, L.~M. and {Siopis}, C. and {Smith}, M. and {Sozzetti}, A. and {Ulla}, A. and {Utrilla}, E. and {van Leeuwen}, M. and {van Reeven}, W. and {Abbas}, U. and {Abreu Aramburu}, A. and {Accart}, S. and {Aerts}, C. and {Aguado}, J.~J. and {Ajaj}, M. and {Altavilla}, G. and {{\'A}lvarez}, M.~A. and {{\'A}lvarez Cid-Fuentes}, J. and {Alves}, J. and {Anderson}, R.~I. and {Anglada Varela}, E. and {Antoja}, T. and {Audard}, M. and {Baines}, D. and {Baker}, S.~G. and {Balaguer-N{\'u}{\~n}ez}, L. and {Balbinot}, E. and {Balog}, Z. and {Barache}, C. and {Barbato}, D. and {Barros}, M. and {Barstow}, M.~A. and {Bartolom{\'e}}, S. and {Bassilana}, J. -L. and {Bauchet}, N. and {Baudesson-Stella}, A. and {Becciani}, U. and {Bellazzini}, M. and {Bernet}, M. and {Bertone}, S. and {Bianchi}, L. and {Blanco-Cuaresma}, S. and {Boch}, T. and {Bombrun}, A. and {Bossini}, D. and {Bouquillon}, S. and {Bragaglia}, A. and {Bramante}, L. and {Breedt}, E. and {Bressan}, A. and {Brouillet}, N. and {Bucciarelli}, B. and {Burlacu}, A. and {Busonero}, D. and {Butkevich}, A.~G. and {Buzzi}, R. and {Caffau}, E. and {Cancelliere}, R. and {C{\'a}novas}, H. and {Cantat-Gaudin}, T. and {Carballo}, R. and {Carlucci}, T. and {Carnerero}, M.~I. and {Carrasco}, J.~M. and {Casamiquela}, L. and {Castellani}, M. and {Castro-Ginard}, A. and {Castro Sampol}, P. and {Chaoul}, L. and {Charlot}, P. and {Chemin}, L. and {Chiavassa}, A. and {Cioni}, M. -R.~L. and {Comoretto}, G. and {Cooper}, W.~J. and {Cornez}, T. and {Cowell}, S. and {Crifo}, F. and {Crosta}, M. and {Crowley}, C. and {Dafonte}, C. and {Dapergolas}, A. and {David}, M. and {David}, P. and {de Laverny}, P. and {De Luise}, F. and {De March}, R. and {De Ridder}, J. and {de Souza}, R. and {de Teodoro}, P. and {de Torres}, A. and {del Peloso}, E.~F. and {del Pozo}, E. and {Delbo}, M. and {Delgado}, A. and {Delgado}, H.~E. and {Delisle}, J. -B. and {Di Matteo}, P. and {Diakite}, S. and {Diener}, C. and {Distefano}, E. and {Dolding}, C. and {Eappachen}, D. and {Edvardsson}, B. and {Enke}, H. and {Esquej}, P. and {Fabre}, C. and {Fabrizio}, M. and {Faigler}, S. and {Fedorets}, G. and {Fernique}, P. and {Fienga}, A. and {Figueras}, F. and {Fouron}, C. and {Fragkoudi}, F. and {Fraile}, E. and {Franke}, F. and {Gai}, M. and {Garabato}, D. and {Garcia-Gutierrez}, A. and {Garc{\'\i}a-Torres}, M. and {Garofalo}, A. and {Gavras}, P. and {Gerlach}, E. and {Geyer}, R. and {Giacobbe}, P. and {Gilmore}, G. and {Girona}, S. and {Giuffrida}, G. and {Gomel}, R. and {Gomez}, A. and {Gonzalez-Santamaria}, I. and {Gonz{\'a}lez-Vidal}, J.~J. and {Granvik}, M. and {Guti{\'e}rrez-S{\'a}nchez}, R. and {Guy}, L.~P. and {Hauser}, M. and {Haywood}, M. and {Helmi}, A. and {Hidalgo}, S.~L. and {Hilger}, T. and {H{\l}adczuk}, N. and {Hobbs}, D. and {Holland}, G. and {Huckle}, H.~E. and {Jasniewicz}, G. and {Jonker}, P.~G. and {Juaristi Campillo}, J. and {Julbe}, F. and {Karbevska}, L. and {Kervella}, P. and {Khanna}, S. and {Kochoska}, A. and {Kontizas}, M. and {Kordopatis}, G. and {Korn}, A.~J. and {Kostrzewa-Rutkowska}, Z. and {Kruszy{\'n}ska}, K. and {Lambert}, S. and {Lanza}, A.~F. and {Lasne}, Y. and {Le Campion}, J. -F. and {Le Fustec}, Y. and {Lebreton}, Y. and {Lebzelter}, T. and {Leccia}, S. and {Leclerc}, N. and {Lecoeur-Taibi}, I. and {Liao}, S. and {Licata}, E. and {Lindstr{\o}m}, E.~P. and {Lister}, T.~A. and {Livanou}, E. and {Lobel}, A. and {Madrero Pardo}, P. and {Managau}, S. and {Mann}, R.~G. and {Marchant}, J.~M. and {Marconi}, M. and {Marcos Santos}, M.~M.~S. and {Marinoni}, S. and {Marocco}, F. and {Marshall}, D.~J. and {Martin Polo}, L. and {Mart{\'\i}n-Fleitas}, J.~M. and {Masip}, A. and {Massari}, D. and {Mastrobuono-Battisti}, A. and {Mazeh}, T. and {McMillan}, P.~J. and {Messina}, S. and {Michalik}, D. and {Millar}, N.~R. and {Mints}, A. and {Molina}, D. and {Molinaro}, R. and {Moln{\'a}r}, L. and {Montegriffo}, P. and {Mor}, R. and {Morbidelli}, R. and {Morel}, T. and {Morris}, D. and {Mulone}, A.~F. and {Munoz}, D. and {Muraveva}, T. and {Murphy}, C.~P. and {Musella}, I. and {Noval}, L. and {Ord{\'e}novic}, C. and {Orr{\`u}}, G. and {Osinde}, J. and {Pagani}, C. and {Pagano}, I. and {Palaversa}, L. and {Palicio}, P.~A. and {Panahi}, A. and {Pawlak}, M. and {Pe{\~n}alosa Esteller}, X. and {Penttil{\"a}}, A. and {Piersimoni}, A.~M. and {Pineau}, F. -X. and {Plachy}, E. and {Plum}, G. and {Poggio}, E. and {Poretti}, E. and {Poujoulet}, E. and {Pr{\v{s}}a}, A. and {Pulone}, L. and {Racero}, E. and {Ragaini}, S. and {Rainer}, M. and {Raiteri}, C.~M. and {Rambaux}, N. and {Ramos}, P. and {Ramos-Lerate}, M. and {Re Fiorentin}, P. and {Regibo}, S. and {Reyl{\'e}}, C. and {Ripepi}, V. and {Riva}, A. and {Rixon}, G. and {Robichon}, N. and {Robin}, C. and {Roelens}, M. and {Rohrbasser}, L. and {Romero-G{\'o}mez}, M. and {Rowell}, N. and {Royer}, F. and {Rybicki}, K.~A. and {Sadowski}, G. and {Sagrist{\`a} Sell{\'e}s}, A. and {Sahlmann}, J. and {Salgado}, J. and {Salguero}, E. and {Samaras}, N. and {Sanchez Gimenez}, V. and {Sanna}, N. and {Santove{\~n}a}, R. and {Sarasso}, M. and {Schultheis}, M. and {Sciacca}, E. and {Segol}, M. and {Segovia}, J.~C. and {S{\'e}gransan}, D. and {Semeux}, D. and {Shahaf}, S. and {Siddiqui}, H.~I. and {Siebert}, A. and {Siltala}, L. and {Slezak}, E. and {Smart}, R.~L. and {Solano}, E. and {Solitro}, F. and {Souami}, D. and {Souchay}, J. and {Spagna}, A. and {Spoto}, F. and {Steele}, I.~A. and {Steidelm{\"u}ller}, H. and {Stephenson}, C.~A. and {S{\"u}veges}, M. and {Szabados}, L. and {Szegedi-Elek}, E. and {Taris}, F. and {Tauran}, G. and {Taylor}, M.~B. and {Teixeira}, R. and {Thuillot}, W. and {Tonello}, N. and {Torra}, F. and {Torra}, J. and {Turon}, C. and {Unger}, N. and {Vaillant}, M. and {van Dillen}, E. and {Vanel}, O. and {Vecchiato}, A. and {Viala}, Y. and {Vicente}, D. and {Voutsinas}, S. and {Weiler}, M. and {Wevers}, T. and {Wyrzykowski}, {\L}. and {Yoldas}, A. and {Yvard}, P. and {Zhao}, H. and {Zorec}, J. and {Zucker}, S. and {Zurbach}, C. and {Zwitter}, T.},
        title = "{Gaia Early Data Release 3. Summary of the contents and survey properties}",
      journal = {\aap},
         year = 2021,
        month = may,
       volume = {649},
          eid = {A1},
        pages = {A1},
          doi = {10.1051/0004-6361/202039657},
archivePrefix = {arXiv},
       eprint = {2012.01533},
 primaryClass = {astro-ph.GA},
       adsurl = {https://ui.adsabs.harvard.edu/abs/2021A&A...649A...1G}
}

@ARTICLE{jone20,
       author = {{Jones}, G.~C. and {Maiolino}, R. and {Caselli}, P. and {Carniani}, S.},
        title = "{Gas and star formation from HD and dust emission in a strongly lensed galaxy}",
      journal = {\mnras},
         year = 2020,
        month = nov,
       volume = {498},
       number = {3},
        pages = {4109-4118},
          doi = {10.1093/mnras/staa2689},
archivePrefix = {arXiv},
       eprint = {2009.00674},
 primaryClass = {astro-ph.GA},
       adsurl = {https://ui.adsabs.harvard.edu/abs/2020MNRAS.498.4109J}
}

@ARTICLE{wits23,
       author = {{Witstok}, Joris and {Jones}, Gareth C. and {Maiolino}, Roberto and {Smit}, Renske and {Schneider}, Raffaella},
        title = "{An empirical study of dust properties at the earliest epochs}",
      journal = {\mnras},
         year = 2023,
        month = aug,
       volume = {523},
       number = {2},
        pages = {3119-3132},
          doi = {10.1093/mnras/stad1470},
archivePrefix = {arXiv},
       eprint = {2305.09714},
 primaryClass = {astro-ph.GA},
       adsurl = {https://ui.adsabs.harvard.edu/abs/2023MNRAS.523.3119W}
}

@ARTICLE{somm22,
       author = {{Sommovigo}, L. and {Ferrara}, A. and {Pallottini}, A. and {Dayal}, P. and {Bouwens}, R.~J. and {Smit}, R. and {da Cunha}, E. and {De Looze}, I. and {Bowler}, R.~A.~A. and {Hodge}, J. and {Inami}, H. and {Oesch}, P. and {Endsley}, R. and {Gonzalez}, V. and {Schouws}, S. and {Stark}, D. and {Stefanon}, M. and {Aravena}, M. and {Graziani}, L. and {Riechers}, D. and {Schneider}, R. and {van der Werf}, P. and {Algera}, H. and {Barrufet}, L. and {Fudamoto}, Y. and {Hygate}, A.~P.~S. and {Labb{\'e}}, I. and {Li}, Y. and {Nanayakkara}, T. and {Topping}, M.},
        title = "{The ALMA REBELS Survey: cosmic dust temperature evolution out to z   7}",
      journal = {\mnras},
         year = 2022,
        month = jul,
       volume = {513},
       number = {3},
        pages = {3122-3135},
          doi = {10.1093/mnras/stac302},
archivePrefix = {arXiv},
       eprint = {2202.01227},
 primaryClass = {astro-ph.GA},
       adsurl = {https://ui.adsabs.harvard.edu/abs/2022MNRAS.513.3122S}
}

@ARTICLE{roma23,
       author = {{Roman-Oliveira}, Fernanda and {Fraternali}, Filippo and {Rizzo}, Francesca},
        title = "{Regular rotation and low turbulence in a diverse sample of z {\ensuremath{\sim}} 4.5 galaxies observed with ALMA}",
      journal = {\mnras},
         year = 2023,
        month = may,
       volume = {521},
       number = {1},
        pages = {1045-1065},
          doi = {10.1093/mnras/stad530},
archivePrefix = {arXiv},
       eprint = {2302.03049},
 primaryClass = {astro-ph.GA},
       adsurl = {https://ui.adsabs.harvard.edu/abs/2023MNRAS.521.1045R}
}

@ARTICLE{kewl19,
       author = {{Kewley}, Lisa J. and {Nicholls}, David C. and {Sutherland}, Ralph S.},
        title = "{Understanding Galaxy Evolution Through Emission Lines}",
      journal = {\araa},
         year = 2019,
        month = aug,
       volume = {57},
        pages = {511-570},
          doi = {10.1146/annurev-astro-081817-051832},
archivePrefix = {arXiv},
       eprint = {1910.09730},
 primaryClass = {astro-ph.GA},
       adsurl = {https://ui.adsabs.harvard.edu/abs/2019ARA&A..57..511K}
}

@ARTICLE{hsia24,
       author = {{Hsiao}, Tiger Yu-Yang and {{\'A}lvarez-M{\'a}rquez}, Javier and {Coe}, Dan and {Crespo G{\'o}mez}, Alejandro and {Abdurro'uf} and {Dayal}, Pratika and {Larson}, Rebecca L. and {Bik}, Arjan and {Blanco-Prieto}, Carmen and {Colina}, Luis and {P{\'e}rez-Gonz{\'a}lez}, Pablo Guillermo and {Costantin}, Luca and {Prieto-Jim{\'e}nez}, Carlota and {Adamo}, Angela and {Bradley}, Larry D. and {Conselice}, Christopher J. and {Fujimoto}, Seiji and {Furtak}, Lukas J. and {Hutchison}, Taylor A. and {James}, Bethan L. and {Jim{\'e}nez-Teja}, Yolanda and {Jung}, Intae and {Kokorev}, Vasily and {Mingozzi}, Matilde and {Norman}, Colin and {Ricotti}, Massimo and {Rigby}, Jane R. and {Sharon}, Keren and {Vanzella}, Eros and {Welch}, Brian and {Xu}, Xinfeng and {Zackrisson}, Erik and {Zitrin}, Adi},
        title = "{JWST MIRI Detections of H{\ensuremath{\alpha}} and [O III] and a Direct Metallicity Measurement of the z = 10.17 Lensed Galaxy MACS0647‑JD}",
      journal = {\apj},
         year = 2024,
        month = oct,
       volume = {973},
       number = {2},
          eid = {81},
        pages = {81},
          doi = {10.3847/1538-4357/ad6562},
archivePrefix = {arXiv},
       eprint = {2404.16200},
 primaryClass = {astro-ph.GA},
       adsurl = {https://ui.adsabs.harvard.edu/abs/2024ApJ...973...81H}
}

@ARTICLE{sand25,
       author = {{Sanders}, Ryan L. and {Shapley}, Alice E. and {Topping}, Michael W. and {Reddy}, Naveen A. and {Berg}, Danielle A. and {Khostovan}, Ali Ahmad and {Bouwens}, Rychard J. and {Brammer}, Gabriel and {Carnall}, Adam C. and {Cullen}, Fergus and {Dav{\'e}}, Romeel and {Dunlop}, James S. and {Ellis}, Richard S. and {F{\"o}rster Schreiber}, N.~M. and {Furlanetto}, Steven R. and {Glazebrook}, Karl and {Illingworth}, Garth D. and {Jones}, Tucker and {Kriek}, Mariska and {McLeod}, Derek J. and {McLure}, Ross J. and {Narayanan}, Desika and {Oesch}, Pascal A. and {Pahl}, Anthony J. and {Pettini}, Max and {Schaerer}, Daniel and {Stark}, Daniel P. and {Steidel}, Charles C. and {Tang}, Mengtao and {Clarke}, Leonardo and {Donnan}, Callum T. and {Kehoe}, Emily},
        title = "{The AURORA Survey: High-redshift Empirical Metallicity Calibrations from Electron Temperature Measurements at z = 2─10}",
      journal = {\apj},
         year = 2026,
        month = jun,
       volume = {1003},
       number = {2},
          eid = {228},
        pages = {228},
          doi = {10.3847/1538-4357/ae66e2},
archivePrefix = {arXiv},
       eprint = {2508.10099},
 primaryClass = {astro-ph.GA},
       adsurl = {https://ui.adsabs.harvard.edu/abs/2026ApJ..1003..228S}
}

@ARTICLE{kenn98,
       author = {{Kennicutt}, Robert C., Jr.},
        title = "{The Global Schmidt Law in Star-forming Galaxies}",
      journal = {\apj},
         year = 1998,
        month = may,
       volume = {498},
       number = {2},
        pages = {541-552},
          doi = {10.1086/305588},
archivePrefix = {arXiv},
       eprint = {astro-ph/9712213},
 primaryClass = {astro-ph},
       adsurl = {https://ui.adsabs.harvard.edu/abs/1998ApJ...498..541K}
}

@ARTICLE{salp55,
       author = {{Salpeter}, Edwin E.},
        title = "{The Luminosity Function and Stellar Evolution.}",
      journal = {\apj},
         year = 1955,
        month = jan,
       volume = {121},
        pages = {161},
          doi = {10.1086/145971},
       adsurl = {https://ui.adsabs.harvard.edu/abs/1955ApJ...121..161S}
}

@ARTICLE{chab03,
       author = {{Chabrier}, Gilles},
        title = "{Galactic Stellar and Substellar Initial Mass Function}",
      journal = {\pasp},
         year = 2003,
        month = jul,
       volume = {115},
       number = {809},
        pages = {763-795},
          doi = {10.1086/376392},
archivePrefix = {arXiv},
       eprint = {astro-ph/0304382},
 primaryClass = {astro-ph},
       adsurl = {https://ui.adsabs.harvard.edu/abs/2003PASP..115..763C}
}

@ARTICLE{pant07,
       author = {{Panter}, Benjamin and {Jimenez}, Raul and {Heavens}, Alan F. and {Charlot}, Stephane},
        title = "{The star formation histories of galaxies in the Sloan Digital Sky Survey}",
      journal = {\mnras},
         year = 2007,
        month = jul,
       volume = {378},
       number = {4},
        pages = {1550-1564},
          doi = {10.1111/j.1365-2966.2007.11909.x},
archivePrefix = {arXiv},
       eprint = {astro-ph/0608531},
 primaryClass = {astro-ph},
       adsurl = {https://ui.adsabs.harvard.edu/abs/2007MNRAS.378.1550P}
}

@ARTICLE{long09,
       author = {{Longhetti}, M. and {Saracco}, P.},
        title = "{Stellar mass estimates in early-type galaxies: procedures, uncertainties and models dependence}",
      journal = {\mnras},
         year = 2009,
        month = apr,
       volume = {394},
       number = {2},
        pages = {774-794},
          doi = {10.1111/j.1365-2966.2008.14375.x},
archivePrefix = {arXiv},
       eprint = {0811.4041},
 primaryClass = {astro-ph},
       adsurl = {https://ui.adsabs.harvard.edu/abs/2009MNRAS.394..774L}
}

@ARTICLE{dutt10,
       author = {{Dutton}, Aaron A. and {van den Bosch}, Frank C. and {Dekel}, Avishai},
        title = "{On the origin of the galaxy star-formation-rate sequence: evolution and scatter}",
      journal = {\mnras},
         year = 2010,
        month = jul,
       volume = {405},
       number = {3},
        pages = {1690-1710},
          doi = {10.1111/j.1365-2966.2010.16620.x},
archivePrefix = {arXiv},
       eprint = {0912.2169},
 primaryClass = {astro-ph.CO},
       adsurl = {https://ui.adsabs.harvard.edu/abs/2010MNRAS.405.1690D}
}

@ARTICLE{gonz10,
       author = {{Gonz{\'a}lez}, Valentino and {Labb{\'e}}, Ivo and {Bouwens}, Rychard J. and {Illingworth}, Garth and {Franx}, Marijn and {Kriek}, Mariska and {Brammer}, Gabriel B.},
        title = "{The Stellar Mass Density and Specific Star Formation Rate of the Universe at z \raisebox{-0.5ex}\textasciitilde 7}",
      journal = {\apj},
         year = 2010,
        month = apr,
       volume = {713},
       number = {1},
        pages = {115-130},
          doi = {10.1088/0004-637X/713/1/115},
archivePrefix = {arXiv},
       eprint = {0909.3517},
 primaryClass = {astro-ph.CO},
       adsurl = {https://ui.adsabs.harvard.edu/abs/2010ApJ...713..115G}
}

@ARTICLE{beth12,
       author = {{B{\'e}thermin}, M. and {Dor{\'e}}, O. and {Lagache}, G.},
        title = "{Where stars form and live at high redshift: clues from the infrared}",
      journal = {\aap},
         year = 2012,
        month = jan,
       volume = {537},
          eid = {L5},
        pages = {L5},
          doi = {10.1051/0004-6361/201118607},
archivePrefix = {arXiv},
       eprint = {1201.0546},
 primaryClass = {astro-ph.CO},
       adsurl = {https://ui.adsabs.harvard.edu/abs/2012A&A...537L...5B}
}

@ARTICLE{driv13,
       author = {{Driver}, S.~P. and {Robotham}, A.~S.~G. and {Bland-Hawthorn}, J. and {Brown}, M. and {Hopkins}, A. and {Liske}, J. and {Phillipps}, S. and {Wilkins}, S.},
        title = "{Two-phase galaxy evolution: the cosmic star formation histories of spheroids and discs}",
      journal = {\mnras},
         year = 2013,
        month = apr,
       volume = {430},
       number = {4},
        pages = {2622-2632},
          doi = {10.1093/mnras/sts717},
archivePrefix = {arXiv},
       eprint = {1301.0979},
 primaryClass = {astro-ph.CO},
       adsurl = {https://ui.adsabs.harvard.edu/abs/2013MNRAS.430.2622D}
}





\appendix

\section{Astrometric correction}\label{astrosec}
Since we aim to compare the resolved morpho-kinematics of \ha and \cii emission, it is necessary to verify the astrometry of the JWST/NIRSpec IFU data. To do this, we align an archival JWST/NIRCam image to the \textit{Gaia} DR3 reference frame (\citealt{gaia16,gaia21}), then align our NIRSpec data to this frame.

\JJ benefits from a wealth of existing archival data from space-based observatories, including HST WFC3/IR F160W (PID 15410, PI M. Neeleman), JWST/MIRI F770W and NIRCam F150W, F200W, F277W, F444W (PID 3954, PI F. Lelli), and JWST/NIRSpec IFU G395M/F290LP (PID 5761, PI M. Neeleman).
Detailed SED fitting of this source will be performed in other works, and we only use the JWST/NIRCam F444W image to verify the astrometry of our NIRSpec IFU data. We download the calibrated (stage 3) JWST/NIRCam F444W image from MAST, as this is the only archival NIRCam image featuring spectral overlap with our NIRSpec data.

\begin{figure}
    \centering
    \includegraphics[width=0.5\textwidth]{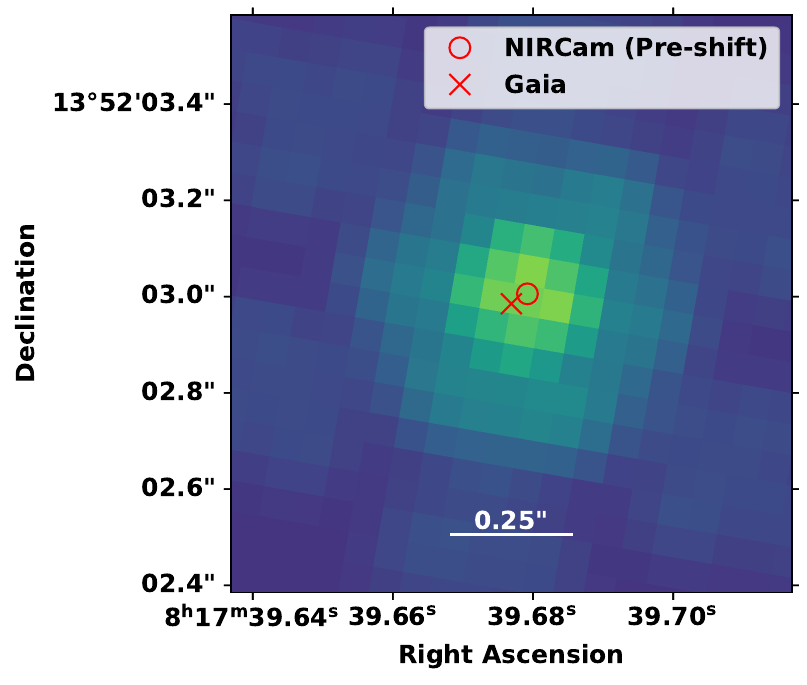}
    \caption{Astrometric verification of JWST/NIRCam F444W image (colour) retrieved from MAST archive. The best-fit centroid is shown by a red circle, while the Gaia DR3 position is shown by a red `x'.}
    \label{gaia0}
\end{figure}

To begin, we search the Gaia DR3 catalogue\footnote{\url{https://gea.esac.esa.int/archive/}} for stars near \JJ. We choose the second closest star (Gaia DR3 650676554525737216), as the closest (Gaia DR3 650676344069432832) features large proper motion uncertainties. The position is corrected for proper motion since 2016, resulting in a shift of $6.2\pm0.1$\,mas (where the error includes uncertainties in the proper motion and initial position). The Gaia star is well-detected in the NIRCam image, so we determine its centroid by fitting a 2D Gaussian with \textlcsc{lmfit}. The resulting position is offset from the Gaia position (see Figure \ref{gaia0}) by $39.2\pm2.3$\,mas (where the error is taken from the uncertainty outputted by \textlcsc{lmfit}). The NIRCam image is shifted to the Gaia astrometric frame by adjusting the values in the image header.

Next, we consider the NIRSpec IFU astrometry. Our R2700 data cube is convolved with the NIRCam F444W throughput curve ($3.9\lesssim\lambda_{\rm obs}/\mu{\rm m}<5.1$), resulting in a pseudo- photometric map that we may compare with the NIRCam image. This comparison is not perfect, however, as the R2700 data feature a spectral gap near the centre of the F444W filter, resulting in a lower flux. While we will not use these data for flux calibration verification for this reason, they are still useful for spatial alignment. 

\begin{figure}
    \centering
    \includegraphics[width=0.5\textwidth]{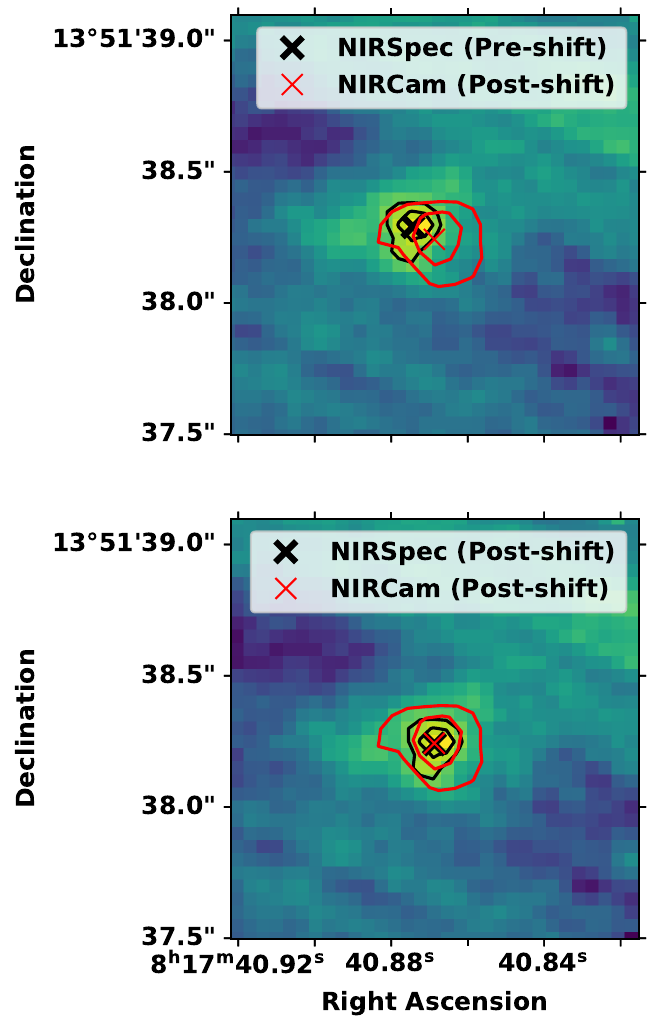}
    \caption{Application of astrometric correction to NIRSpec IFU data. Each panel shows the NIRCam F444W data (after alignment to the Gaia reference frame; red contours) compared to the convolved NIRSpec data (colourmap and black contours). In the top panel, we show the native NIRSpec data, while we align the NIRSpec data to the NIRCam data in the lower frame. Contours are shown at $80\%,90\%$ of the maximum value in each map, and the best-fit centroid of each map is shown as a coloured cross.}
    \label{gaia1}
\end{figure}

The convolved NIRSpec image is shown in Figure \ref{gaia1} (colourmap and black contours), compared to the NIRCam data (red contours). Before spatial alignment, the two maps feature a spatial offset of $95\pm9\,$mas, which is comparable to the pointing uncertainty of NIRSpec with no target acquisition (\citealt{rigb23}). We align the NIRSpec data to the NIRCam reference frame by editing the image header. By combining all relevant uncertainties (i.e., Gaia star proper motion, Gaia star position, NIRCam image centroid, and NIRSpec image centroid) in quadrature, we find a total astrometric uncertainty of 6\,mas.

From equation 10.7 of the ALMA technical handbook\footnote{\url{https://almascience.eso.org/documents-and-tools/cycle12/alma-technical-handbook}}, the positional uncertainty of ALMA is a function of the resolution and peak S/N. Using the values of the \cii moment 0 map yields a positional uncertainty of $\sim14\,$mas, or $\sim0.3$\,spaxels in the NIRSpec IFU cube. 

\section{PSF treatment}\label{psf_stuff}
While the JWST/NIRSpec IFU allows for high-resolution characterization of galaxies, it is vital to account for the wavelength-dependent behaviour of the PSF when comparing emission at different wavelengths. As noted by other works (e.g., \citealt{jone25b}), the NIRSpec PSF features both a central peak (which may be approximated by a Gaussian) and extended, low-level wings. In this Appendix, we use the code \textlcsc{stpsf}\footnote{\url{https://stpsf.readthedocs.io/en/latest/}} to model the PSF of our data.

\begin{figure}
    \centering
    \includegraphics[width=0.5\textwidth]{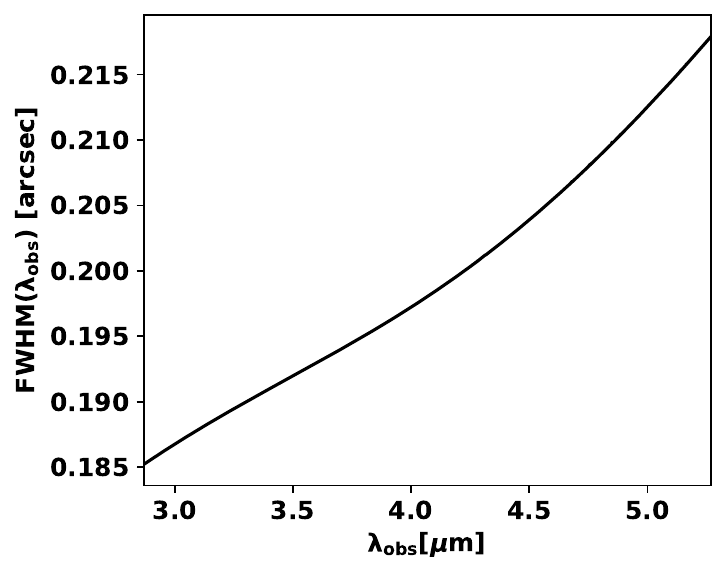}
    \caption{FWHM of the NIRSpec IFU PSF over the G395H/F270LP wavelength range, determined from a 2D Gaussian fit to the model PSF of \textlcsc{stpsf}.}
    \label{psf_fwhm}
\end{figure}

First, we use \textlcsc{stpsf} to create a model data cube of the PSF over the wavelength range of our data, fit a 2-D Gaussian model to the PSF at each wavelength, and record the geometric mean of the best fit FWHM of the major and minor axes (Figure \ref{psf_fwhm}). It is clear that the spatial resolution is degraded at redder wavelengths, with a variation of $\sim7.5\%$ from the mean ($\sim0.2''$).

\begin{figure}
    \centering
    \includegraphics[width=0.5\textwidth]{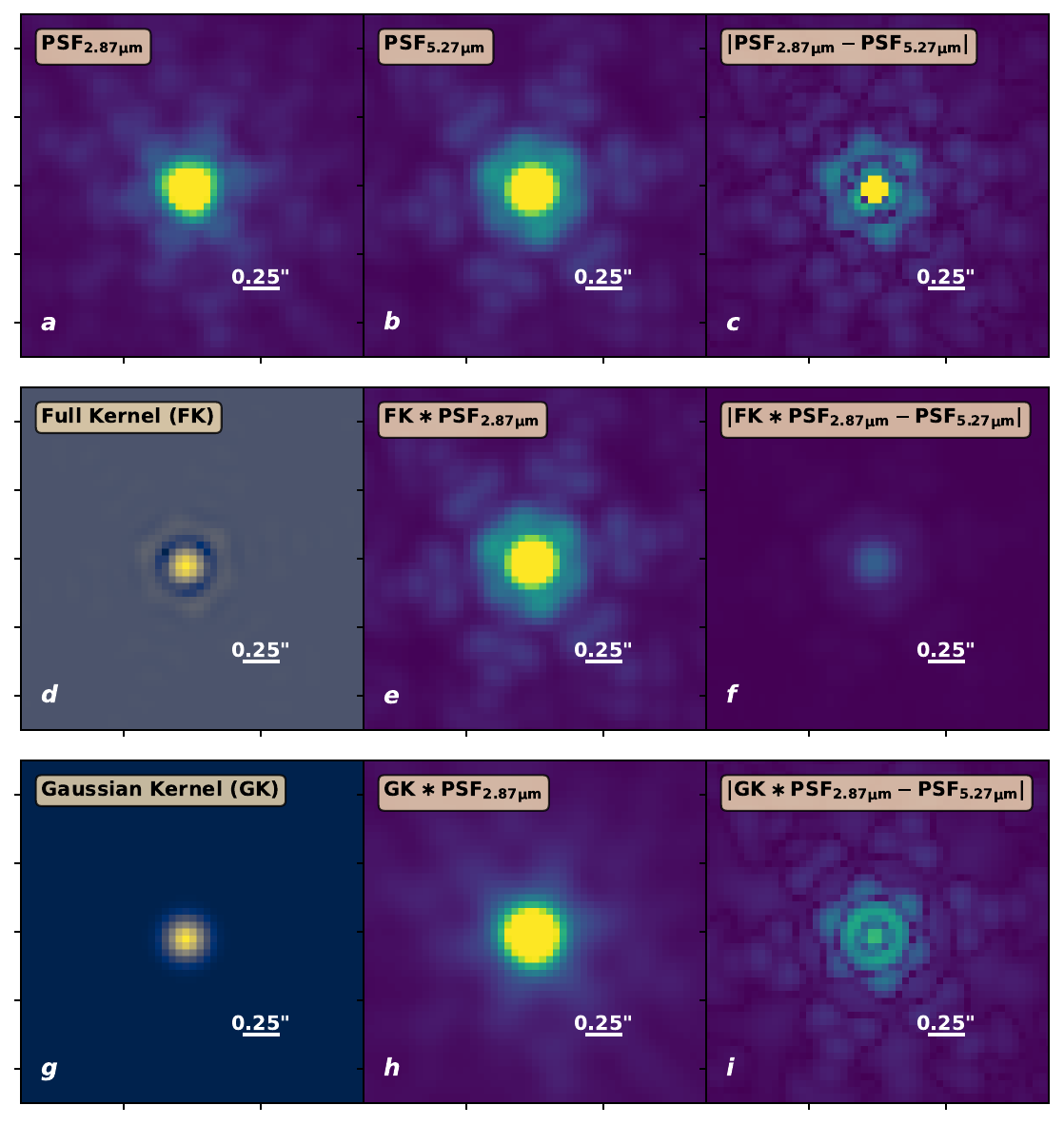}
    \caption{Demonstration of varying PSF size across our data cube, and our method of forcing a common PSF through custom convolution. The top row (\textit{a-c}) shows the finest PSF, coarsest PSF, and the difference between them. The next two rows show kernels derived to match the two PSFs, using \textlcsc{phoutils} create\_matching\_kernel (`Full kernel', or FK; second row) and a Gaussian approximation of the resulting kernel (`Gaussian kernel', or GK; lower row). For each of these, we show the kernel, (\textit{d} \& \textit{g}) the result of convolving the kernel with the finest PSF (\textit{e} \& \textit{h}), and the residuals (\textit{f} \& \textit{i}).}
    \label{psf_comp}
\end{figure}

To examine this in a different way, the top row of Figure \ref{psf_comp} displays the finest PSF ($\lambda_{\rm obs}=2.87\,\mu$m, panel \textit{a}), coarsest PSF ($\lambda_{\rm obs}=5.27\,\mu$m, panel \textit{b}), and the difference the two (panel \textit{c}). It is clear that assuming a constant PSF will result in residuals at all scales.

In order to ensure a fair comparison across all wavelengths of our data, we force a common PSF ($\rm PSF_{5.27\mu m}$) across the data cube. For each wavelength of the cube, the necessary kernel is derived using the \textlcsc{phoutils} task create\_matching\_kernel (adopting a top hat window with $\beta=0.315$), and we use the astropy task \textit{convolve} to perform the convolution. As an example, we show the derived kernel to transform $\rm PSF_{2.87\mu m}$ to $\rm PSF_{5.27\mu m}$ in panel \textit{d}. The resulting convolved PSF (panel \textit{e}) is similar to $\rm PSF_{5.27\mu m}$, resulting in low residuals (panel \textit{f}).

We emphasise that the convolution kernel is non-Gaussian, and that this method is preferable to a Gaussian convolution. Indeed, while the core of the convolution kernel is well described by a Gaussian kernel (panel \textit{g}), if we convolve $\rm PSF_{2.87\mu m}$ with a Gaussian kernel, the resulting convolved map lacks the wings of the real PSF (panel \textit{h}), resulting in large residuals (panel \textit{i}). 

\section{Posterior distributions of ISM Conditions}\label{cornerplots}

\begin{figure}
    \centering
    \includegraphics[width=0.3\textwidth]{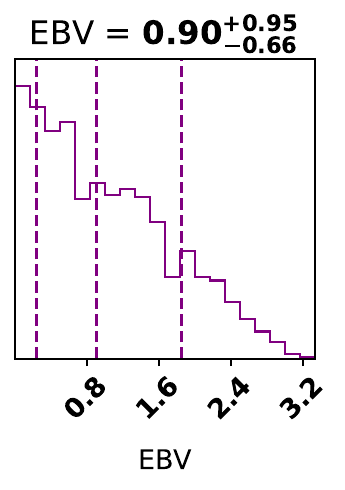}
    \caption{Posterior distribution for $E(B-V)$, as found through the procedure outlined in Section \ref{ISMSec}.}
    \label{cornerplotsfig}
\end{figure}

\section{Azimuthal normalisation}\label{azapp}
In the main work, we present \BB fits to the \cii and \ha data cubes of \JJ, using flexible flux normalisation (i.e., `LOCAL'). However, this approach allows for flux distributions that deviate strongly from the azimuthal symmetry expected from relaxed rotating disk galaxies (e.g., \citealt{wisn15}). As an alternative, we present the same analysis assuming that the flux of \JJ features azimuthal symmetry (i.e., `AZIM' normalisation in \BB).

\begin{figure*}
\centering
\includegraphics[width=0.49\textwidth]{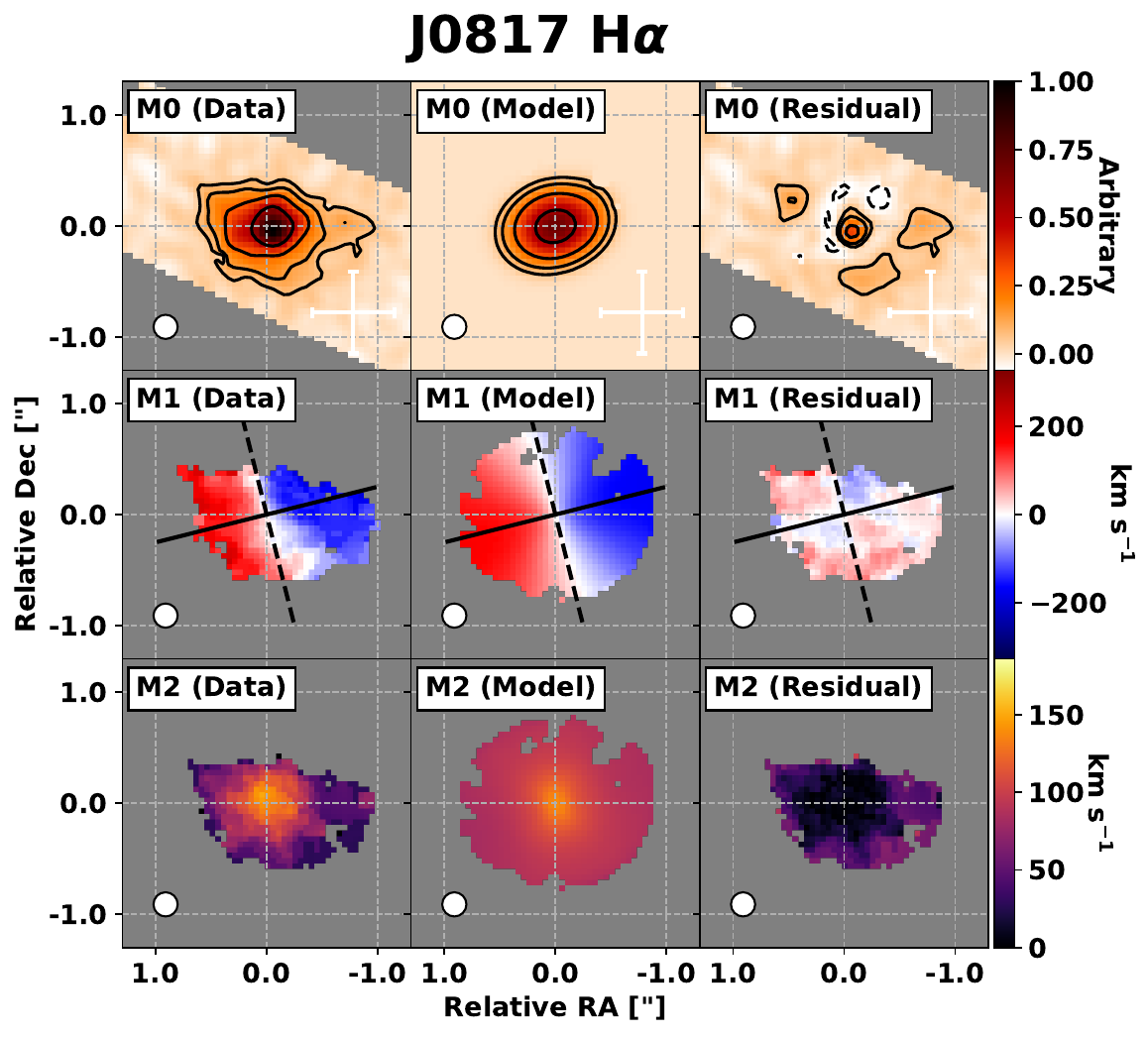}
\includegraphics[width=0.49\textwidth]{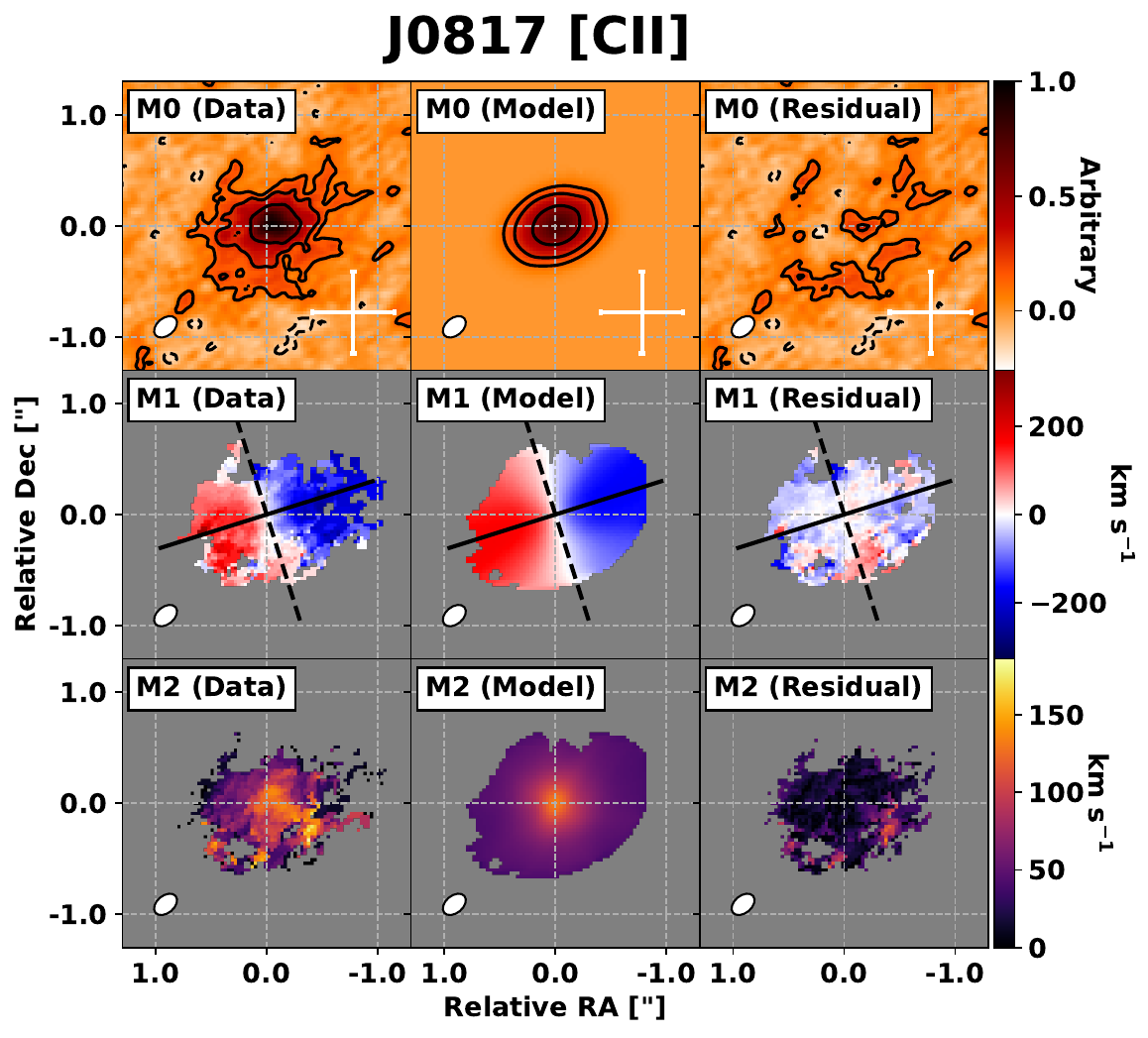}
\\ \hspace{-11.5mm}\includegraphics[width=0.33\textwidth]{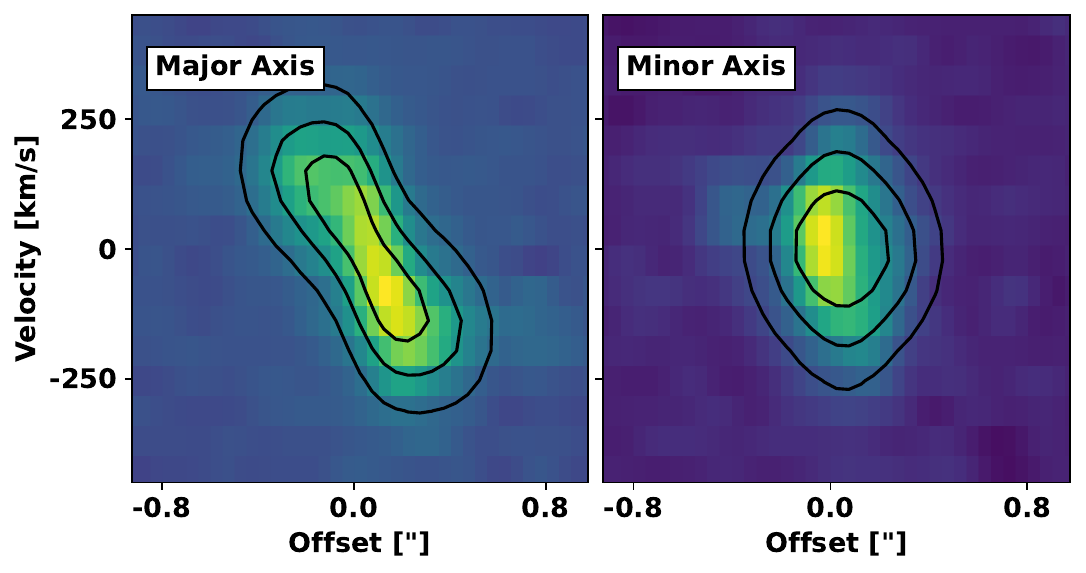}
\hspace{28.5mm}\includegraphics[width=0.33\textwidth]{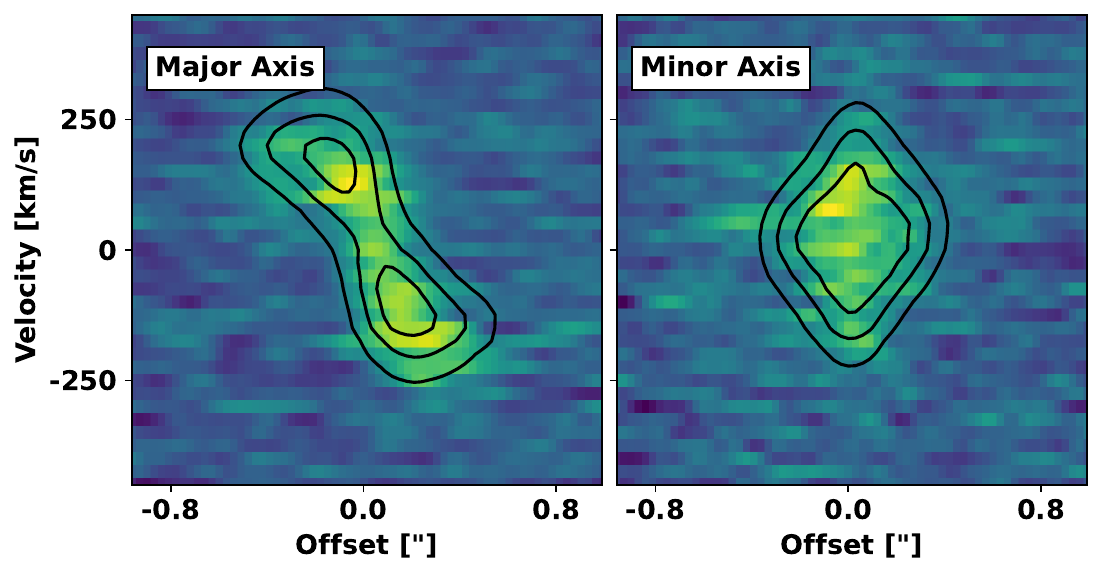}
\caption{Comparison of \BB fits to the \ha (left) and \cii (right) data cubes assuming `AZIM' normalisation. For each, we present the moment 0 (first row), moment 1 (second row), and moment 2 (third row) for the observed data (left column), model (centre column), and the difference between the two (column). The white crosses in the top row show a scale of $\rm5\,kpc\times5\,kpc$. The bottom row includes PVDs extracted from the observed data cube (colours) and model (contours) along the major (lower left) and minor (lower right) axes.}
\label{mommaps_A}
\end{figure*}

\begin{figure*}
\centering
\includegraphics[width=\textwidth]{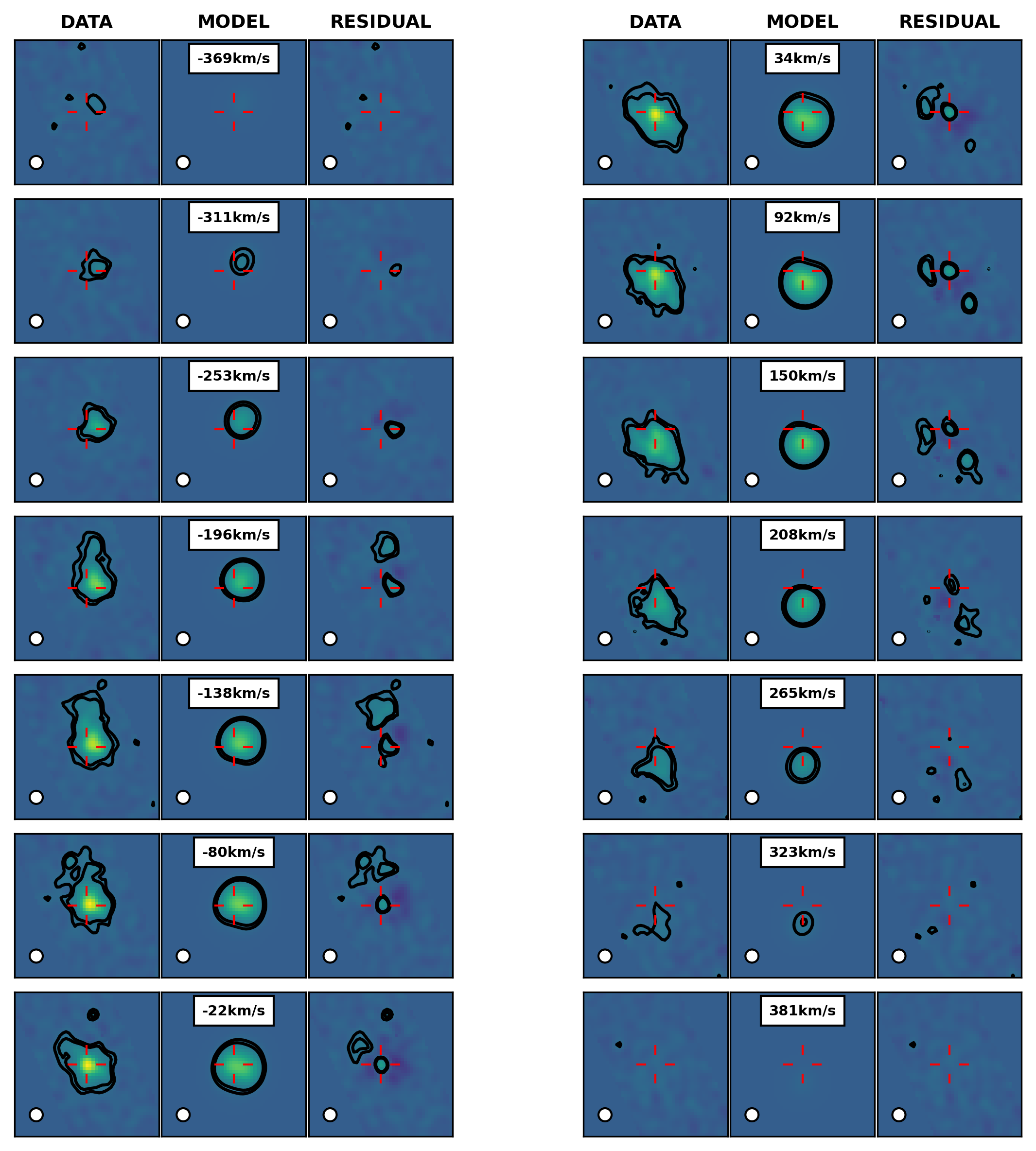}
\caption{Channel maps of the observed \ha data (left columns), the best-fit model (central columns) and the residual (right columns) for all channels used in the \BB fit assuming `AZIM' normalisation. Each panel features the same colourmap limits and contours (3 and $5\times$ the mean channel RMS noise level). The PSF is included as a white ellipse to the lower left, and the velocity of the channel is included in the `MODEL' panel.}
\label{cm_ha_A}
\end{figure*}

\begin{figure*}
\centering
\includegraphics[width=\textwidth]{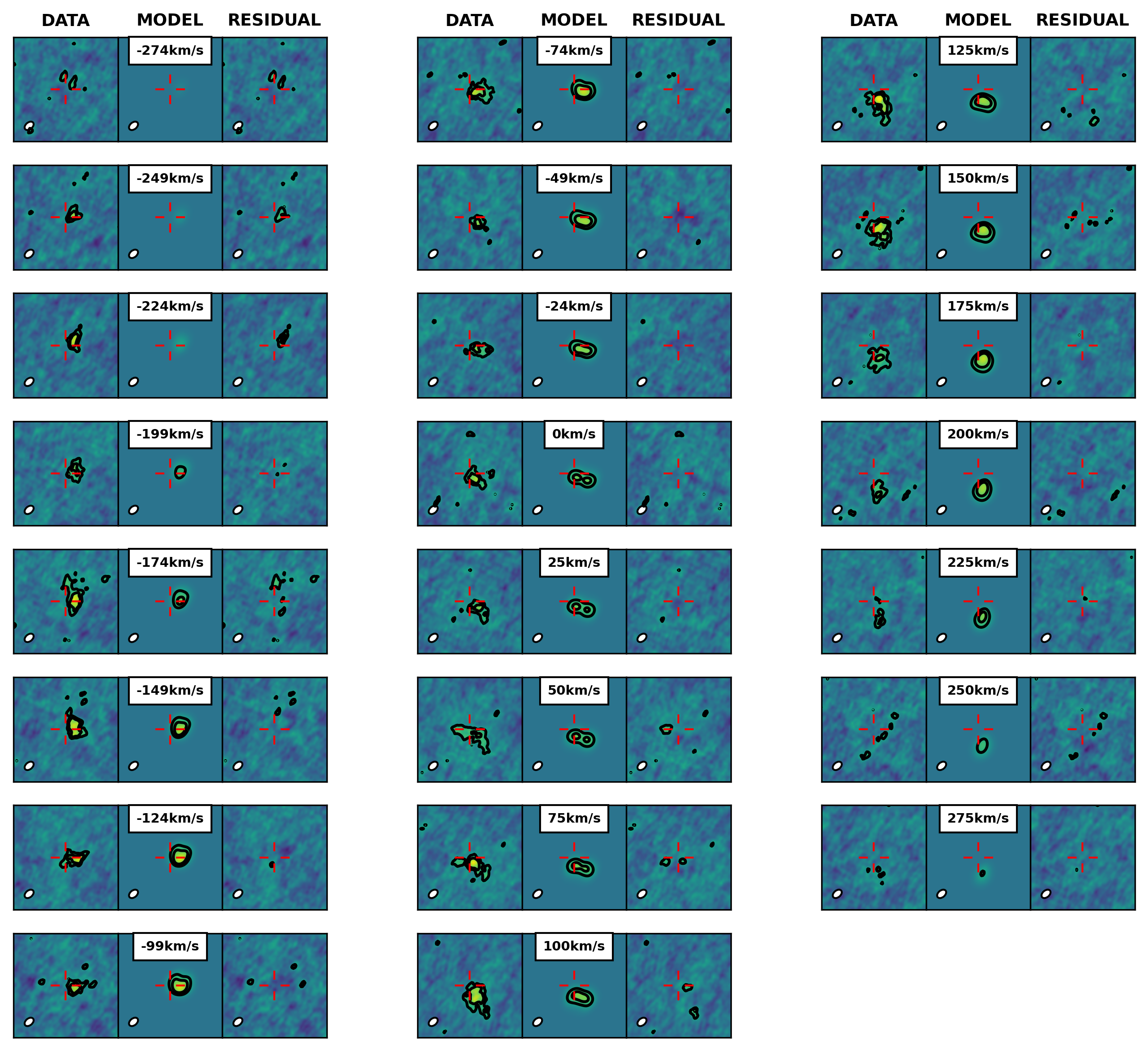}
\caption{Channel maps of the observed \cii data (left columns), the best-fit model (central columns) and the residual (right columns) for all channels used in the \BB fit assuming `AZIM' normalisation. Each panel features the same colourmap limits and contours (3 and $5\times$ the mean channel RMS noise level). The PSF is included as a white ellipse to the lower left, and the velocity of the channel is included in the `MODEL' panel.}
\label{cm_cii_A}
\end{figure*}

While the moment 1 and 2 maps of the AZIM models (Figure \ref{mommaps_A}) are similar to those of the LOCAL models (Figure \ref{mommaps}), the moment 0 maps and PVDs differ. The requirement of azimuthal symmetry results in significant M0 residuals for the \ha fit, revealing that the peaks of the intensity and velocity dispersion are spatially offset. This is also seen in the PVDs for this fit, where the model (contours) is not a perfect match for the data (colours). The channel maps for these data and model (Figure \ref{cm_ha_A}) show this mismatch more clearly, as well as residuals to the north and east. The \cii data is not as strongly affected by the change in normalisation (see low residuals in channel maps of Figure \ref{cm_cii_A}), but the M0 residuals and PVD mismatch are also increased.

The resulting rotational velocity and velocity dispersion profiles are in agreement with the LOCAL model (i.e., within $1\sigma$), resulting in identical estimates of $M_{\rm dyn}$. But this highlights the reality that while \JJ is a relaxed rotating disk galaxy, it does not feature the ideal azimuthal symmetry seen in low-redshift disks. This may partially be explained by variable dust attenuation across the galaxy, which would result in an asymmetric \ha distribution and a more ideal \cii distribution (as observed). But a more likely explanation is this source features extra-planar gas from minor mergers or filaments, as seen in simulations of disk galaxies at $z\gtrsim4$ (e.g, \citealt{pall17,koha19}). Thus, while the underlying kinematics can be disk-like, the total line emission may feature contributions from other sources. But in the case of \JJ, these other contributions are subdominant.


\section{FIR SED constraints}\label{FIRSED}

\JJ has been observed with ALMA in bands 3 and 4 (2023.1.00976.S; PI J. Prochaska), and in band 7 (2015.1.01564.S, 2017.1.01052.S; PI: M. Neeleman). This allows us to examine the dust properties of this source through FIR SED modelling. Here, we perform a preliminary inspection of the available data to gain a rough estimate of the dust mass in \JJ, and defer a more in-depth analysis to future works.

\begin{figure}
    \centering
    \includegraphics[width=0.5\textwidth]{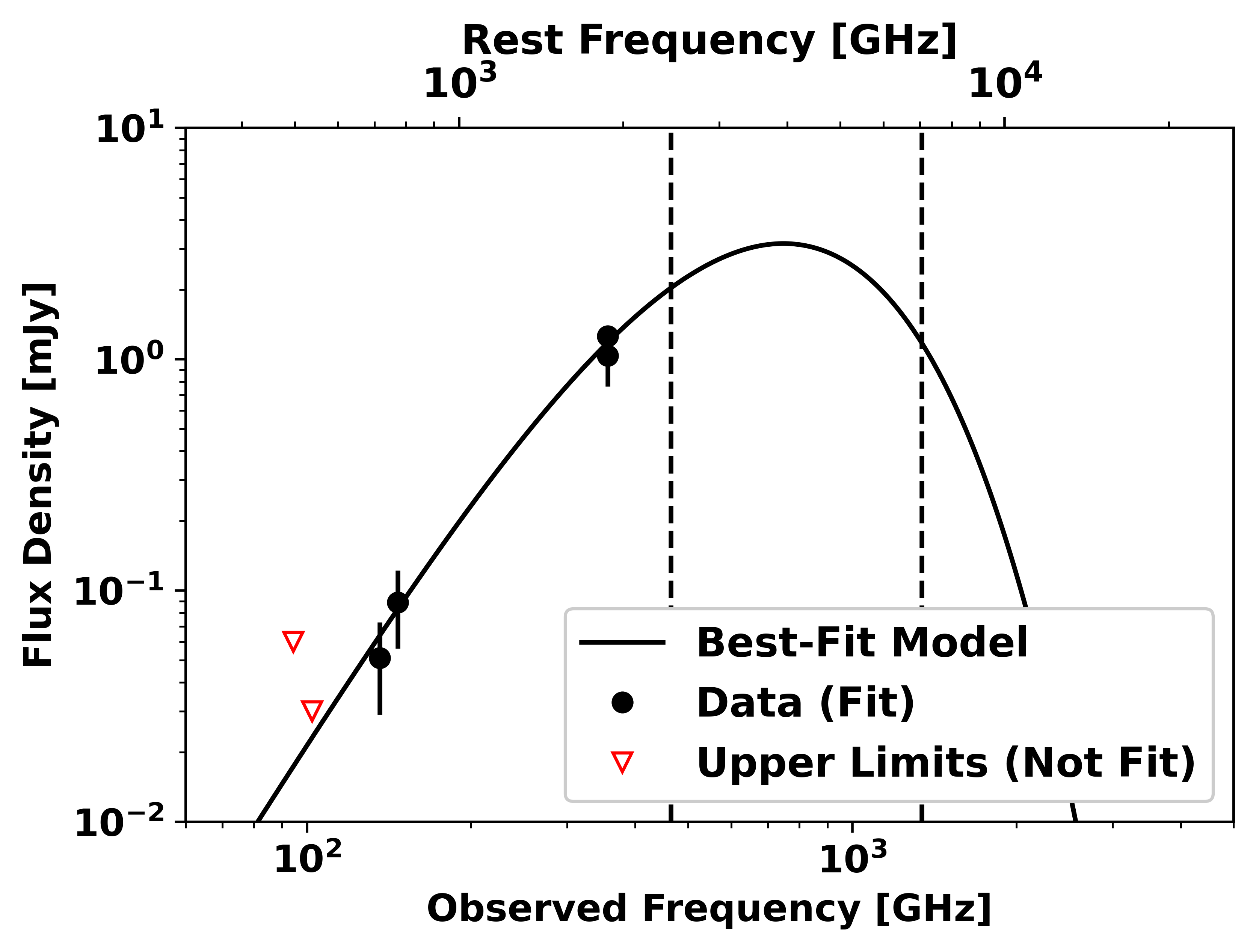}
    \caption{The FIR SED for \JJ, including detections in bands 4 and 7 (black points) and non-detections in band 3 (red triangles). The best fit modified blackbody model (assuming $T_{\rm dust}=40$\,K, $\beta_{\rm IR}=1.8$) is shown in black. The frequency range commonly used to measure $L_{\rm FIR}$ is denoted by the vertical black dashed lines.}
    \label{firsedfig}
\end{figure}

First, we download the primary beam-corrected continuum images for each dataset from the ALMA archive, and measure the flux density of each in CARTA through 2D Gaussian fits. This results in strong detections in bands 4 and 7, and non-detections in band 3. Since the band 3 images feature large synthesized beams ($FWHM\sim2''$), we assume that \JJ is unresolved and estimate upper limits on the flux density of $3\times$ the RMS noise level. The resulting SED is shown in Figure \ref{firsedfig}. 

To analyse this SED, we use a modified blackbody model (see \citealt{jone20} for details) with variable dust temperature ($T_{\rm dust}$), dust emissivity index ($\beta_{\rm IR}$), and dust mass ($M_{\rm dust}$). Due to the low number of SED points, we do not proceed with a full modelling analysis at this time. Instead, we fix $T_{\rm dust}=40$\,K (based on the dust temperature evolution of \citealt{jone23}) and $\beta_{\rm IR}=1.8$ (based on the findings of \citealt{wits23}). Using \textlcsc{ultranest}, we find that the data are best fit with a dust mass of $\log_{10}(M_{\rm dust}/M_{\odot})=7.98\pm0.04$ (Figure \ref{firsedfig}).

While constraints on these properties can be found through a more detailed analysis (i.e., combination of ALMA data in each band to produce optimal continuum images, improved flux density measurements, blackbody model fitting), we use our results to demonstrate that the FIR continuum emission of \JJ requires a substantial dust mass, which lies between SFGs ($\sim10^7$\,M$_{\odot}$; e.g., \citealt{somm22}) and SMGs ($\sim10^{7-9}$\,M$_{\odot}$; e.g., \citealt{wits23}). 

\bsp	
\label{lastpage}
\end{document}